\documentclass[twocolumn,showpacs,preprintnumbers,amsmath,amssymb,prb,floatfix]{revtex4}

\usepackage{graphicx,tabularx}
\usepackage{amssymb}
\usepackage{dcolumn}
\usepackage{color}
\usepackage{verbatim} 
\usepackage[mathcal]{euscript}
\def\bsu{$\beta_{6u}$}

\definecolor{gray0}{gray}{0.0}
\definecolor{gray64}{gray}{0.25}
\definecolor{gray128}{gray}{0.5}
\definecolor{gray192}{gray}{0.75}
\definecolor{gray255}{gray}{1.0}

\begin{document}
\title{Pt-induced nanowires on Ge(001): a DFT study.}
\author{Danny E. P. Vanpoucke}
\affiliation{Computational Materials Science, Faculty of Science
and Technology and MESA+ Institute for Nanotechnology, University
of Twente, P.O. Box 217, 7500 AE Enschede, The Netherlands}
\author{Geert Brocks}
\affiliation{Computational Materials Science, Faculty of Science
and Technology and MESA+ Institute for Nanotechnology, University
of Twente, P.O. Box 217, 7500 AE Enschede, The Netherlands}

\date{\today}
\begin{abstract}
We study formation of the nanowires formed after deposition of Pt on a Ge(001) surface. The nanowires form spontaneously after high temperature annealing. They are thermodynamically stable, only one atom wide and up to a few hundred atoms long. \textit{Ab initio} density functional theory calculations are performed to identify possible structures of the Pt-Ge (001) surface with nanowires on top. A large number of structures is studied. With nanowires that are formed out of Pt or Ge dimers or mixed Pt-Ge dimers. By comparing simulated scanning tunneling microscopy images with experimental ones we model the formation of the nanowires and identify the geometries of the different phases in the formation process. We find that the formation of nanowires on a Pt-Ge(001) surface is a complex process based on increasing the Pt density in the top layers of the Ge(001) surface. Most remarkably we find the nanowires to consist of \emph{germanium} dimers placed in troughs lined by mixed Pt-Ge dimer rows.
\end{abstract}

\pacs{68.35.-p, 73.20.At, 62.23.Hj} 
\maketitle
\section{Introduction}
Ever since Gordon Moore in $1965$ observed the doubling of processing power every year, electronics industry has been driven forward by Moore's law, making it a self-fulfilling prophecy.\cite{fn:Moore,Moore:elec65}
The exponential growth in processing power has been made possible by producing ever smaller devices. This progress, with regard to further miniaturization, is steadily coming up to its ultimate and final limit: devices of atomic sizes connected by atomic wires. Besides this industrial point of view, nanowires (NWs) are also important at the fundamental theoretical and experimental level because of their inherent one-dimensional ($1$D) nature. $1$D electronic systems present exotic physical phenomena such as Peierls instabilities, charge density waves and Luttinger liquid behavior.\cite{Lutt:jmp63,Yeom:prl99,Yao:nat99,Gamb:nat02} Also the effects of the dimensionality on, for example, the magnetic properties can be studied in such systems.\cite{Shen:prb97,DoranD:prl98,Gamb:nat02,Nilius:sc02,Crain:sc05,Lagoute:prb06,LimDK:nano07,Hong:prb07} NWs can be produced in many different ways, resulting in a large range of sizes and properties. This goes from monatomic Au wires created in break junctions to NWs grown on imprinted surfaces, from atomic chains build one atom at a time using a scanning tunneling microscope (STM) tip to atomic Co chains grown on Pt(997) step edges and self-assembled wires and stripes.\cite{Eigler:nat90,Yanson:nat98,Gurlu:apl03,Wang:prb04,Wang:ss05,Barth:nat05,Eames:prb06} Because of its high conductance and general resistance to corrosion and oxidation, Au is a favored metal to use in the creation of atomic wires, both as free standing wires and as reconstructed wires, chains or stripes on semiconductor surfaces.\cite{Yanson:nat98,Wang:prb04,Wang:ss05} Recently metal/Ge structures attracted interest because of the observation of self-assembled NWs in these systems. For instance, the deposition of Ho on Ge(111) results in stripe-like wires, while deposition of submonolayer amounts of Pt or Au on Ge(001) induces the formation of monatomic wires, hundreds of nanometers long.\cite{Eames:prb06,Gurlu:apl03,Schafer:prl2008} In $2003$, G\"url\"u \textit{et al.} created arrays of self-assembled NWs by deposition of a quarter monolayer
of Pt on a Ge(001)-surface and annealing it at $1050$ K.\cite{Gurlu:apl03,Schafer:prb06} These NWs are defect and kink free, and only one atom wide. Their length is only limited by the underlying Pt reconstructed Ge(001)-surface. This reconstructed surface contains $0.25$ monolayer (ML) of Pt in its top layer forming a checkerboard pattern of Ge-Ge and Pt-Ge dimers, as we have shown previously.\cite{vanpoucke:prb2008R,Vanpoucke:prb09beta}\\
\indent In this work we present possible geometries of the different phases during the NW formation on a Pt modified Ge surface, called the  $\beta$-terrace.\cite{Gurlu:apl03,Vanpoucke:prb09beta} Based on comparison of simulated STM images to experimental STM images we will show that the NWs consist of \emph{germanium atoms} in a trough lined with Pt atoms. Furthermore, we will propose geometries for the experimentally observed widened trough (WT), a precursor to the NW formation, and
possible structural evolution of the WT to a NW.\\
\indent This paper is organized as follows: In Sec.~\ref{sc:theormeth} we describe the theoretical methods used for the total energy calculations and the generation of calculated STM images (which we call pseudo-STM images). In Sec.~\ref{sc:results} we present our results for different surface models with increasing Pt density. In Sec.~\ref{sc:discussion} we propose possible formation paths linking the calculated geometries to the experimentally observed structures. We examine the NWs and WT geometries in more detail, and compare them to the models present in literature. Finally, in Sec.~\ref{sc:conclusion} the conclusions are presented.
\section{Theoretical method}\label{sc:theormeth}
Density functional theory (DFT) calculations are carried out using the projector augmented waves (PAW) method and the Ceperley-Alder local density approximation (LDA) functional, as implemented in the VASP program.\cite{Blochl:prb94,Kresse:prb99,Kresse:prb93,Kresse:prb96} A plane wave basis set with kinetic energy cutoff of $345$ eV is applied. The surface is modeled by periodically repeated slabs of $12$ layers of Ge atoms. The slabs are mirrored in the $z$-direction with reconstructions on both surfaces, in which specific Ge atoms are replaced by Pt atoms. A vacuum region of $\sim15.5$ \AA\ is used to separate the slabs along the $z$-axis. Extensive convergence tests showed no advantage in time/accuracy for an asymmetric Ge slab with one H-passivated surface over the symmetric Ge slab. Based on the advantages of the higher symmetry of the symmetric slab, such as for example the presence of only a single type of surface, we choose to use the later. The Brillouin zone (BZ) of the $(2\times4)$ surface unit cell is sampled using a $8\times4$ Monkhorst-Pack special $k$-point mesh.\cite{Monkhorst:prb76} To optimize the geometry of the slabs we use the conjugate gradient method and keep the positions of the Ge atoms in the center two layers fixed as to represent bulk Ge.\\
\indent Pseudo-STM images are calculated using the Tersoff-Hamann method, which states that the tunneling current in an STM experiment is proportional to the local density of states (LDOS) integrated from the Fermi level to the bias.\cite{Tersoff:prb85} It was implemented in its most basic formulation, approximating the STM tip by an infinitely small point source. The integrated LDOS is calculated as $\overline{\rho}(\mathbf{r},\varepsilon) \propto \int_{\varepsilon}^{\varepsilon_{\mathrm{F}}} \rho(\mathbf{r},\varepsilon^{\prime})\mathrm{d}\varepsilon^{\prime}$, with $\varepsilon_{\mathrm{F}}$ the Fermi energy. An STM in constant current mode follows a surface of constant current, which translates into a surface of constant integrated LDOS $\overline{\rho}(x,y,z,\varepsilon)=\mathrm{C}$, with C a real constant. For each C, this construction returns a height $z$ as a function of the position $(x,y)$. This heightmap is then mapped linearly onto a gray scale. We choose C such that the isosurface has a height $z$ between $2$ and $4$ \AA\ above the highest atom of the surface.
\section{Results}\label{sc:results}
\subsection[Experimental background]{The experimental background}\label{ssc:NWRes_IntroExp}
In this section we present the results of our calculations on the NWs for a variety of different surface structures. A discussion of the emerging models for the NWs and their integration in possible formation paths is the subject of Sec.~\ref{sc:discussion}.We will start this subsection by summarizing the relevant experimental data.\\
\ \\
\indent The formation of Pt induced NWs was first observed by G\"url\"u \textit{et al.} in $2003$ and later by Sch\"afer \textit{et al.}\cite{Gurlu:apl03,Schafer:prb06} After deposition of $0.25$ ML of Pt on a clean Ge(001) surface, and subsequent annealing at $1050$ K, G\"url\"u \textit{et al.} observed the formation of one atom thick NWs and hundreds of nanometers long. These wires are defect and kink free, and are observed both as solitary wires and arrays of wires with a constant spacing of $1.6$ nm. The NWs are located in the troughs between the QDRs of a Pt modified Ge(001) surface. The $1.6$ nm spacing in the arrays means that only in every second trough a NW can be present. The Zandvliet group also discovered that although the NW-arrays are conducting the conduction is likely caused by surface states confined between the wires, rather than by the wires themselves.\cite{Oncel:prl05,Vriesde:apl2008}\\
\begin{figure}[!tbp]
\begin{center}
  \includegraphics[width=8.5cm,keepaspectratio]{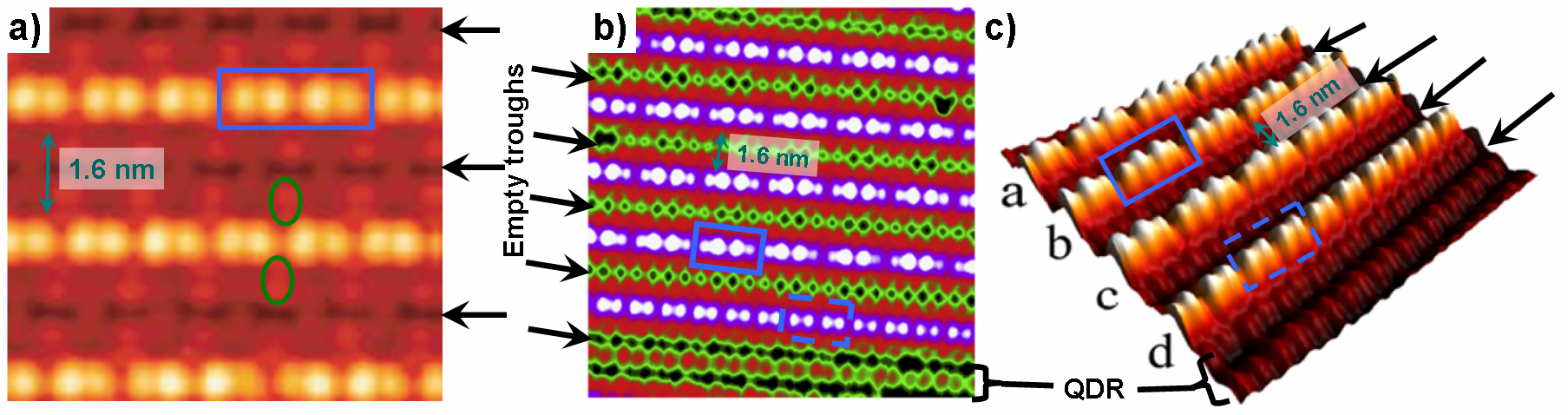}\\
\end{center}
    \caption{(color online) Three typical experimental STM images.\cite{Oncel:prl05,Houselt:ss08,Zandvliet:privComm}
    a) An empty state STM image of several NWs at $77$ K using a bias of $0.15$ V and a tunnel current of $0.437$ nA.\cite{Oncel:prl05} b) and c) $2$D and $3$D filled state STM images recorded at $4.7$ K, using a bias voltage of $-1.5$ V and a tunneling current of $0.5$ nA.\cite{Houselt:ss08} The ellipses indicate the symmetric bulges, while the rectangles indicate two NW dimers showing $(4\times1)$ periodicity. The dashed rectangles show the $(2\times1)$ periodicity for edge NWs. Empty troughs are indicated with the arrows and have a width of $1.6$ nm. In (b) and (c) also a QDR of the $\beta$-terrace is visible.
  }\label{fig:5_c_ExpNWImg}
\end{figure}
\indent Figure \ref{fig:5_c_ExpNWImg} shows some experimental STM images of the NWs. We indicate typical features that should be present in a calculated STM image. i) The NW image should be dimerised (solid and dashed rectangles). ii) The NW dimer images should be doubly peaked for the filled state images, and for the empty state images at small bias, see Fig.~\ref{fig:5_c_ExpNWImg}a. The double peaks should merge into one peak for the empty state images at a large bias. iii) Bulges, symmetric to both sides of the NW should be present, and should be located between the NW dimer images (ellipses). iv) A NW image should be present only  every second trough. And v) the NW dimer images in an array should have a $(4\times1)$ periodicity along the NW with two NW dimers per unit cell.\\
\indent The last feature, the $(4\times1)$ periodicity, has been linked to the presence of a Peierls instability by van Houselt \textit{et al.}\cite{Houselt:ss08} Since a Peierls instability is a subtle low temperature effect we neglect the $(4\times1)$ periodicity initially and start from a smaller unit cell, containing only a single NW dimer. In the following sections we will present a series of surface models and adsorption geometries with increasing Pt density in the top layers. This Pt density will in general refer to the amount of Pt present in the substrate, the atoms of the NW will be mentioned separately.\\
\indent In Sec.~\ref{ssc:model1_onb6u}, we start from a clean Ge(001) surface, or a Pt modified Ge(001) surface, the $\beta$-terrace,\cite{Gurlu:apl03,Vanpoucke:prb09beta} and consider NWs of Pt atoms.
Various adsorption geometries for Pt adatoms and addimers are studied and it will be shown that none of the calculated STM images for these structures resemble the experimental STM images at all. Two important observations will be made: the preference of Pt to build into the substrate forming Pt-Ge surface dimers, and the fact that Ge atoms bound to Pt atoms show up brightly in the calculated STM images, in contrast to the Pt atoms which show up very dark.\\
\indent In Sec.~\ref{ssc:model1_5_intermediate} we investigate the exchange between Pt and Ge atoms in the substrate and the incorporation of Pt in the in the  substrate. This leads to models of the substrate with $0.5$ ML of Pt in the top layers. The adsorption of both Pt and Ge dimers on these substrates is investigated in Sec.~\ref{ssc:model2_onb4as}, and it is shown that Pt NWs are invisible, while Ge dimers present bright NW dimer images. Also the symmetric bulges are identified. Although the model presented in Sec.~\ref{ssc:model2_onb4as} shows reasonable agreement with the experiment, the double peaked NW dimer images in the filled state pictures are missing. The observation that Ge dimers bound on top of Pt dimers produce such double peaked images leads to a third substrate model in Sec.~\ref{ssc:model3_onb4as_ptL3}. In this model $0.75$ ML of Pt is present in the top layers. Again the adsorption of both Pt and Ge NWs is investigated, and it is shown that a Ge NW on this substrate model shows excellent agreement with the experimental STM images, presenting all features  with the exception of the $(4\times1)$ periodicity due to the restricted size of the unit cell. In Sec.~\ref{ssc:model4_npu} we investigate the experimentally observed $(4\times1)$ periodicity. We do not observe a Peierls distortion. However, by placing additional atoms in the trough of the NW, we will show that a surface containing $0.8125$ ML of Pt can generate the experimentally observed $(4\times1)$ periodicity, and thus contains all typically observed features of the NWs.
\subsection[First simple model]{Pt-adatoms and
wires on the Ge(001) surface and the $\beta$-terrace}\label{ssc:model1_onb6u}
\begin{figure}[!tbp]
\begin{center}
  \includegraphics[width=5cm,keepaspectratio]{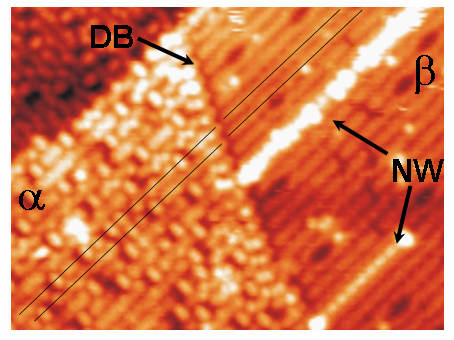}\\
\end{center}
    \caption{(color online) The DB between an $\alpha$- and
    $\beta$-terrace.\cite{Zandvliet:privComm} Although many vacancies are
    present in the $\alpha$-terrace, the dimer rows can still be
    recognized. Lines show two neighboring QDRs on the $\beta$-terrace, and
    two neighboring dimer rows on the $\alpha$-terrace.
  }\label{fig:5_b_NWdomainBoundary}
\end{figure}
Experimental STM images of the Pt modified Ge surface with NWs show the NWs to be build of dimers positioned in the troughs between the QDRs of the $\beta$-terrace (\textit{cf.}\ Fig.~\ref{fig:5_c_ExpNWImg}).\cite{Gurlu:apl03} Figure~\ref{fig:5_b_NWdomainBoundary} shows an experimental STM image of the domain boundary (DB) between an $\alpha$ and a $\beta$-terrace. The latter is a Pt modified surface with $0.25$ ML Pt incorporated in the top surface layer, whereas the $\alpha$-terrace is thought to be essentially a Ge(001) surface containing a substantial amount of defects. The QDRs of the $\beta$-terrace show perfect alignment with the dimer rows of the $\alpha$-terrace, so the dimer row structure of the Ge(001) surface is maintained in the $\beta$-terrace. Based on this observation, we assume that the formation of the $\beta$-terrace and the NWs does not modify the dimer row periodicity of the $\beta$-terrace (or Ge(001)) reconstruction dramatically, except for the possible substitution of atomic species. This means that the NWs should be positioned at the center of the trough between the QDRs. Furthermore, experimental observations also show that Pt atoms deposited at room temperature initially move to the subsurface, which is shown theoretically to be an energetically favorable situation.\cite{Vanpoucke:prb09beta} After high temperature annealing the Pt atoms are assumed to pop up
again from the bulk as Pt dimers.\cite{Fischer:prb07}\\
\indent Since the NWs are inherently connected to the $\beta$-terrace, this geometry is a good starting point for a simple model of the NWs. Based on the experimental observations two scenarios can be envisaged. In a first possible scenario the top layer of the $\beta$-terrace remains as it is, and extra subsurface Pt atoms are ejected onto the surface to form the NWs. In the experiment roughly $0.25$ ML of Pt is deposited. After annealing the surface shows different terraces with different reconstructions. So we can easily assume that these patches contain different platinum densities, such that the local platinum density can be more than $0.25$ ML.
Since the $\beta$-terrace appears before the NWs during the NW formation process, a second scenario consists of the ejection onto the surface of the Pt atoms from the $\beta$-terrace.These ejected Pt atoms then form Pt dimers on the surface, while Ge atoms from the bulk take their positions in the dimer rows. This way a Ge(001) surface is formed with Pt ad-dimers. We use the Ge(001) b($2\times1$) geometry for the clean Ge(001) surface and the $\beta_{6u}$-geometry, we found for the $\beta$-terrace,\cite{Vanpoucke:prb09beta} and place Pt atoms on specific adsorption sites on the dimer rows and in the troughs of these surfaces. In the following we will refer to the $\beta_{6u}$-geometry simply as $\beta_{6}$-geometry.
\begin{figure}[!tbp]
\begin{center}
  \includegraphics[width=8cm,keepaspectratio]{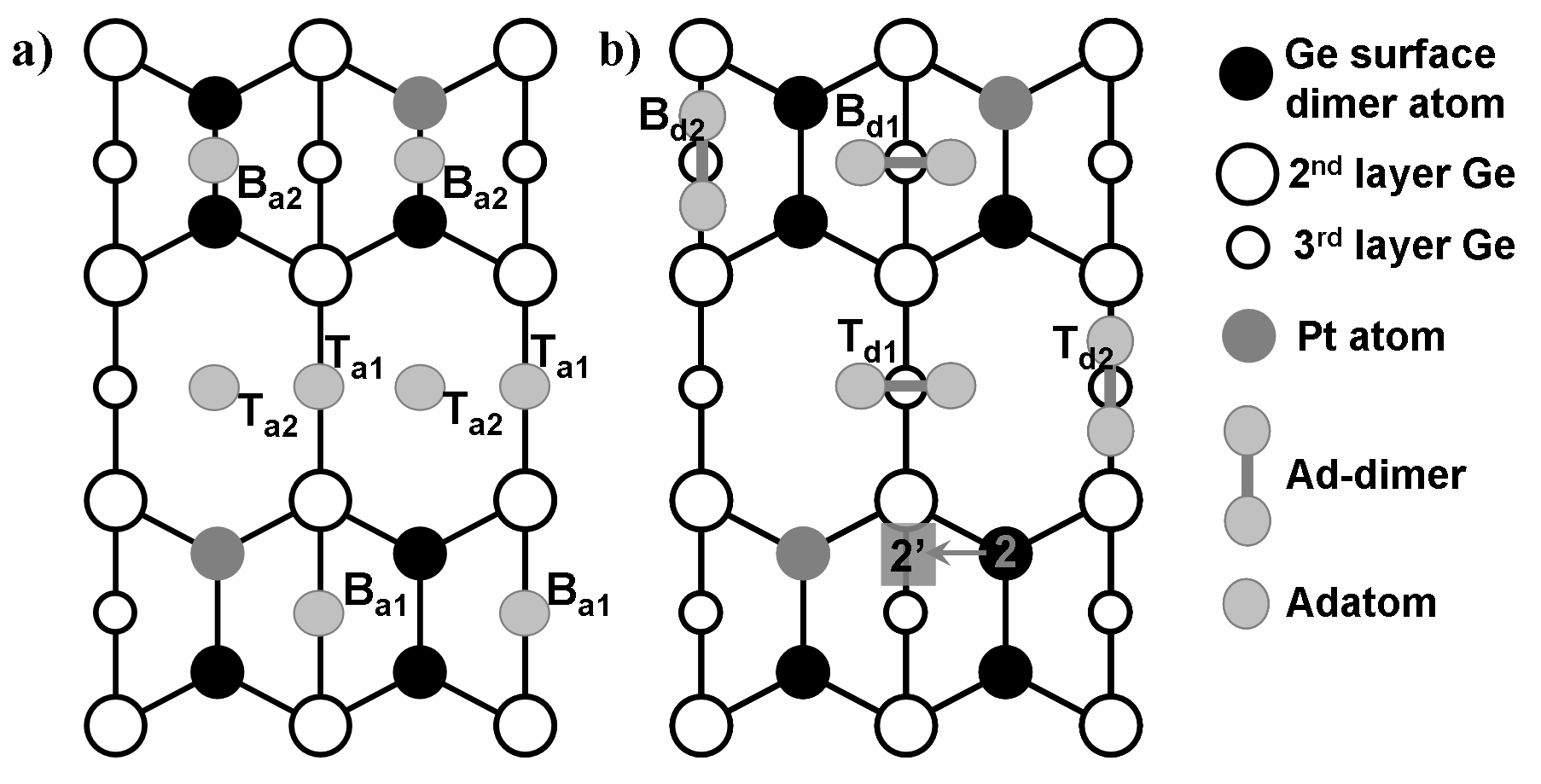}\\
\end{center}
  \caption{Adsorption sites before relaxation.\newline
  a) Adsorption sites for Pt adatoms on the
  $\beta_{6}$-geometry.\cite{Vanpoucke:prb09beta} b) Adsorption sites of Pt ad-dimers on the $\beta_{6}$ (shown here) and
  Ge(001) $b($2$\times$1$)$ surface geometry.}\label{fig:1geomonbeta6andge}
\end{figure}
Figure~\ref{fig:1geomonbeta6andge}a shows the initial positions of Pt ad-atoms on the $\beta_{6}$-geometry and Fig.~\ref{fig:1geomonbeta6andge}b of ad-dimers on the $\beta_{6}$- and Ge(001)-surface. After relaxation we calculated formation energies for these geometries, see Table~\ref{table:1formEfirstset}.
\begin{table}[!t] \center{\textbf{Formation and adsorption
energy.}\\}
\begin{ruledtabular}
\begin{tabular}{l|rlrlrl}
   & \multicolumn{2}{c}{$E_{\mathrm{f}}$}
   & \multicolumn{2}{c}{$E_{\mathrm{ad}}$}
   & \multicolumn{2}{c}{$r_{\mathrm{ad}}$}\\
   & \multicolumn{2}{c}{(meV)} & \multicolumn{2}{c}{(meV)} &
   \multicolumn{2}{c}{(\AA)} \\
  \hline
  $\beta_{6}$ & $-74$& & - & &
  - & \\
  $\beta_{6}$ B$_{\mathrm{a}1}$ & $-473$ & & $-200$ & & 2.957 & \\
  $\beta_{6}$ B$_{\mathrm{a}2}$ & 3128 & & 1601 & & 4.019 & \\
  $\beta_{6}$ T$_{\mathrm{a}1}$ & 658 & & 366 & & 2.661 & \\
  $\beta_{6}$ T$_{\mathrm{a}2}$ & 161 & & 117 & & 4.146 & \\
  \hline
  $\beta_{6}$ B$_{\mathrm{d}1}$ & 1776 & & 925 & & 2.552 & \\
  $\beta_{6}$ B$_{\mathrm{d}2}$ & 120 & & 97 & & 3.421 & \\
  $\beta_{6}$ T$_{\mathrm{d}1}$ & $-1405$ & & $-665$ & & 4.001 & \\
  $\beta_{6}$ T$_{\mathrm{d}2}$ & 1303 & & 688 &  & 2.623 & \\
  \hline
  $\mathrm{Ge\ b}(2\times1)\ \mathrm{B_{d1}}$ & 2077 & & 1038 & & 2.655 &  \\
  $\mathrm{Ge\ b}(2\times1)\ \mathrm{B_{d2}}$ & 2262 & & 1131 & & 2.538 & \\
  $\mathrm{Ge\ b}(2\times1)\ \mathrm{T_{d1}}$ & 359 & & 180 & & 3.980 & \\
  $\mathrm{Ge\ b}(2\times1)\ \mathrm{T_{d2}}$ & 176 & & 88 & & 2.851 & \\
\end{tabular}
\end{ruledtabular}
\caption{Formation and adsorption energies for adsorption of adatoms and addimers on the $\beta_{6}$ and Ge(001) $b(2\times1)$ surface reconstructions shown in Fig.~\ref{fig:1geomonbeta6andge}. $r_{\mathrm{ad}}$ is the distance between the adatoms after relaxation, which indicates whether bonds are formed between the adsorbed atoms.}\label{table:1formEfirstset}
\end{table}
The surface formation energy, $E_{\mathrm{f}}$, was calculated with regard to the Ge(001) b($2\times1$) geometry using the expression:
\begin{equation}
E_{\mathrm{f}} = \big(E_{\mathrm{NW}} - E_{\mathrm{Ge(001)}} - \Delta \mathrm{N}_{\mathrm{Pt}}\mathrm{E}_{\mathrm{Pt}} - \Delta \mathrm{N}_{\mathrm{Ge}}\mathrm{E}_{\mathrm{Ge}}\big)/2,
\end{equation}
with $E_{\mathrm{NW}}$ the total energy of the relaxed structure, $E_{\mathrm{Ge(001)}}$ the total energy of the Ge(001) b($2\times1$) surface reconstruction, $\mathrm{E}_{\mathrm{Pt}}$ and $\mathrm{E}_{\mathrm{Ge}}$ the bulk energy for a Pt and a Ge atom respectively and $\Delta \mathrm{N}_{\mathrm{Pt}}$($\Delta \mathrm{N}_{\mathrm{Ge}}$) the difference in number of Pt(Ge) atoms between the relaxed structure and the Ge(001) b($2\times1$) geometry. The division by $2$ is because of the symmetry of the system, giving two modified surfaces on the slab. Negative values of $E_{\mathrm{f}}$ indicate a stable structure compared to segregation in a Ge(001) surface and Pt bulk. Since different substrate reconstructions contribute differently to this formation energy, we want to separate the  contribution to the formation energy of the substrate from that of the NW. This would allow us to compare the stability of the NW or other adstructures on different surfaces. For this purpose we introduce an adsorption energy, $E_{\mathrm{ad}}$. It is calculated using the expression,
\begin{equation}
E_{\mathrm{ad}} = \big(E_{\mathrm{NW}} - E_{\mathrm{subs}} - \Delta \mathrm{N}_{\mathrm{Pt}}\mathrm{E}_{\mathrm{Pt}} - \Delta \mathrm{N}_{\mathrm{Ge}}\mathrm{E}_{\mathrm{Ge}}\big)/\mathrm{N}_{\mathrm{adatom}},
\end{equation}
where $E_{\mathrm{subs}}$ is the total energy of the surface without adstructure and without reconstructions induced by the adstructure, $\Delta\mathrm{N_{Ge}}$ ($\Delta\mathrm{N_{Pt}}$) the difference in number of Ge (Pt) atoms between the system containing the adstructure and the substrate system without the adstructure, and $\mathrm{N}_{\mathrm{adatom}}$ is the number of adatoms used for the adstructure. Negative values of $E_{\mathrm{ad}}$ indicate that the adatoms
form a stable structure on the surface. Although the relation between the values of $E_{\mathrm{f}}$ and $E_{\mathrm{ad}}$ is quite trivial in case of Table~\ref{table:1formEfirstset}, this will not be the case in the following sections. The values presented here are to allow easy comparison later on.\\
\indent In the following we will discuss the adsorption geometries shown in Fig.~\ref{fig:1geomonbeta6andge}. Starting with the first scenario, Table~\ref{table:1formEfirstset} shows that Pt atoms on a $\beta_{6}$-surface in general create an unstable structure. Only the B$_{\mathrm{a}1}$- and T$_{\mathrm{d}1}$-geometry relax into a stable structure. In the B$_{\mathrm{a}1}$- and T$_{\mathrm{a}1}$-geometry the Pt adatoms move toward one another forming dimers with a length of $2.957$ \AA\ and $2.661$ \AA\ respectively. In case of the B$_{\mathrm{a}1}$-structure the stability of the structure is improved because the Pt dimer breaks up the surface Ge dimer (this bond is weaker than the Pt-Ge dimer bond), which allows the Pt adatoms to sink into the subsurface where they each can create four Pt-Ge bonds, which are energetically favorable over Pt-Pt and Ge-Ge bonds. The Pt dimer in the T$_{\mathrm{a}1}$ configuration, on the other hand, sinks into the trough, increasing the coordination number of the Pt adatoms to three.\\
\indent The B$_{\mathrm{a}2}$ geometry is the most unstable geometry. After relaxation, the Pt atoms are still located on top of the Pt-Ge dimer and the Ge-Ge dimer putting them \mbox{$\sim2$ \AA\ }above the dimer row (\textit{cf.}\ Fig.~\ref{fig:1geomonbeta6andge}). The bond length of the three Pt-Ge bonds of the Pt adatoms is about $2.3$ \AA\ while the bond length between the Pt adatom and the Pt atom in the surface is $2.5$ \AA. This gives the impression that the Pt adatom on top of the Pt-Ge dimer is slightly tilted toward the Ge atom while on the Ge-Ge dimer the Pt adatom is centered. The Ge-Ge dimers on which adatoms are located are almost flat, with a tilt angle of $<1^{\circ}$, much less than the usual tilt angle of $\sim20^{\circ}$ found for Ge-Ge dimers on a Ge(001) c$(4\times2)$ or $\beta_6$ surface reconstruction.\\
\indent The adatoms in the T$_{\mathrm{a}2}$-geometry sink deep into the trough creating five Pt-Ge bonds per adatom. In spite of the large number of Pt-Ge bonds of the adatoms, this structure turns out to be unstable (\textit{cf.}\ Table~\ref{table:1formEfirstset}) because of the
distortion of the lattice in the surface layers.\\
\indent When Pt dimers are placed on the $\beta_{6}$-geometry we see in Table~\ref{table:1formEfirstset} that only the T$_{\mathrm{d}1}$-structure is stable. The most unstable structures are B$_{\mathrm{d}1}$ and T$_{\mathrm{d}2}$. After relaxation the B$_{\mathrm{d}1}$-structure is very similar to the B$_{\mathrm{a}2}$-structure, but in this case the Pt adatom on top of the Ge-Ge dimer is tilted toward the second Pt adatom to form $2$ extra Pt-Pt bonds, one with the second adatom and one with the surface Pt atom, explaining the increase in stability when going from the B$_{\mathrm{a}2}$- to the B$_{\mathrm{d}1}$-structure. The Pt ad-dimer in the B$_{\mathrm{d}1}$-structure is stretched by a small amount to a length of $2.552$ \AA, while it is tilted over an angle of $16.5^{\circ}$. After relaxation, the T$_{\mathrm{d}2}$-structure shows a Pt dimer in a bridge position. This dimer has a tilt angle of roughly $1^{\circ}$ and a length of $2.62$ \AA. Due to the Pt-Pt bond on one side of the trough the dimer makes an angle of $8.7^{\circ}$ with the direction perpendicular to the dimer row. Each adatom also forms an extra Pt-Ge bond with a second-layer Ge atom, while breaking a Ge-Ge bond between the first and second layer.\\
\indent In the B$_{\mathrm{d}2}$-geometry, the Pt dimer breaks up during relaxation. The adatoms cause drastic deformations of the quasi-dimer row (QDR) to form up to five Pt-Ge bonds. Both backbonds of one of the Ge atoms of the Ge-Ge surface-dimer are broken and the atom is pushed up, while a Pt adatom takes its place. Now one side of what previously was a QDR has become a Pt-Ge zigzag-chain, orthogonal to the surface plane, with the
Ge atoms a the high points and the Pt atoms at the low points.\\
\indent The most stable structure is the T$_{\mathrm{d}1}$-structure. Although the initial geometry is very simple (\textit{cf.}\ Fig.~\ref{fig:1geomonbeta6andge}b), reconstructions during relaxation cause large deformations, and the resulting calculated STM images do not resemble the experimental STM images at all (\textit{cf.}\ Fig.~\ref{fig:22_STM_Td1}b). Again, we notice that the Pt atoms try to move into subsurface positions. The ad-dimer breaks up and one of the Pt atoms replaces the Ge atom at position $2$ (\textit{cf.}\ Fig.~\ref{fig:1geomonbeta6andge}b) pushing it to position $2^{\prime}$ on top of the surface. This ejected Ge atom, indicated by a yellow disc in Fig.~\ref{fig:22_STM_Td1}b, gives rise to a large round bulge, in the calculated STM images. This exchange also causes the QDR to transform into a dimer row containing Pt atoms at one side and Ge atoms at the other side, what we previously called a $\gamma_1$-geometry.\cite{Vanpoucke:prb09beta}
This $\gamma_1$-geometry is the most stable surface reconstruction containing $0.5$ ML of Pt atoms in the top layer, explaining the observed improved stability for the T$_{\mathrm{d}1}$ structure.
From this, we draw an important conclusion for further surface models: the exchange of a Ge atom for a Pt atom in a surface Ge-Ge dimer greatly improves the stability of the structure by creating a straight row of Pt atoms. This is also probably the reason why the B$_{\mathrm{d}2}$-geometry, where also a straight row of Pt atoms is created, is so much more stable than the B$_{\mathrm{d}1}$-geometry.\\
\indent None of the structures discussed up to this point result in a calculated STM image that resembles the experimentally observed NWs. Because the $\beta$-terrace contains $0.25$ ML of Pt, and in STM experiments the NWs show a structure containing $0.25$ ML of adatoms, the second scenario assumes the Pt atoms of the $\beta$-terrace are ejected onto the surface and form the NWs. The surface itself would then reconstruct to a normal Ge(001) surface using bulk Ge atoms. Calculations for Pt NWs on a clean Ge b($2\times1$) surface were carried out, to check this second scenario. The adsorption geometries considered are shown in Fig.~\ref{fig:1geomonbeta6andge}b and the formation energies are listed in Table~\ref{table:1formEfirstset}.\cite{fn:Beta4Ge} The formation energies clearly show the known trend for
Pt on Ge.\cite{Niranjan:prb07} The further the Pt atoms stick out of the surface the more unstable the structure turns out to be, again indicating a preference of Pt atoms to move into the surface.\\
\indent Pt dimers on a Ge surface are in general unstable, and when placed in the trough they break up and sink in, increasing their coordination and increasing the stability of the system by roughly $2$ eV per dimer. In this process Ge atoms from the top most layers are pushed up and stick out of the
surface higher than before.\\
\begin{figure}[!tbp]
\begin{center}
  \includegraphics[width=8cm,keepaspectratio]{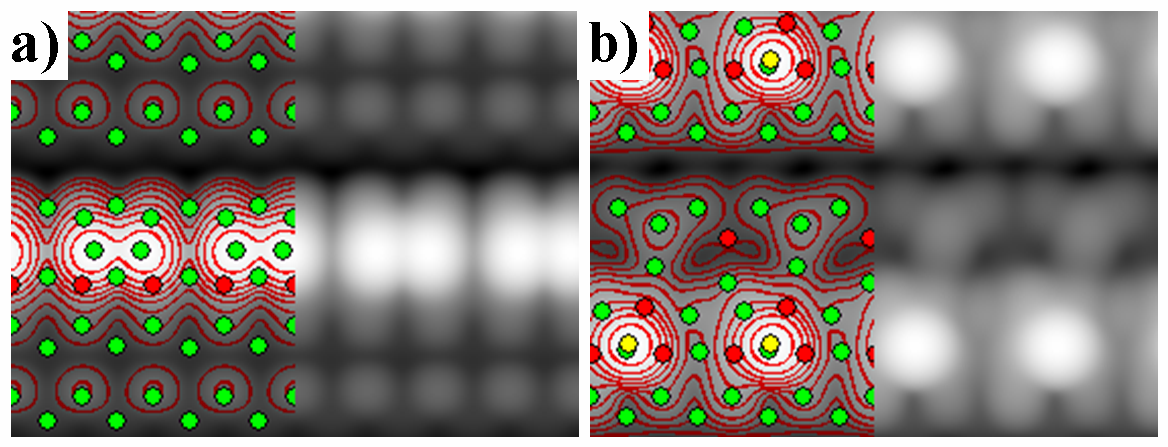}\\
\end{center}
  \caption{(color online) Calculated STM images for a Pt dimer at the T$_{\mathrm{d1}}$ site of the Ge(001) (a) and $\beta_6$ (b) geometries, after relaxation. For these filled state images a simulated bias of $-1.5$ V, and $z=3.0$ \AA\ is used. Green/red (light/dark gray) discs indicate the locations of the Ge/Pt atoms in the two top layers. The yellow (white) discs in (b) indicate the position of the ejected Ge atom at position $2^{\prime}$ (\textit{cf.}\ Fig.~\ref{fig:1geomonbeta6andge}). Contours, separated $0.3$ \AA, are added to guide the eye.
  }\label{fig:22_STM_Td1}
\end{figure}
\indent Although none of these structures turn out to be stable, the optimized T$_{\mathrm{d}1}$ adsorption structure on the Ge(001) substrate gives a hint of where to look for the NWs. Although this structure is not thermodynamical stable in comparison to the $\beta$-terrace, calculated STM images show a remarkable bright chain of double peaked dimer images, as can be seen in Fig.~\ref{fig:22_STM_Td1}a. These NWs consist of Ge dimers, but their STM images are similar to the experimental images of NWs. There are some problems though. First of all the NW is not located in the trough between the QDR, as observed experimentally, but rather on top of one.
The Ge dimers forming the NW were originally the bottom side atoms (in Fig.~\ref{fig:22_STM_Td1}a) of the Ge dimer row, causing the NW to be located at the extension of a Ge dimer row (or a QDR on the $\beta$-terrace) in contradiction to the experimental observation.\cite{Gurlu:apl03,Fischer:prb07} 
Also, the symmetric bulges at the sides of the NW are missing and when moving to a positive bias the double peak should disappear,\cite{Houselt:thesis}
which does not happen for the calculated STM images of this structure. Here the double peak feature remains very pronounced. Although this structure can be discarded as geometrical structure for the observed NWs on grounds of their formation energy and comparison of the STM images, the calculated STM images show us an important fact: Ge dimers should be considered as possible building blocks for the experimentally observed NWs.\\
\indent It is interesting to note that the T$_{\mathrm{d}1}$ structure shows, after relaxation, roughly the same geometry Stekolnikov \textit{et al.}\cite{stekolnikov:prl08} propose as NW geometry. In Sec~\ref{ssc:compar} we will address this point in more detail.
\subsection[Intermediate geometries]{Nanowires on modified $\beta$-reconstructions} \label{ssc:model1_5_intermediate} In the previous paragraph it was shown that the adsorption of Pt atoms and dimers in general does not lead to stable structures. The Pt atoms try to move into the subsurface, causing large distortions in the original surface structure. This leads to the natural conclusion that different building blocks will be needed to generate the experimentally observed NWs. Two options come to mind. The optimized structure starting from Pt at the T$_{\mathrm{d}1}$ site on Ge(001) showed that Ge-Ge dimers could be a viable candidate to show up as a NW in pseudo-STM images. Another possibility are Pt-Ge dimers, which could be used to explain the observed asymmetry between the NW-dimer atoms in the observed $4\times1$ periodicity of the NWs. Because experiments show the NW to be growing in the troughs between the QDRs, and the above calculations also show a preference for adsorption in the troughs, mainly trough configurations will be considered from this point on.
\begin{figure}[!tbp]
\begin{center}
  \includegraphics[width=8cm,keepaspectratio]{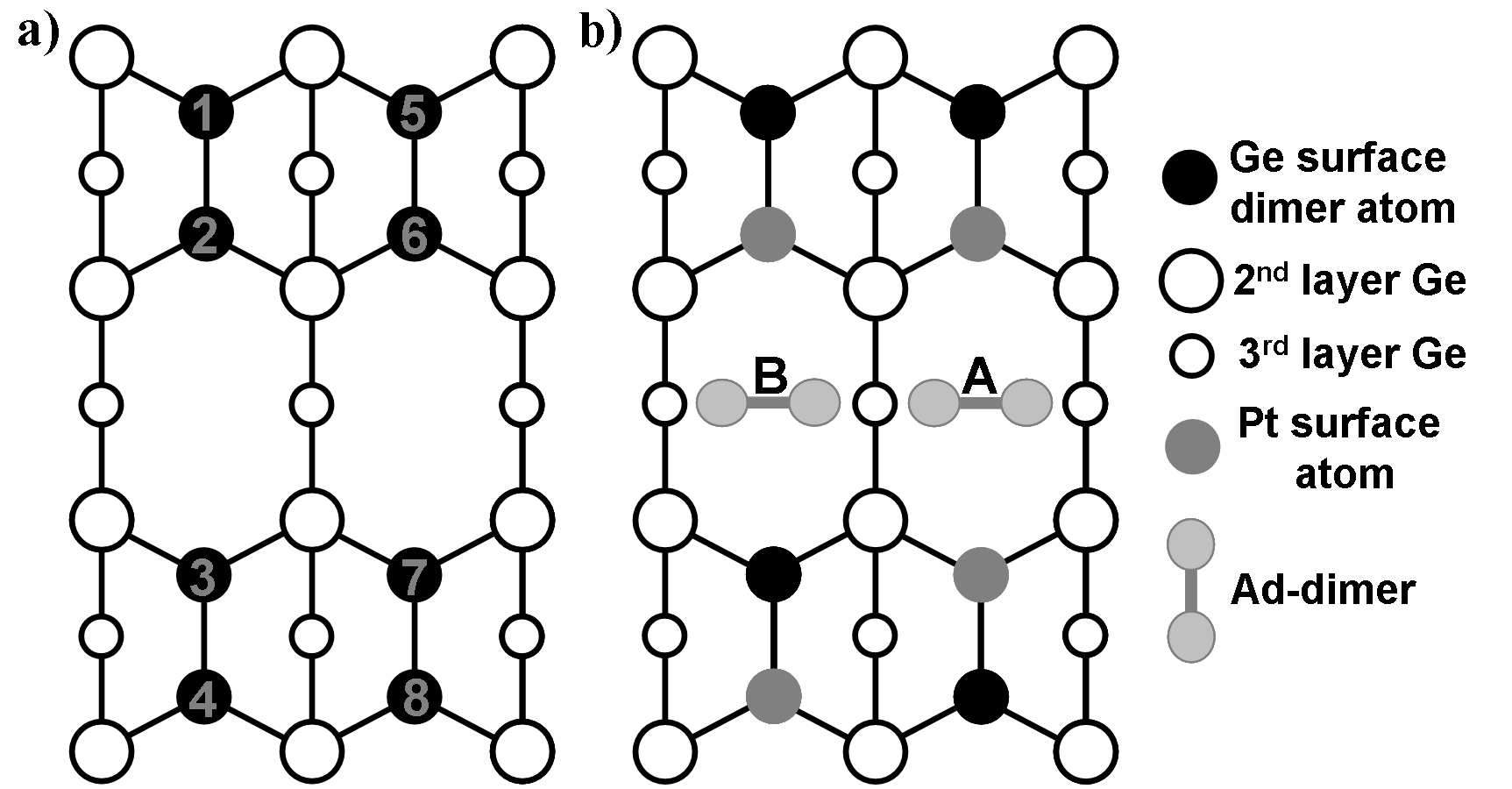}\\
\end{center}
  \caption{a) Possible position indices of Pt atoms in the top layer. b) Adsorption sites of ad-dimers, here shown on a surface
  containing Pt atoms at positions $2$, $4$, $6$, and $7$. Site A is
  positioned between two Pt atoms, while site B is positioned
  between a Pt and a Ge atom.}\label{fig:2indexjes}
\end{figure}
Due to the large variety of possible surface structures, these will be defined by indicating the positions of the Pt atoms in the top layer, as shown in Fig.~\ref{fig:2indexjes}a. The adsorption
sites are shown in Fig.~\ref{fig:2indexjes}b.\\
\indent The stable T$_{\mathrm{d}1}$ on $\beta_{6u}$ structure in Sec.~\ref{ssc:model1_onb6u} showed the exchange of one Pt-dimer atom with a Ge-surface atom. It could now be imagined that the second Pt atom behaves similarly and takes the position of a Ge atom on the opposite side of the trough. The structure obtained would now contain Pt atoms at positions $2$, $4$, $6$, and $7$: Two Pt atoms at positions $4$ and $6$ from the $\beta_6$ geometry, a Pt atom at position $2$ from the T$_{\mathrm{d}1}$ relaxation, as described in the previous subsection, and, the second Pt atom from the original Pt addimer at position $7$. This last position is chosen because it is known that Pt-dimers build into the dimer rows are very unstable (Ref.~\onlinecite{Vanpoucke:prb09beta}), giving preference to position $7$ over position $3$. During this process two Ge atoms are expelled onto the surface. These could bind and form a dimer, with a length of $2.558$ \AA, bridging the trough, with the dimer parallel to the dimers in the dimer rows.
This dimer has a formation energy $E_{\mathrm{f}}=-568$ meV. Because of the asymmetry between the two sides of the trough it is buckled with angle of $17.9^{\circ}$. Conversely, if the two expelled Ge atoms do not dimerise but maximize the number of Pt-Ge bonds the formation energy $E_{\mathrm{f}}$ is calculated to be $-768$ meV, roughly $0.2$ eV more stable than a dimer. Calculated STM images of this last structure show that due to the Ge atoms on top of the QDRs the two dimer images of the underlying dimers are replaced by a single image centered between the two dimers, as can be seen in Fig.~\ref{fig:3STM_trog_nwa}c. In experiments a widened trough WT is observed before the formation of the NWs. One significant feature of this WT is the fact that the dimer images of the QDRs are replaced by images which are two dimers wide, symmetrically around the trough. The structure described above could be a candidate for this observed structure. The Ge atoms could then during annealing dimerise (dimers bridging the WT are also experimentally observed) and rotate from their bridge position into a position
parallel to the QDRs, forming the observed NWs.\\
\indent Using a surface containing Pt atoms at positions $2$, $4$, $6$, and $7$, Ge dimers are placed at sites A and B and a Pt dimer at site A. After relaxation we find the Ge dimer that is placed at site B to have broken up and, due to the periodic boundary conditions and the surface unit cell size, moved toward the site A configuration. The formation energy $E_{\mathrm{f}}$ for this B-site structure is calculated to be $-609$ meV, while that of the A
configuration is slightly more stable, $-713$ meV.\\
\begin{figure}[!t]
\begin{center}
  \includegraphics[width=8cm,keepaspectratio]{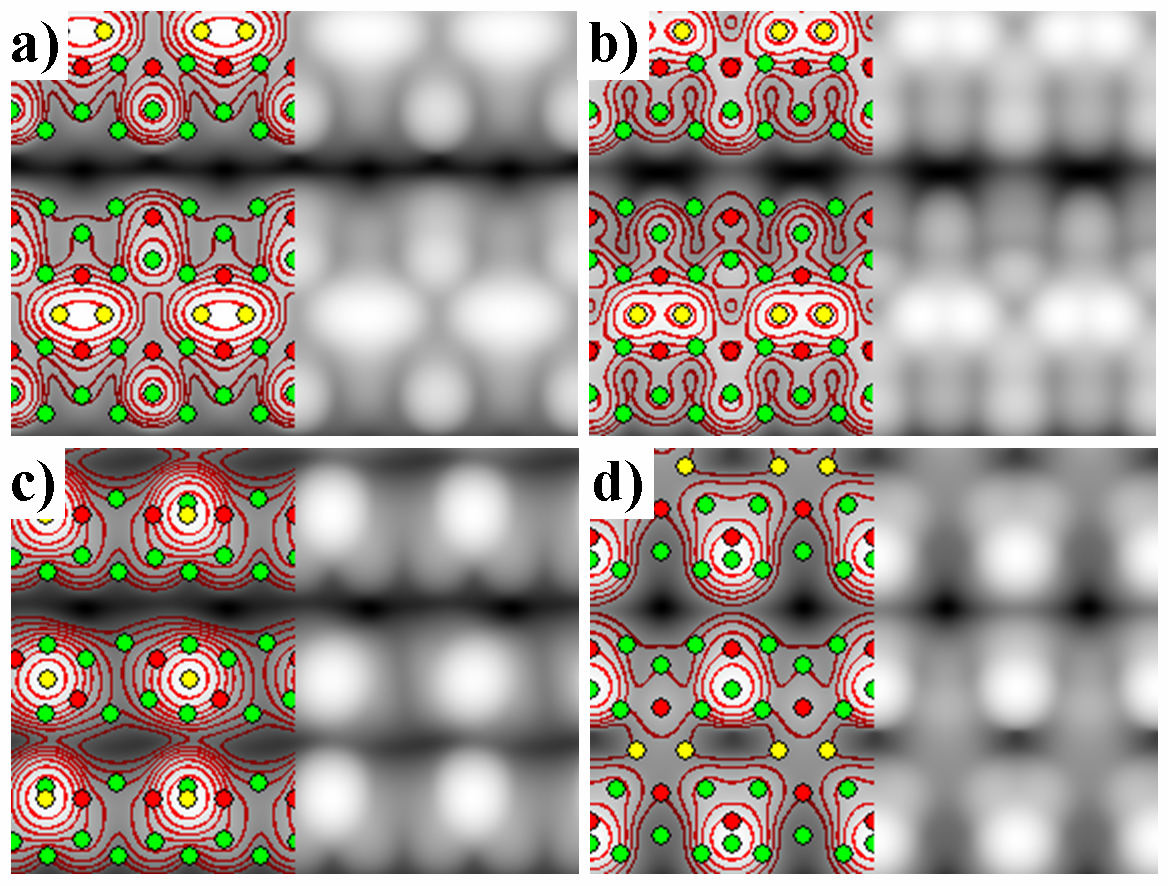}\\
\end{center}
  \caption{(color online) Calculated STM images of a Ge NW (a and b), Ge adatoms (c), and a Pt NW (d) on a surface containing Pt atoms at positions $2$, $4$, $6$, and $7$ (\textit{cf.}\ Fig.~\ref{fig:2indexjes}). The
  green/red (light/dark gray) discs show the positions of Ge/Pt atoms in
  the surface, and the yellow (white) discs show the positions of
  the Ge NW atoms(a and b), Ge adatoms (c), or Pt NW atoms (d). Contours are added to guide the eye. (a, c, and d) Calculated filled state STM images, for $z=3.0$ \AA\ and a simulated bias of $-1.50$ V. (b) Calculated empty state STM image, for $z=3.0$ \AA\ and a simulated bias of $+1.50$ V.
  }\label{fig:3STM_trog_nwa}
\end{figure}
\indent The Ge atoms in the A configuration are spaced by
$2.577$ \AA, which is about $6$\% larger than a normal Ge bulk bond. The Ge atoms are also bound to the Pt atoms at positions $6$ and $7$ forming a diamond shape structure. Pt-Ge cells with this shape appeared often during relaxations and seemed to represent a relatively stable substructure. Calculated STM images for this structure are shown in Fig.~\ref{fig:3STM_trog_nwa}a and b. They show a good resemblance to the experimentally observed images, but some important discrepancies exist. Although the general structure of the NW and the signatures of the bulges symmetric around the wire are present, the asymmetry of the latter is not in agreement with experiment. However, this can be easily solved by using a more symmetric surface (\textit{cf.}\ Sec.~\ref{ssc:model2_onb4as}). Also, the images of the wire itself are not entirely consistent with experiment, both at positive and negative bias. Where the filled state image (Fig.~\ref{fig:3STM_trog_nwa}a) should show a double peaked feature and the empty state image (Fig.~\ref{fig:3STM_trog_nwa}b) a single peak, we observe the opposite behavior.\\
\indent The Pt dimer at site A gives some surprising results. Contrary to what might be expected from the previous subsection this structure seems stable, with a formation energy $E_{\mathrm{f}}=-1270$ meV, making it about twice as stable as the Ge structures above. Furthermore, the pseudo-STM images also show something comparable to the WT which was experimentally observed (\textit{cf.}\ Fig.~\ref{fig:3STM_trog_nwa}d). This time a pair of dimer images is replaced by a single bright image centered `\emph{on}' a dimer, instead of in between. Most interesting however is the fact that the Pt ad-dimer itself is invisible in the trough, giving further indication that the experimentally observed wires are most likely not formed of Pt atoms.\\
\indent In general we find that calculations with Ge ad-dimers in the troughs of surfaces with different Pt stoichiometries show a serious improvement in formation energy when Ge atoms can bridge the trough and bind to Pt atoms on both sides of the trough. We also see that the Ge atoms are
drawn toward the Pt atoms imbedded in the surface.\\
\indent Calculations using Pt-Ge dimers indicate similar behavior. Although their pseudo-STM images show very asymmetric dimer images, it also shows them not to be compatible with the assumption that the different atom types could cause the in experiment observed asymmetry of the NW dimers. So Pt-Ge dimers can be ruled out as origin of the experimentally observed $4\times1$ periodicity.
\subsection[Second model]{Pt and Ge nanowires on a $\gamma$-reconstructed surface}\label{ssc:model2_onb4as}
The previous subsection showed some crucial points for further development of the NW model. Firstly, there is the possibility that a NW consists of Ge atoms instead of Pt atoms, as was always assumed in experiments. Secondly, the surface supporting the NWs is not a simple $\beta$-surface with one quarter ML of Pt imbedded, but rather a modified $\beta$-terrace containing at least $0.5$ ML Pt in the top layer. Thirdly, the experimentally observed symmetry around the wire, which is electronic, should also appear in the surface geometry.\\
\begin{figure}[!tbp]
\begin{center}
  \includegraphics[width=8cm,keepaspectratio]{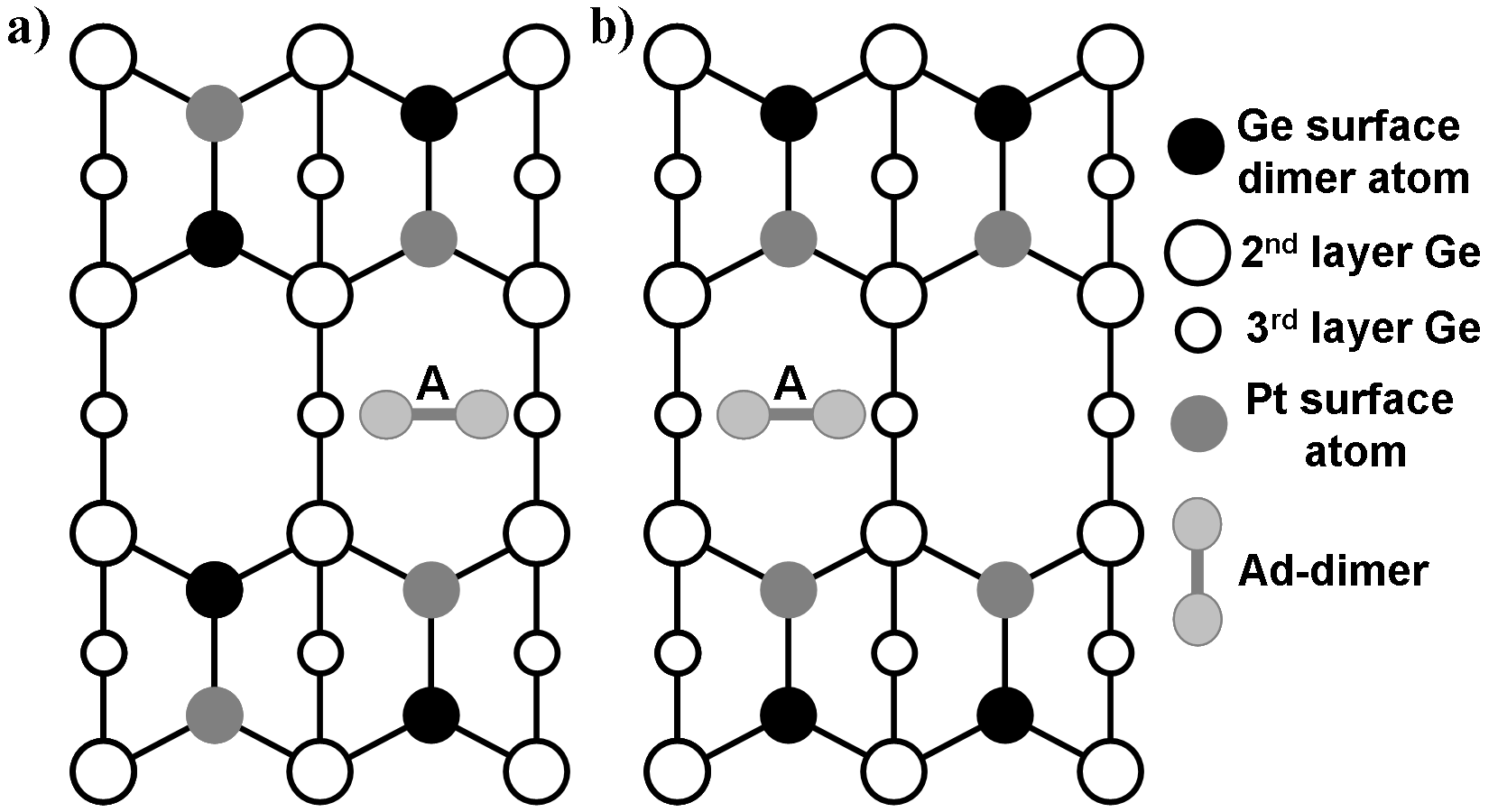}\\
\end{center}
  \caption{Schematic representation of the $\gamma_{c}$ (a) and
  $\gamma_{as}$ (b) surface \mbox{geometries,} with the NW dimer
  adsorption site A.}\label{fig:4geom_set2}
\end{figure}
\indent Fig.~\ref{fig:4geom_set2} shows the two surface geometries we will use in the following. Both contain $0.5$ ML of Pt in the top layer. The first one has Pt at positions $1$, $4$, $6$, and $7$ and will be referred to as $\gamma_{c}$.\cite{Vanpoucke:prb09beta} It has a c($4\times2$) structure and all troughs are equivalent. The second surface geometry contains Pt atoms at positions $2$, $3$, $6$, and $7$, and will be referred to as $\gamma_{as}$. It contains alternating Pt lined and Ge lined troughs. Unlike the $\gamma_{c}$ surface the troughs between the dimer rows are not equivalent. Because of what we learned in the previous paragraph, ad-structures are only placed in the Pt-lined trough. Both Pt and Ge dimers are used as NW at site A in both structures, and after relaxation their formation energies are calculated and are shown in Table~\ref{table:2formE_b4as}.\\
\begin{table}[!] \center{\textbf{Formation and adsorption
energy.}\\}
\begin{ruledtabular}
\begin{tabular}{l|rlrl}
   & \multicolumn{2}{c}{\makebox[2.5cm]{$E_{\mathrm{f}}$}}
   & \multicolumn{2}{c}{\makebox[2.5cm]{$E_{\mathrm{ad}}$}} \\
   & \multicolumn{2}{c}{(meV)} & \multicolumn{2}{c}{(meV)}  \\
  \hline
  $\gamma_{c}$ Ge NW & \makebox[1.6cm][r]{$-365$} & & \makebox[1.6cm][r]{$-298$} &
  \\ 
  $\gamma_{c}$ Pt NW & $-199$ & & $-215$ & \\ 
  $\gamma_{c}$ Ge NW$\times2$ & $-322$ & & $-138$ & \\ 
  $\gamma_{c}$ Pt NW$\times2$ & $289$ & & $14$ & \\ 
  \hline
  $\gamma_{as}$ Ge NW & $-1003$ & & $-379$ & \\ 
  $\gamma_{as}$ Pt NW & $-371$ & & $-63$ & \\ 
  $\gamma_{as2}$ bare & $387$ & & - & \\ 
  $\gamma_{as2}$ Pt NW & $-1495$ & & $-625$ & \\ 
  \hline
  $\gamma_{as}$ Pt NW + Ge NW & $-1397$ & & $-513$ & \\ 
  $\gamma_{as}$ Pt NW + Pt NW & $1027$ & & $699$ & \\ 
  $\gamma_{as2}$ Pt NW + Ge NW & $-1029$ & & $233$ & \\ 
  $\gamma_{as2}$ Pt NW + Pt NW & $-443$ & & $526$ & \\ 
\end{tabular}
\end{ruledtabular}
\caption{Formation and adsorption energies of Pt and Ge NWs adsorbed on the $\gamma_{c}$ and $\gamma_{as}$ surface geometry, shown in Fig.~\ref{fig:4geom_set2}. The adsorption energies for the Pt NW + X NW geometries, are the adsorption energies only for the `X NW', and are calculated with regard to the surfaces `$\gamma_{as}$ Pt NW' and `$\gamma_{as2}$ Pt NW'.}\label{table:2formE_b4as}
\end{table}

\indent Table~\ref{table:2formE_b4as} shows that on the $\gamma_{c}$ surface both Pt and Ge NWs are stable, with the highest stability for the Ge NWs. It also shows that both structures are more stable on a $\gamma_{as}$ surface.
This is partly because the formation energy is calculated with regard to the Ge(001) surface, and thus includes the formation energy of the substrate, which shows a higher stability for the $\gamma_{as}$ than for the $\gamma_{c}$-geometry (\textit{cf.}\ Table~I in Ref.~\onlinecite{Vanpoucke:prb09beta}). It is important to note that the experimentally observed NWs do not appear in adjacent troughs. In a single array of NWs there is always one trough spacing between two NWs.
This means that the growth of NWs in adjacent troughs must somehow be prohibited. In the $\gamma_{as}$ geometry this can easily be understood. Since only every second trough is lined with Pt atoms, and only the Pt lined troughs allow for stable binding of wires, only every second trough a wire can form.\\ 
\indent For the $\gamma_c$ geometry there is no such a substrate related restriction and the geometry allows for NWs to be present in both troughs of the unit cell. Table~\ref{table:2formE_b4as} shows that for systems with Ge NWs in each trough ($\gamma_c$ NW $\times2$)
the adsorption energy $E_{\mathrm{ad}}$ is more than a factor of $2$ smaller, so the binding of each NW to the surface is much weaker. So doubling the number of Pt NWs makes the structure unstable. Even in case this would inhibit the growth of NWs in adjacent troughs it would not prevent the existence of patches where NWs randomly switch between adjacent troughs, littering all troughs with NW segments. Such a terrace would look like patch of dots and dashes instead of the clean NW-arrays that are observed in experiment.\\
\indent The reason such disordered patches of NWs are not observed can be understood from a comparison between the $\gamma_{c}$ and $\gamma_{as}$ geometry. The $\gamma_{c}$-reconstruction is less stable than the $\gamma_{as}$-reconstruction. This is due to the strain between antiparallel Pt-Ge dimers in the QDRs. Flipping from antiparallel to parallel geometries the system gains $\sim0.3$ eV per Pt-Ge dimer pair. Since the $\gamma_{c}$-geometry only consists of these anti-parallel pairs, combined with the fact that the stabilizing effect of the adsorbed wires (\textit{i.e.}\ their adsorption energy) is smaller than on the $\gamma_{as}$-geometry, we can assume that the $\gamma_{c}$-structure will not be dominantly present in the structure of the
NWs.\\
\begin{figure}[!tbp]
\begin{center}
  \includegraphics[width=5cm,keepaspectratio]{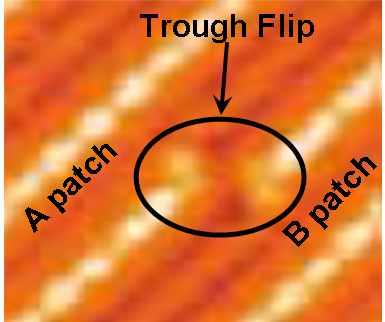}\\
\end{center}
  \caption{(color online) An experimental example of a trough flip
  observed at the boundary of two different NW patches.\cite{Fischer:prb07} Note that two NWs are spaced at least one empty trough in the
  lateral direction. 
  }\label{fig:5troughflip}
\end{figure}
\indent However, in some experimental images at the boundary between different NW patches, the boundary NW can be seen to flip from one trough to an adjacent trough (\textit{cf.}\ Fig.~\ref{fig:5troughflip}).\cite{fn:throughflip} The actual NW is always in between two empty troughs on one side and one empty trough at the other. This shows that inside NW patches the presence of NWs is uniquely connected to the underlying surface geometry, which prevents the presence of NWs in neighboring troughs. A trough flip as seen in Fig.~\ref{fig:5troughflip} would entail a QDR where at one side (bottom half Fig.~\ref{fig:5troughflip}) the Pt atoms are on the left side of the QDR, while at the other side (top half of Fig.~\ref{fig:5troughflip}) the Pt atoms are on the right side. At the position of the flip an antiparallel pair of Pt-Ge dimers exists, inducing a local strain on the system.\\
\indent Table~\ref{table:2formE_b4as} shows the adsorption energy for the Ge NW on the $\gamma_{as}$ surface to be $\sim 0.3$ eV larger than that of the Pt NW. The latter structure is only a local minimum. The Pt NW has sunk into the trough, about $0.7$ \AA\ below the average top layer position, and bound to second layer Ge atoms. These Ge atoms can be pushed up to the surface allowing for the Pt atoms at positions $2$ and $3$ (\textit{cf.}\ Fig.~\ref{fig:2indexjes}a) to dive into the subsurface increasing the tilt angle of these Pt-Ge dimers from $4.8^{\circ}$ to $-61.2^{\circ}$. The two remaining Pt-Ge dimers (with Pt at positions $6$ and $7$) stay almost flat with a tilt angle of $6.3^{\circ}$, while being lowered beneath the pushed up Ge atoms of the second layer. We will refer to this new surface reconstruction (without the Pt NW atoms) as $\gamma_{as2}$ in what follows. As can be seen in Table~\ref{table:2formE_b4as}, this reconstruction increases the formation energy of the Pt NW by over $1$ eV, making $E_{\mathrm{f}}=-1495$ meV.
\begin{figure}[!t]
\begin{center}
  \includegraphics[width=8cm,keepaspectratio]{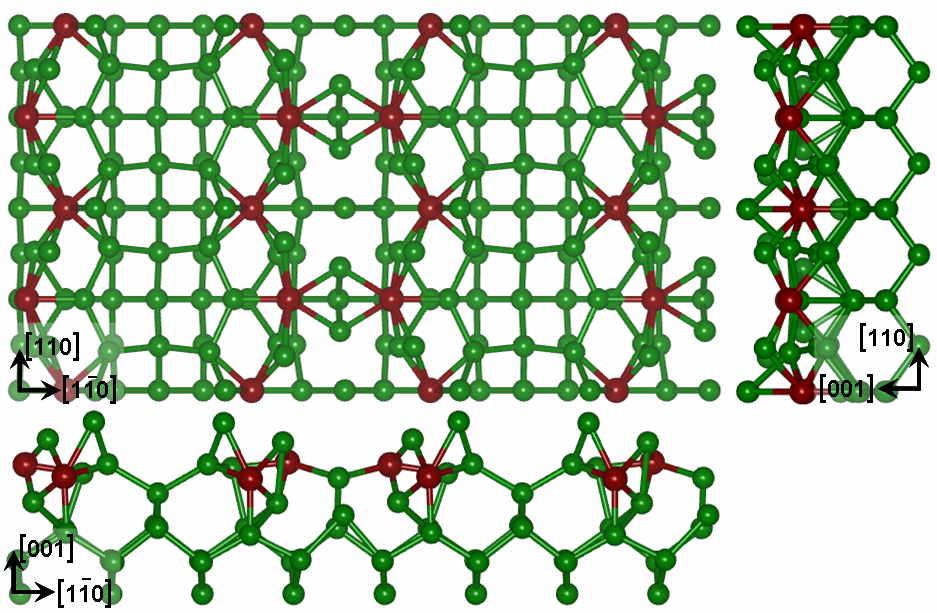}\\
\end{center}
  \caption{(color online) Ball-and-stick model of the $\gamma_{as2}$-surface reconstruction after relaxation. The top left image shows a top view of the structure, while the top right and bottom images show side views along the [$1\overline{1}0$] and [110] directions respectively. Green and red balls show the positions of Ge and Pt atoms respectively. The top left figure shows $2\times2$-surface unit cells, with three troughs in the $x$-direction. The middle trough is lined with Pt atoms and is the location of NW
  adsorption in the calculations. The bottom image clearly shows the large tilting angles of some of the Pt-Ge surface dimers. It also shows the second layer Ge atoms to be pushed up to positions higher than the Pt atoms in the
  substrate. 
  If the third layer Ge atoms are replaced with Pt atoms, the
  $\gamma_{as2}^{\star}$ surface is generated.
  }\label{fig:4bis_gamma_as2}
\end{figure}
Figure~\ref{fig:4bis_gamma_as2} shows a ball and stick model of the bare $\gamma_{as2}$-surface. It is obtained after removal of the Pt NW atoms and additional relaxation of the system. This $\gamma_{as2}$-surface reconstruction is found to be unstable with a formation energy $E_{\mathrm{f}}=387$ meV. This means that this reconstruction is induced by the presence of the Pt NW in the trough.\\
\indent The pseudo-STM images of the Pt NW on $\gamma_{as}$ show no signature of a NW image. The images of the Pt-Ge dimers in between (with Pt at positions $2$ and $3$) on the other hand show up more brightly and remind of the symmetric bulges seen in the experiment. On the $\gamma_{as2}$-surface the Pt NW remains invisible but the bright Pt-Ge dimer images become larger, and there is no signature left of the other Pt-Ge dimers. For every two Pt-Ge dimers only one bright image remains, centered on a Pt-Ge dimer. This picture
again greatly resembles the experimentally observed WT.\\
\begin{figure}[!t]
\begin{center}
  \includegraphics[width=8cm,keepaspectratio]{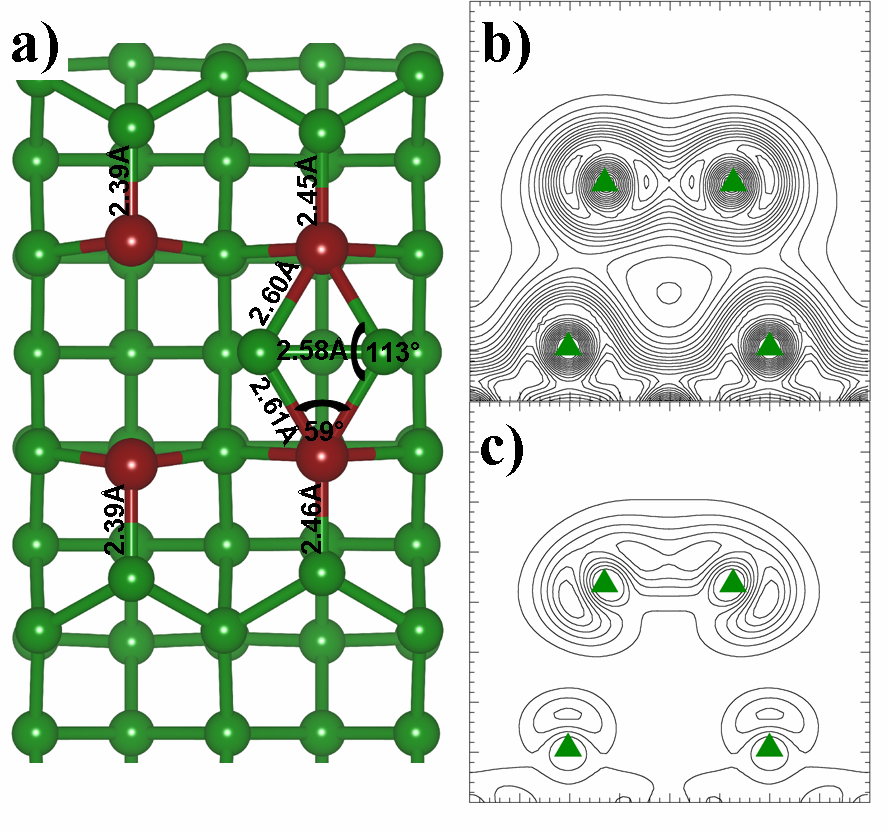}\\
\end{center}
  \caption{(color online) a) Geometric structure of the Ge NW on the
  $\gamma_{as}$-surface after relaxation. Pt atoms are shown in red,
  Ge atoms in green. b) Contour plot of the
  total charge density of the adsorbed Ge NW, in the vertical plane along
  the NW. c) Same as b) but now for the partial charge density,
  for all energy levels from $0.7$ eV below the Fermi level up to the Fermi
  level. Green triangles indicate the position of Ge atoms in the
  plane.The contours are separated $0.03$ ($0.01$) $e/$\AA$^3$ for the
  total (partial) charge density plot. Major tics are separated $1$
  \AA.}\label{fig:6trog_nwa_b4_geom_en_cont}
\end{figure}
\indent Contrary to the Pt NW, the Ge NW on $\gamma_{as}$ shows after relaxation a nicely symmetric structure with the Ge NW at the center of the trough, in a slightly elevated position $0.7$ \AA\ above the top layer. The calculated STM images show some of the main features of the experimentally observed NWs. The symmetric bulges are present, as is a clear NW signature, for both filled and empty state images. However the NW dimer images are not double peaked as is observed in the experiments. Examination of the total charge distribution shows that the Ge NW atoms dimerise forming a $\sigma$-bond, as can be seen in Fig.~\ref{fig:6trog_nwa_b4_geom_en_cont}b. From this one might expect to see a double peak feature in the STM images. However, an STM only sees energy levels close to the Fermi-level, and the partial charge density near the Fermi level (Fig.~\ref{fig:6trog_nwa_b4_geom_en_cont}c) shows the existence of a $\pi$-bond above the Ge-dimer, causing the NW dimer image to show up as a single peak.\\
\indent The Ge NW binds to the Pt atoms in the top layer through $\sigma$-bonds, forming the diamond shaped structure already noted in Sec.~\ref{ssc:model1_onb6u}, as can be seen in Fig.~\ref{fig:6trog_nwa_b4_geom_en_cont}. These bonds with the Pt surface atoms pull the Ge NW slightly into the trough, stretching the Pt-Ge bonds of the dimers in the QDR about $2.5$\%. The dimer length of the
NW dimers is $2.575$ \AA, only $6$\% longer than a normal Ge bulk bond.\\
\indent In Sec.~\ref{ssc:model1_onb6u} we saw that a chain of double peaked dimer images emerges when Ge dimers are placed on top of Pt atoms. Combined with the fact, that the Ge NW stay above the trough while the Pt NW sinks in, one could imagine the existence of a double layered NW combining these properties. We therefore investigated some systems where a second NW dimer of Ge or Pt atoms was placed on top of the first NW dimer on both the $\gamma_{as}$ and $\gamma_{as2}$ surfaces. These calculations show that for both surfaces the extra Ge NW is most stable (\textit{cf.}\ Table~\ref{table:2formE_b4as}). In case of the extra NW on the $\gamma_{as}$ surface we find an adsorption energy $E_{\mathrm{ad}}=-513$ meV for the Ge NW and $E_{\mathrm{ad}}=699$ meV for the Pt NW, which means the second Pt wire does not stick to the surface while the Ge wire does. However, we showed that the Pt NW on $\gamma_{as}$ induces a reconstruction to $\gamma_{as2}$ creating a much more stable subsurface. On the $\gamma_{as2}$ surface however both the extra NWs do not stick. They have an adsorption energy $E_{\mathrm{ad}}=233$ meV and $526$ meV for the extra Ge and Pt NW respectively. Furthermore, inspection of the geometries show that the double layered NWs fall over sideways in the trough, giving rise to a NW image in the calculated STM pictures which is asymmetric in the trough. Despite all this, these pseudo-STM images show the required doubly peaked NW dimer images, which means the stacking of Ge on top of Pt atoms needs to be somehow maintained if we want to observe doubly peaked NW dimer images in the calculated STM images.
\subsection[Third model]{Pt and Ge nanowires on the
$\gamma_{as}^{\star}$-surface}\label{ssc:model3_onb4as_ptL3}

It was shown in the previous subsection that the Pt NW atoms on the $\gamma_{as}$-surface sink into the trough. Going even further, one could imagine the Pt atoms to exchange positions with the third layer Ge atoms at the bottom of the trough, thus increasing their coordination and increasing the Pt density to $0.75$ ML of Pt in the top layers of the surface. We will use the $\star$-superscript to indicate the presence of Pt atoms at the bottom positions of one trough of a system.
\begin{figure}[!tbp]
\begin{center}
  \includegraphics[width=8cm,keepaspectratio]{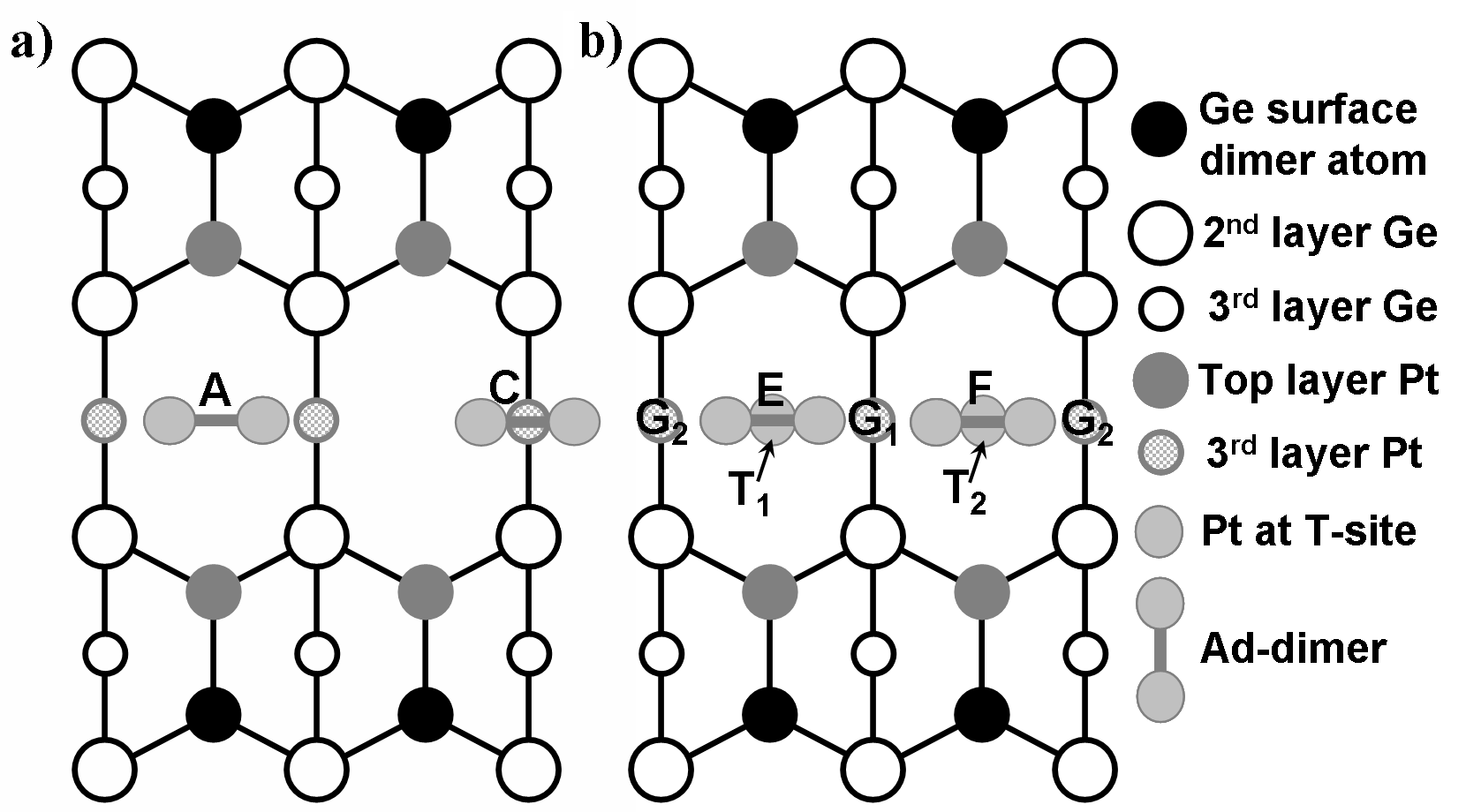}\\
\end{center}
  \caption{Schematic representation of the $\gamma_{as}^{\star}$-structure
  with adsorption sites discussed in the text.}\label{fig:7geom_ptL3}
\end{figure}
Figure~\ref{fig:7geom_ptL3}a shows a schematic representation of the geometry of the $\gamma_{as}^{\star}$-structure. It contains $0.75$ ML of Pt, with Pt atoms at positions $2$, $3$, $6$, $7$, see Fig.~\ref{fig:2indexjes}and the two positions at the bottom of the Pt-lined trough. Keeping in mind the surface deformation from $\gamma_{as}$ to $\gamma_{as2}$ induced by a Pt NW, we also build a $\gamma_{as2}^{\star}$-surface in similar fashion (\textit{cf.}\ Fig.~\ref{fig:4bis_gamma_as2}), by replacing the two Ge atoms at the bottom of the Pt lined trough of the $\gamma_{as2}$-surface. Pt and Ge dimers are adsorbed at sites A and C (\textit{cf.}\ Fig.~\ref{fig:7geom_ptL3}a) for the $\gamma_{as}^{\star}$-surface and at site A for the $\gamma_{as2}^{\star}$-surface.
\begin{table}[!tb] \center{\textbf{Formation and adsorption
energy.}\\}
\begin{ruledtabular}
\begin{tabular}{l|rlrl}
   & \multicolumn{2}{c}{\makebox[2.5cm]{$E_{\mathrm{f}}$}}
   & \multicolumn{2}{c}{\makebox[2.5cm]{$E_{\mathrm{ad}}$}} \\
   & \multicolumn{2}{c}{(meV)} & \multicolumn{2}{c}{(meV)}  \\
  \hline
  $\gamma_{as}^{\star}$ bare & \makebox[1.6cm][r]{$-1011$} & &
  \makebox[1.6cm][r]{-} & \\ 
  $\gamma_{as2}^{\star}$ bare & $-1991$ & & - & \\ 
  \hline
  $\gamma_{as}^{\star}$ Ge NW A & $-2055$ & & $-522$ & \\ 
  $\gamma_{as}^{\star}$ Pt NW A & $-962$ & & $25$ & \\ 
  $\gamma_{as}^{\star}$ Ge NW C & $-3364$ & & $-1176$ & \\ 
  $\gamma_{as}^{\star}$ Pt NW C & $-2877$ & & $-933$ & \\ 
  $\gamma_{as2}^{\star}$ Ge NW A & $-2488$ & & $-248$ & \\ 
  $\gamma_{as2}^{\star}$ Pt NW A & $-2312$ & & $-160$ & \\ 
  \hline
  $\gamma_{as2}^{\star}$ Ge NW + Ge NW & $-2281$ & & $103$ & \\ 
  $\gamma_{as2}^{\star}$ Ge NW + Pt NW & $-1597$ & & $445$ & \\ 
  $\gamma_{as2}^{\star}$ Pt NW + Ge NW & $-2448$ & & $-68$ & \\ 
  $\gamma_{as2}^{\star}$ Pt NW + Pt NW & $728$   & & $1520$&  \\
\end{tabular}
\end{ruledtabular}
\caption{Formation and adsorption energies of Pt and Ge NWs adsorbed on the $\gamma_{as}^{\star}$ and $\gamma_{as2}^{\star}$ surface geometry, shown in Fig.~\ref{fig:7geom_ptL3}. The adsorption energies for the X NW + Y NW geometries, are the adsorption energies only for the `Y NW', and are calculated with regard to the same surface with the X NW adsorbed. The X and Y NWs are each time adsorbed at site A.}\label{table:3formE_b4as_ptL3}
\end{table}
Table~\ref{table:3formE_b4as_ptL3} shows that both the bare $\gamma_{as}^{\star}$- and $\gamma_{as2}^{\star}$-surfaces are stable reconstructions, with the $\gamma_{as2}^{\star}$-surface roughly $1.0$ eV more stable than the $\gamma_{as}^{\star}$-surface. For both surfaces the adsorption of a Ge NW at site A is more favorable than the adsorption of a Pt NW. In case of the $\gamma_{as}^{\star}$-surface, the Pt NW even has a small positive adsorption energy, making it unstable. Although the transition from $\gamma_{as}^{\star}$ to $\gamma_{as2}^{\star}$ shows an improvement in formation energy for the Ge NW system, the main contribution comes from the surface itself, decreasing the adsorption energy of the wire.\\
\begin{figure}[!t]
\begin{center}
  \includegraphics[width=8cm,keepaspectratio]{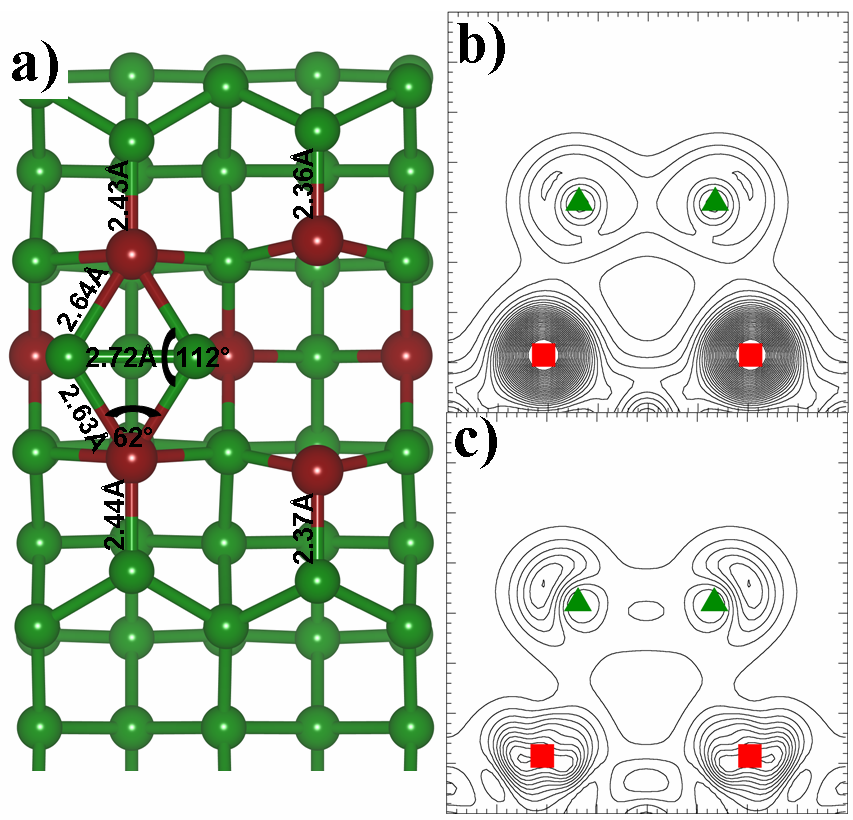}\\
\end{center}
  \caption{(color online) a) Geometric structure of the Ge NW on the
  $\gamma_{as}^{\star}$-surface after relaxation. Pt atoms are shown in red,
  Ge atoms in green. b) Contour plot of the
  total charge density of the adsorbed Ge NW, in the vertical plane along
  the NW. c) Same as b) but now for the partial charge density,
  for all energy levels going from $0.7$ eV below the Fermi level up to the
  Fermi level. Green triangles (red squares) indicate the position of Ge (Pt)
  atoms in the plane. The contours are separated $0.1$($0.01$) $e/$\AA$^3$
  for the total (partial) charge density plot.}\label{fig:8b4as_ptL3_nwge}
\end{figure}
\indent On the $\gamma_{as}^{\star}$-surface, the Ge NW dimers in the A configuration  is stretched to a length of $2.72$ \AA\ (\textit{cf.}\ Fig.~\ref{fig:8b4as_ptL3_nwge}a) and is located $0.52$ \AA\ above
the average height of the surface atoms. In contrast, the Pt NW dimer sinks into the trough $0.06$ \AA, almost at level with the Pt
atoms in the top layer of the surface. Contrary to the Ge NW on the $\gamma_{as}$-surface, the Ge NW on the $\gamma_{as}^{\star}$ is dimerised only through a $\sigma$-bond, located close to the Fermi level. The electrons used to fill the $\pi$-bond are now used in bonds with the Pt atoms at the bottom of the trough, as can be seen from the contour plots in Fig.~\ref{fig:8b4as_ptL3_nwge}b and c. The strong bonds with the two Pt atoms at positions $2$ and $3$ remain. The Pt NW on the other hand binds to the Pt atoms at the bottom of the trough, the surface Pt atoms at positions $2$ and
$3$ and to the four second layer Ge atoms in the trough. On the $\gamma_{as2}^{\star}$-surface the Ge atoms of the second layer are pushed upward to above the average top
layer level, and form bonds with the NW atoms in both cases.\\
\begin{figure}[!tb]
\begin{center}
  \includegraphics[width=8cm,keepaspectratio]{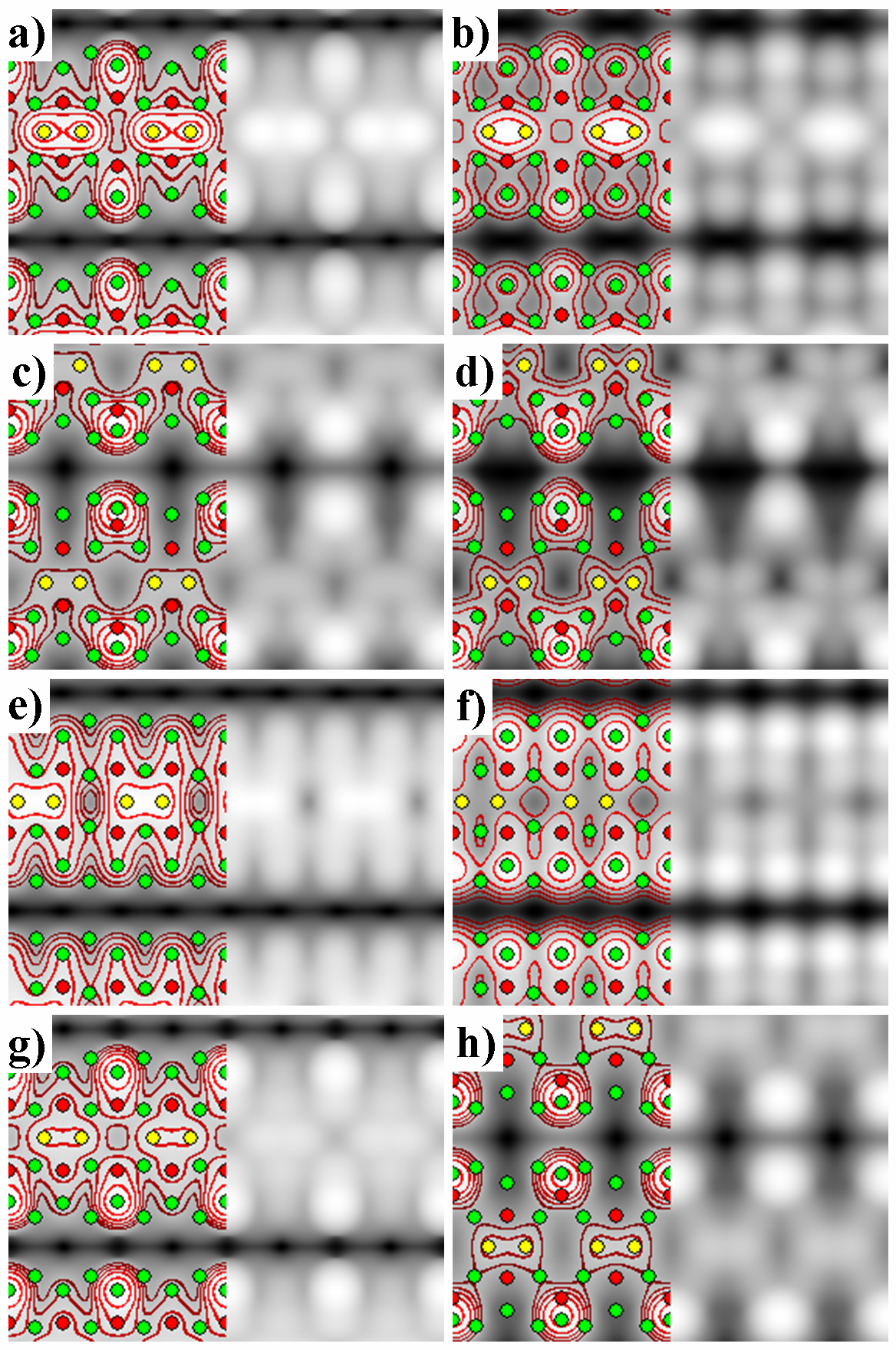}\\
\end{center}
  \caption{(color online) Calculated STM images of a Ge NW adsorbed at the A (a and b) and C (e and f) site on a  $\gamma_{as}^{\star}$-surface and on the A site of a $\gamma_{as2}^{\star}$- (c and d) surface, and calculated filled state STM images of a Pt NW adsorbed at the A site of a $\gamma_{as}^{\star}$- (g) and $\gamma_{as2}^{\star}$-surface (h). The green/red (light/dark gray) discs show the positions of Ge/Pt atoms in the surface, and the yellow (white) discs show the positions of the Ge/Pt NW atoms. Contours are added to guide the eye. (a, c, e, g, and h) Calculated filled state STM images, for $z=3.0$ \AA\ and a simulated bias of $-1.50$ V. (b, d, and f) Calculated empty state STM images, for $z=3.0$ \AA\ and a simulated bias of $+1.50$ V.}\label{fig:9stm_b4as_b4as2_ptL3}
\end{figure}
\indent Figures~\ref{fig:9stm_b4as_b4as2_ptL3}(a--d) show calculated STM images of the Ge NW adsorbed at site A of the $\gamma_{as}^{\star}$- and $\gamma_{as2}^{\star}$-surface. Although the latter has a better formation energy, the pseudo STM images of the first show much better agreement with the experimentally observed images. The pseudo-STM images of the Ge NW on the $\gamma_{as2}^{\star}$-surface on the other hand show great resemblance to the experimentally observed WT. In Sec.~\ref{sc:discussion} we will look into the different structures which present a WT in their calculated STM images in more detail, and discard them or give them a place in the larger picture of the NW formation.\\
\indent For the Ge NW on $\gamma_{as}^{\star}$ we can identify all the experimentally observed features: the symmetric bulges at the sides of the NW are present and the NW dimer images show a double peaked feature in the filled state image (Fig.~\ref{fig:9stm_b4as_b4as2_ptL3}a) and a single peaked one in the empty state image (Fig.~\ref{fig:9stm_b4as_b4as2_ptL3}b).
The Pt NWs on the other hand are invisible on both surface geometries, as is shown in Fig.~\ref{fig:9stm_b4as_b4as2_ptL3}g and h, making it possible to exclude them again from the realm of possibly observed NW atoms.\\
\indent Because the $\gamma_{as2}^{\star}$-surface has a better formation energy, and symmetric bulges are readily present, the $\gamma_{as2}^{\star}$-surface might appear a suitable surface for NWs. However, although the Ge NW on the $\gamma_{as2}^{\star}$-surface reconstruction has a better formation energy, the average height of the surface atoms (the bulges in particular) makes it hard to observe any NW in the trough. Therefore we again consider a double layered NW, assuming the second layer of the NW will stick out of the (widened) trough far enough to be visible. For all these stacked NWs the formation energy, given in Table~\ref{table:3formE_b4as_ptL3}, is less than this of the Ge NW on the $\gamma_{as2}^{\star}$-surface. Also the adsorption of the second NW dimer is shown in Table~\ref{table:3formE_b4as_ptL3} to be unstable to only marginally stable in the best case. Furthermore, it again shows the preference for Pt atoms to be buried underneath Ge atoms (Compare $\gamma_{as2}^{\star}$ Ge NW $+$ Pt NW to $\gamma_{as2}^{\star}$ Pt NW $+$ Ge NW in Table~\ref{table:3formE_b4as_ptL3}). Exchanging the Pt and Ge NWs in the stack improves the formation energy by $0.85$ eV. Furthermore, Table~\ref{table:3formE_b4as_ptL3} shows that a Pt NW as top wire of the stack always has a large positive adsorption energy, meaning it would not stick to the surface.\\
\indent The pseudo-STM images show bright NW images in each case. The NW peaks are much higher in comparison to the surface than the ones discussed before. Unlike the structures with Pt NWs presented above, here the Pt NWs (\textit{i.e.}\ the top wire in the stack) show up as bright images, this is because in this reconstruction the extra Pt NW sticks far out of the surface after relaxation. The stacked NWs however tend to topple over sideways in the trough, causing the pseudo STM pictures to show NW images shifted toward the QDR, in contradiction to the experimentally observed NWs. Based on their adsorption energies and calculated STM images it allows us to discard these stacked NWs on the $\gamma_{as2}^{\star}$-surface.\\
\indent Table~\ref{table:3formE_b4as_ptL3} also shows the formation and adsorption energy of Pt and Ge NWs adsorbed at the C site on the $\gamma_{as}^{\star}$-surface. It shows that adsorption at the C site is at least $1$ eV more stable than at
the A site, making the adsorbed NW at the A site metastable configurations. After relaxation, the Ge NW stays centered in the trough at the C site. Contrary to the A site adsorption all Pt atoms in the top layer have a bond with the Ge NW. The difference in geometry also shows up in the pseudo-STM images, as  can be seen in Fig.~\ref{fig:9stm_b4as_b4as2_ptL3}e and f. Although double peaked dimer images are present in the filled state image (Fig.~\ref{fig:9stm_b4as_b4as2_ptL3}e), the symmetric bulges are missing, and the dimer image is out of phase half a dimer length in comparison to the experimentally observed NWs. Since the NW dimers are now strongly bound to four Pt atoms in the top layer instead of two (in case of the A site), the previously empty states are now filled with extra charge from the two extra Pt atoms. Because of this, the empty state images (Fig.~\ref{fig:9stm_b4as_b4as2_ptL3}f) do not show a NW image, where in experiment it is clearly observed.\\
\indent An adsorbed Pt NW sinks in the trough binding tetrahedrally to four neighboring Pt atoms, two in the top layer and two at the bottom of the trough. With an adsorption energy $E_{\mathrm{ad}}=-933$ meV, these Pt atoms are bound very strongly in the trough. In the pseudo-STM images however the Pt NW remains invisible.
\subsection[Nec plus ultra]{Stabilizing the nanowires}\label{ssc:model4_npu}
\begin{figure}[!t]
\begin{center}
  \includegraphics[width=8cm,keepaspectratio]{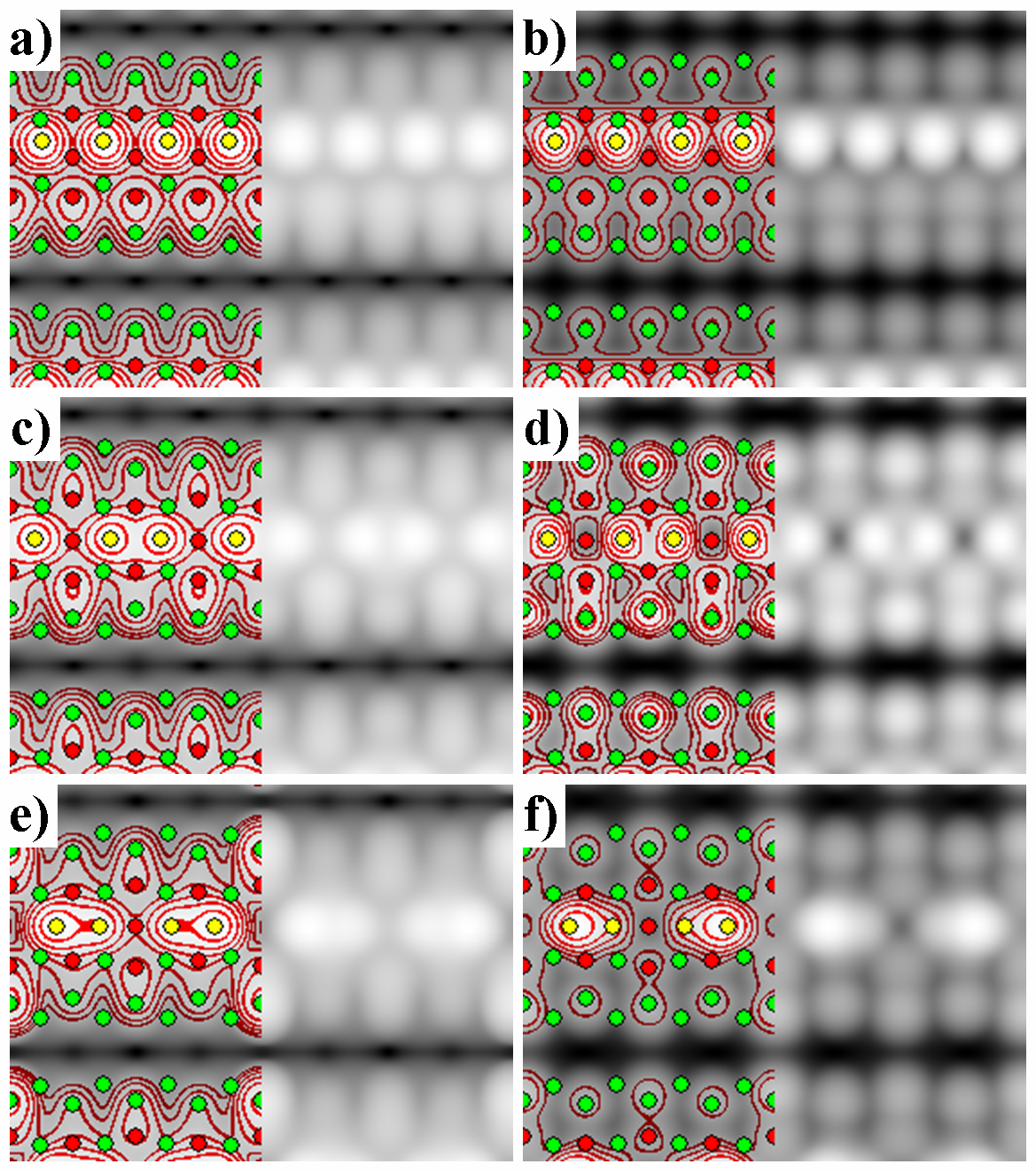}\\
\end{center}
  \caption{(color online) Calculated STM images of a Ge NW adsorbed on a
  $\gamma_{as}^{\star}$-surface with
  extra Pt atoms adsorbed at the T-sites of the trough. a and b:
  Pt atoms are adsorbed at both T-sites of the $(4\times2)$-surface cell
  (2T), c and d: only one T-site is occupied (1T), e and f: only one T-site per two surface-unit-cells is occupied (0.5T). The green (light gray) discs show the positions of Ge atoms in the two top layers of the surface.
  The red (dark gray) discs show the positions of the Pt atoms in the top layer and the Pt atoms at the T-sites in the trough, and the yellow (white) discs
  show the positions of the Ge NW atoms. Contours are added to guide the eye.
  a, c and e: Calculated filled state STM images, for $z=3.0$ \AA\ and a
  simulated bias of $-1.50$ V. b, d and f: Calculated empty state STM images, for $z=3.0$ \AA\ and a simulated bias of $+1.50$ V. }\label{fig:11stm_b4as_ptL3_p2cX_GeNW}
\end{figure}

In the previous section it was shown that Ge NWs at the A site of the $\gamma_{as}^{\star}$-surface are in good agreement with the experimentally observed  NWs, based on calculated STM images. It was however also shown that this structure is, although quite stable, not the most stable configuration with that specific stoichiometry. This raises the question if it is possible to stabilize the NW geometry without changing the electronic structure observed in STM experiments to much. Although experimentally $0.25$ ML of Pt was deposited, it is reasonable to assume local gradients in the distribution.\\
\indent In this section we will pursue two scenarios to improve our model. The first scenario is to try and stabilize the geometry by adding extra Pt or Ge atoms to the trough of the NW. Since the local Pt density at this point is approaching a full ML, the second alternative scenario tries to minimize the amount of Pt in the surface without loss of stability or STM agreement. For this we will investigate ${\star}$-geometries based on surface structures with a local Pt density of no more than $0.5$ ML.\\
\begin{table}[!tb] \center{\textbf{Formation and adsorption
energy.}\\}
\begin{ruledtabular}
\begin{tabular}{l|rlrl}
   & \multicolumn{2}{c}{\makebox[2.5cm]{$E_{\mathrm{f}}$}}
   & \multicolumn{2}{c}{\makebox[2.5cm]{$E_{\mathrm{ad}}$}} \\
   & \multicolumn{2}{c}{(meV)} & \multicolumn{2}{c}{(meV)}  \\
  \hline
  \multicolumn{5}{l}{With a sunken Ge atom at each T-site of the
  trough.(2T)}\\
  \hline
  $\gamma_{as}^{\star}$ Ge NW & \makebox[1.6cm][r]{$-2911$} & &
  \makebox[1.9cm][r]{$316$} & \\ 
  $\gamma_{as}^{\star}$ Pt NW & $-2256$ & & $643$ & \\ 
  $\gamma_{as}^{\star}$ Pt NW $^a$& $-3689$ & & $-74$ & \\ 
  $\gamma_{as2}^{\star}$ Ge NW & $-2784$ & & $634$ & \\ 
  $\gamma_{as2}^{\star}$ Pt NW & $-4308$ & & $-127$ & \\ 
  \hline
  \multicolumn{5}{l}{With sunken Pt atoms at the T-sites of the trough.}\\
  \hline
  $\gamma_{as}^{\star}$ $2$T Ge NW  & $-4763$ & & $-943$ & \\
  $\gamma_{as}^{\star}$ $1$T Ge NW  & $-3696$ & & $-860$ & \\
  $\gamma_{as2}^{\star}$ $1$T Ge NW  & $-3490$ & & $-757$ & \\
  $\gamma_{as}^{\star}$ $0.5$T Ge NW  & $-2914$ & & $-710$ & \\
\end{tabular}
\end{ruledtabular}
\makebox[0.8\textwidth][l]{$^a$ \footnotesize{with ejected Ge atom}} \caption{Formation and adsorption energies of Pt and Ge NWs adsorbed on the $\gamma_{as}^{\star}$- and $\gamma_{as2}^{\star}$-surface geometry with extra Pt or Ge atoms sunken into the trough at the T-sites shown in Fig.~\ref{fig:7geom_ptL3}b. $\mathrm{x}$T indicates the number of T sites per $(4\times2)$ surface cell that are occupied.}\label{table:4formE_pnu}
\end{table}
\ \\
\indent It was shown that Pt atoms sink into the trough while Ge atoms have a stable position above it. However, if the Ge atoms of the NW would sink in the trough more Pt-Ge bonds could be formed. In this case, each of the sunken Ge atoms binds to $4$ Pt atoms in a tetrahedral configuration (at the T-sites shown in Fig.~\ref{fig:7geom_ptL3}b) improving the formation energy by about $1.5$ eV on both the $\gamma_{as}^{\star}$- and $\gamma_{as2}^{\star}$-surfaces. This again shows the Ge NW on $\gamma_{as}^{\star}$-reconstruction to be only metastable. However, Table~\ref{table:4formE_pnu} shows that the adsorption on a $\gamma_{as}^{\star}$-surface with sunken Ge atoms has a positive adsorption energy of $316$ and $643$ meV per NW adatom for the Ge and Pt NW respectively. This means that both wires will not stick to the surface. For the equivalent situation on the $\gamma_{as2}^{\star}$-surface only the Pt NW shows a small negative adsorption energy of $-127$ meV making it marginally stable. Furthermore, for all structures, the pseudo-STM images of the NWs do not show the expected signatures. The Ge NW dimer images are not double peaked, and only on the $\gamma_{as2}^{\star}$-surface the side bulges are observed. In case of the Pt NW on $\gamma_{as}^{\star}$ with sunken Ge atoms, equidistant peaks are observed for each Pt NW dimer, but this structure tends to stabilize by ejecting a sunken Ge atoms on top of the Pt wire (\textit{cf.}\ Table~\ref{table:4formE_pnu}). On the $\gamma_{as2}^{\star}$-surface with sunken Ge atoms this does not happen and the Pt NW remains invisible. Based on these results it can be concluded that
the NWs are not stabilized by Ge atoms sunken into the trough.\\
\indent Similar to the previous idea, extra Pt atoms can be introduced in the trough. This assumption can be founded on the stability and geometry after relaxation of the Pt NW adsorbed at site C of the $\gamma_{as}^{\star}$-structure. With an adsorption energy $E_{\mathrm{ad}}=-923$ meV per Pt NW atom, these Pt atoms are bound very strongly. During relaxation the Pt dimer has broken up (the distance between the two Pt NW atoms after relaxation is $3.98$ \AA) and the Pt atoms are adsorbed at both T sites (Fig.~\ref{fig:7geom_ptL3}b), having a
tetrahedral binding to the four surrounding Pt atoms.\\
\indent If Pt atoms are adsorbed at all T-sites, the Ge NW dimers break up and the Ge NW atoms are spaced equidistantly along the trough.
Figure~\ref{fig:11stm_b4as_ptL3_p2cX_GeNW}a and b show the filled and empty state pseudo STM images for this structure. The NW has again toppled over, which is clearly seen in the asymmetric position of the NW image in the trough. Furthermore, in both pictures the NW image consists of atom images separated roughly $4$ \AA. No dimer images, as observed in experiment, are seen here. This shows that there must be less than two Pt atoms per $(4\times2)$ surface cell present. The reason is twofold, firstly to allow the dimerization of the Ge NW atoms and secondly to keep the NW centered in the trough.\\
\indent We now place $1$ Pt atom at the T$_1$ site (leaving T$_2$ empty) of the $\gamma_{as}^{\star}$- and $\gamma_{as2}^{\star}$-surface and then adsorb a single Ge NW at sites E and F. For the Ge NW on the modified $\gamma_{as}^{\star}$ we see after relaxation that both structures converge to the same structure with Ge atoms moving toward the center sites G$_1$ and G$_2$. There is a small offset with regard to the G$_1$ and G$_2$ sites, away from the adsorbed Pt atom at site T$_1$. Again the Ge atoms form a diamond shape reconstruction with $2$ Pt atoms lining the trough. Contrary to the previous NW structures the $2$ Ge NW atoms do not dimerise (the distance between these two Ge atoms is $3.27$ \AA). Instead they form a strong bond with the sunken Pt atom at site T$_1$. The configuration on the $\gamma_{as}^{\star}$-surface, contrary to the systems studied up until now, has a better formation energy than the configuration on the $\gamma_{as2}^{\star}$-surface. This is because in the latter case the Ge NW atoms bind to Ge atoms pushed up out of the surface instead of surface Pt atoms. The Ge NW on the $\gamma_{as}^{\star}$-surface with one sunken Pt atom is also $0.7$ eV more stable than the Ge NW on the same surface with sunken Ge atoms. With an adsorption energy $E_{\mathrm{ad}}=-860$ meV per NW adatom, this wire is roughly $\sim0.3$ eV more stable than the one on a clean $\gamma_{as}^{\star}$-surface. Figures~\ref{fig:11stm_b4as_ptL3_p2cX_GeNW}c and d show the calculated STM-images for this structure. In the filled state image (Fig.~\ref{fig:11stm_b4as_ptL3_p2cX_GeNW}c), bright large double peaked dimer images and symmetric bulges at the sides of the wire are seen. The empty state image (Fig.~\ref{fig:11stm_b4as_ptL3_p2cX_GeNW}d)
however also shows a double peaked dimer image.\\
\indent At this point it is important to remember the limitations of our applied method for generating STM images. The method used is implemented to simulate a `point'-like STM tip with infinite resolution. In reality however an STM tip has a finite width causing it to show two features, which are to close to one-another, as a single one. In our case the two peaks observed in the filled state image are bigger and spaced more widely than the ones seen in the empty state image, so it is reasonable to assume that in case of the empty state image a real STM might be able to resolve the peaks in the filled state
image but not in the empty state image. In case of the $\gamma_{as2}^{\star}$-structure, the pseudo STM images show the already familiar signature of a WT. Here the Ge NW atoms bind to the Pt atoms at the bottom of the trough, the sunken Pt atom and the second layer Ge atoms which are pushed up by the structure.\\
\begin{figure}[!tbp]
\begin{center}
  \includegraphics[width=8cm,keepaspectratio]{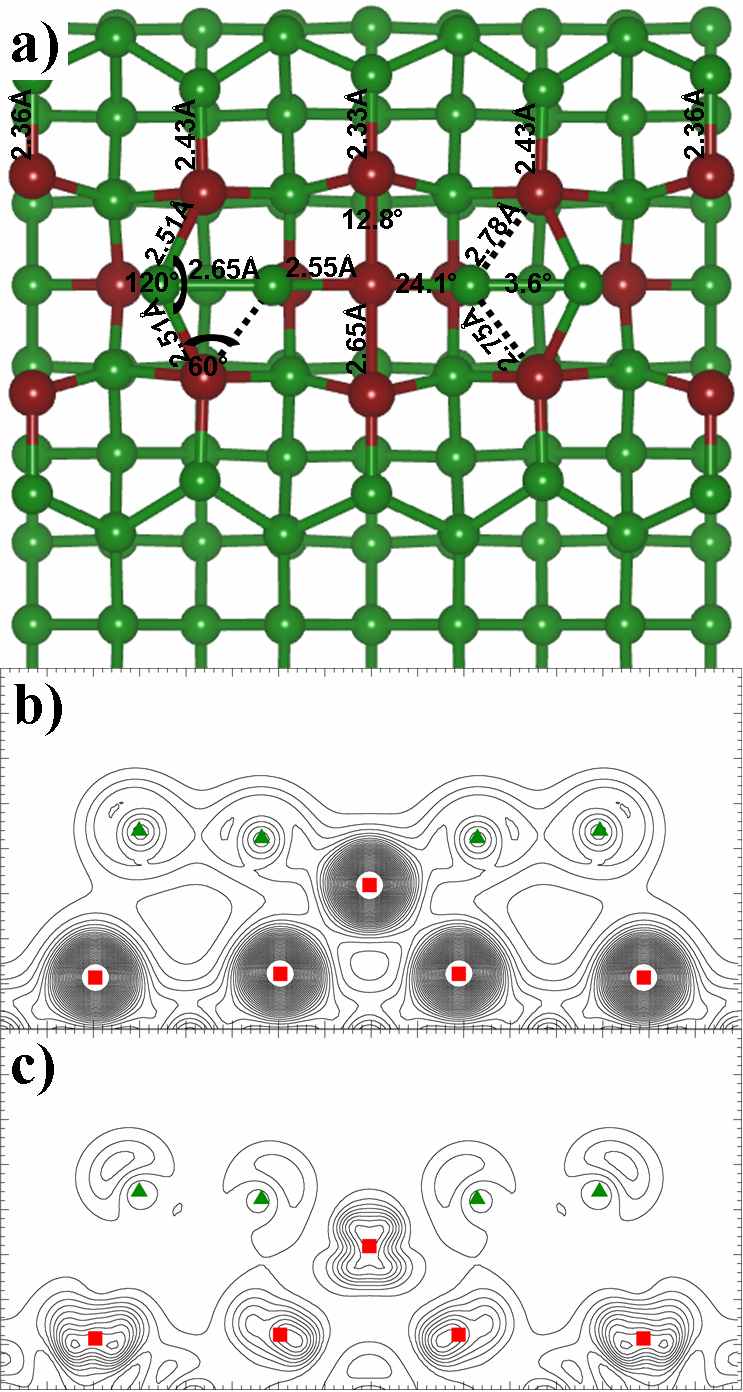}\\
\end{center}
  \caption{(color online) a) Geometric structure after relaxation of the Ge NW on the $\gamma_{as}^{\star}$-surface containing one extra Pt atom in the trough per two $(4\times2)$ surface unit cells (NW2 structure). Pt atoms are
  shown in red, Ge atoms in green. Angles shown on the bonds are with regard
  to the $xy$-plane. b) Contour plot of the total charge density of the adsorbed Ge NW, in the vertical plane along the NW. c) Same as b) but now for the partial charge density, for all energy levels going from $0.7$ eV below the Fermi level up to the Fermi level. Green triangles (red squares) indicate the position of Ge (Pt) atoms in the plane. The contours are separated $0.1$ ($0.01$) $e/$\AA$^3$ for the total (partial) charge density plot. }\label{fig:12b4as_ptL3_p2c025_nwge}
\end{figure}
\indent Because both the Ge NW on the $\gamma_{as}^{\star}$-surface and on the $\gamma_{as}^{\star}$-surface with one sunken Pt atom in the trough per surface cell show good agreement with experiment, and we want the Pt content to be as low as possible, a system with only one sunken Pt atom per two unit cells is examined (shown in Fig.~\ref{fig:12b4as_ptL3_p2c025_nwge}). After relaxation of this $4\times4$ surface cell a formation energy $E_{\mathrm{f}}=-2.914$ eV per surface unit was found and an adsorption energy per NW atom of $-0.710$ eV, in both cases roughly the average of Ge NW on the $\gamma_{as}^{\star}$-surface and on the $\gamma_{as}^{\star}$-surface with one sunken Pt atom in the trough per unit cell. There is however one important improvement over the smaller structures, the observed NW are double peaked and the two peaks are of a different height (Fig.~\ref{fig:11stm_b4as_ptL3_p2cX_GeNW}e) at negative simulated bias. The calculated empty state images in Fig.~\ref{fig:11stm_b4as_ptL3_p2cX_GeNW}f on the other hand shows the single peak image as observed in the experiment. In low temperature STM experiments the NWs in the NW-patches were observed to have a $4\times1$ periodicity along the NW. This periodicity however is absent for solitary NWs and the NWs at the edge of a NW-patch. The presence of this $4\times1$ periodicity was linked to the possible presence of a Peierls instability.\cite{Houselt:ss08} In Sec.~\ref{sc:discussion} we will show that this geometry models the NWs in NW arrays, and we refer the reader to that section for further details. At this point structures with up to one full ML of Pt incorporated have been examined and the results indicate a steady increase toward the formation of a platinum-germanide around the trough region. Further increase of the Pt content will only move further into that direction, increasing the overall stability of the system, but also moving away from the NW arrays, \textit{i.e.}\ the calculated STM images start to diverge from the experimental STM images.\\
\ \\
\indent As alternatives to increasing the Pt density in the Pt lined troughs, we consider the following ideas: Would it be possible to remove some of the Pt atoms in the top layer while maintaining good agreement with the experiment in light of reducing the necessary gradient in the Pt density? Could/should more Pt be located at the bottom of the empty trough lined with Ge atoms?
\begin{table}[!tb] \center{\textbf{Formation and adsorption
energy.}\\}
\begin{ruledtabular}
\begin{tabular}{l|rcrcrcrc}
   & \multicolumn{2}{c}{Bare}  & \multicolumn{4}{c}{With adsorbed} &
   \multicolumn{2}{c}{Formation energy}\\
   & \multicolumn{2}{c}{structure}  & \multicolumn{4}{c}{Ge NW} &
   \multicolumn{2}{c}{of Pt-$\star$ atoms.}\\
   & \multicolumn{2}{c}{\makebox[2.0cm]{$E_{\mathrm{f}}$}}
   & \multicolumn{2}{c}{\makebox[1.5cm]{$E_{\mathrm{f}}$}}
   & \multicolumn{2}{c}{\makebox[1.5cm]{$E_{\mathrm{ad}}$}}
   & \multicolumn{2}{c}{\makebox[2.0cm]{$E_{\mathrm{f}}$}}\\
   & \multicolumn{2}{c}{(meV)} & \multicolumn{2}{c}{(meV)} &
   \multicolumn{2}{c}{(meV)} & \multicolumn{2}{c}{(meV)}
\\
  \hline
  Ge$^{\star}$ & \makebox[1.25cm][r]{194} & & \makebox[1.15cm][r]{503} & &
  \makebox[1.15cm][r]{154} & & \makebox[1.6cm][r]{97} & \\
  $\beta_{6}^{\star}$ & $-269$ & & $-459$ & & $-94$ & & $-98$ &\\
  $\beta_{3}^{\star}$ & $-390$ & & $-1290$ & & $-450$ & & $-229$ &\\
  $\gamma_{as}^{\star}$ & $-1011$ & & $-2055$ & & $-522$ & & $-383$ &\\
  $\gamma_{as}^{\star as}$ & $-236$ & & - & & - & & 5 &\\
\end{tabular}
\end{ruledtabular}
\caption{Formation and adsorption energies of bare and Ge NW adsorbed structures containing Pt atoms at the bottom of one Pt-lined trough (the Pt-$\star$ atoms). The $\gamma_{as}^{\star as}$-structure refers to a $\gamma_{as}^{\star}$-structure but with Pt atoms at the bottom of the Ge-lined trough instead of the Pt-lined trough. The right column gives the formation energy per Pt atom of the Pt atoms at the $\star$-positions.}\label{table:5Why_not_second_trough}
\end{table}
\indent To answer these questions we investigate the stability of some modified geometries and compare them to the $\gamma_{as}^{\star}$-geometry. The Ge$^{\star}$-geometry, is just a plain Ge(001) surface with Pt atoms at the bottom positions of every second trough. The same holds for the $\beta_{6}^{\star}$-geometry. The $\beta_{3}^{\star}$-geometry has Pt atoms at the bottom trough positions of the trough lined by the Pt atoms in the top layer.\cite{Vanpoucke:prb09beta} And finally, the $\gamma_{as}^{\star as}$-surface is a $\gamma_{as}$-surface with Pt atoms at the bottom trough positions of the Ge lined trough (contrary to the $\gamma_{as}^{\star}$-surface where the Pt atoms are in the Pt lined trough).\\
\indent Table~\ref{table:5Why_not_second_trough} shows the formation and adsorption energy for these structures with and without adsorbed Ge NWs. The formation energies of the bare surfaces show a trend of stabilization with increasing Pt density. Contrary to the surface without Pt in the top layer (Ge$^{\star}$), surfaces containing Pt in the top layer stabilize when a Ge NW is adsorbed. Comparison of the $\beta_6^{\star}$- and $\beta_3^{\star}$-structures with adsorbed Ge NW shows for the latter a more symmetric structure. The large increase in stability between these $2$ systems is caused by different number of formed Pt-Ge bonds. For the $\beta_6^{\star}$-surface both Ge NW atoms only bind to a single Pt atom. These Pt-Ge bonds have a bond length of $2.51$ \AA\ while the distance of the Ge NW atoms to the Ge surface dimer atom opposite to this Pt atom is $4.20$ \AA, \textit{i.e.}\ the Ge NW dimer is only bound to the side of the trough containing Pt atoms. This places the NW asymmetric at one side of the trough. In the $\beta_3^{\star}$-geometry, the Ge NW dimer is bound to two opposing Pt surface atoms, with Pt-Ge bonds of $2.60\pm0.01$ \AA. It shows that to have a NW centered in the trough, the Pt atoms need to be present at both sides of this trough in such a configuration that bonds can be formed with the Ge NW dimer.\\
\indent The influence of these Pt-Ge NW bonds becomes even clearer when comparing the adsorption energies. For the Ge$^{\star}$-surface no such bonds appear while one set is present for the $\beta_6^{\star}$-structure and two for the $\beta_3^{\star}$ and $\gamma_{\mathrm{as}}^{\star}$, resulting in a comparable adsorption energy for the last two. Neglecting for a while the difference in formation energy between the bare $\beta_3^{\star}$- and $\gamma_{\mathrm{as}}^{\star}$-surfaces on grounds of different Pt stoichiometry, one might consider the $\beta_3^{\star}$ as a low Pt density replacement for the $\gamma_{\mathrm{as}}^{\star}$-surface.
\begin{figure}[!tbp]
\begin{center}
  \includegraphics[width=8cm,keepaspectratio]{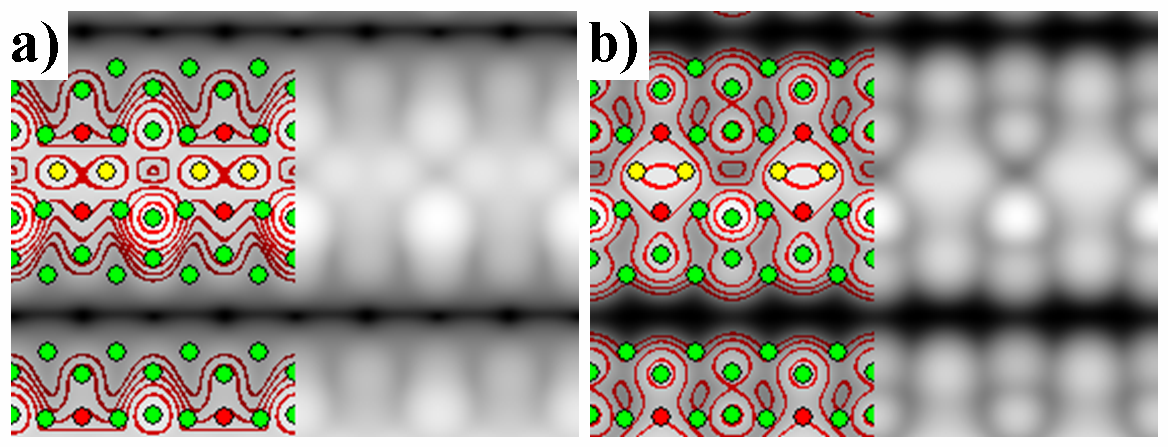}\\
\end{center}
  \caption{(color online) Calculated STM images of a Ge NW adsorbed on a $\beta_{3}^{\star}$-surface. The green/red (light/dark gray) discs show the positions of Ge/Pt atoms in the top layers of the surface. The yellow (white) discs show the positions of the Ge NW atoms. Contours are added to guide the eye. a: Calculated filled state STM image, for $z=3.0$ \AA\ and a simulated bias of $-1.50$ V. b: Calculated empty state STM image, for $z=3.0$ \AA\ and a simulated bias of $+1.50$ V.}\label{fig:15stm_b3_ptL3_GeNW}
\end{figure}
Calculated STM images for a Ge NW on the $\beta_3^{\star}$-surface are shown in Fig.~\ref{fig:15stm_b3_ptL3_GeNW}. Comparison to Fig.~\ref{fig:9stm_b4as_b4as2_ptL3}a and b however shows where the $\beta_3^{\star}$-reconstruction fails. The typical symmetric bulges around the wire show up slightly asymmetric. The NW STM image in both the empty and filled state images is not even half the height of the side bulges. From this we conclude that no Pt can be removed from the surface without loosing agreement with the experiment.\\
\indent The second concern entails why Pt in the $\star$-geometries should only be located in the Pt lined troughs. Tackling this concern however is rather simple. Combining the fact that the Ge$^{\star}$ structure is almost $0.5$ eV per $4\times2$ surface unit cell less stable than the $\beta_6^{\star}$ structure, with the trend of increasing stability with increasing Pt density in the top layer shown in Table~\ref{table:5Why_not_second_trough}, indicates it to be preferable for Pt to build into the bottom of the trough lined with Pt atoms. Looking specifically at the Ge$^{\star}$-structure we see that the Pt atoms substituting Ge atoms at the bottom of the trough actually destabilize the structure while the structures are stabilized in the other cases. Also the comparison of the $\gamma_{\mathrm{as}}^{\star}$- to the $\gamma_{\mathrm{as}}^{\star as}$-structure, with the Pt $\star$-atoms at the bottom of the Ge lined trough, show a preference for the first structure, and no stabilization of the $\gamma_{as}$-structure due to these Pt atoms. This allows us to conclude that no Pt needs to be added to the Ge lined trough.
\section{Discussion}\label{sc:discussion}
\subsection[Nanowires and formation paths]{Possible formation paths of nanowires}\label{ssc:discuss_modeling}
\begin{figure}[!tb]
\begin{center}
  \includegraphics[width=8cm,keepaspectratio]{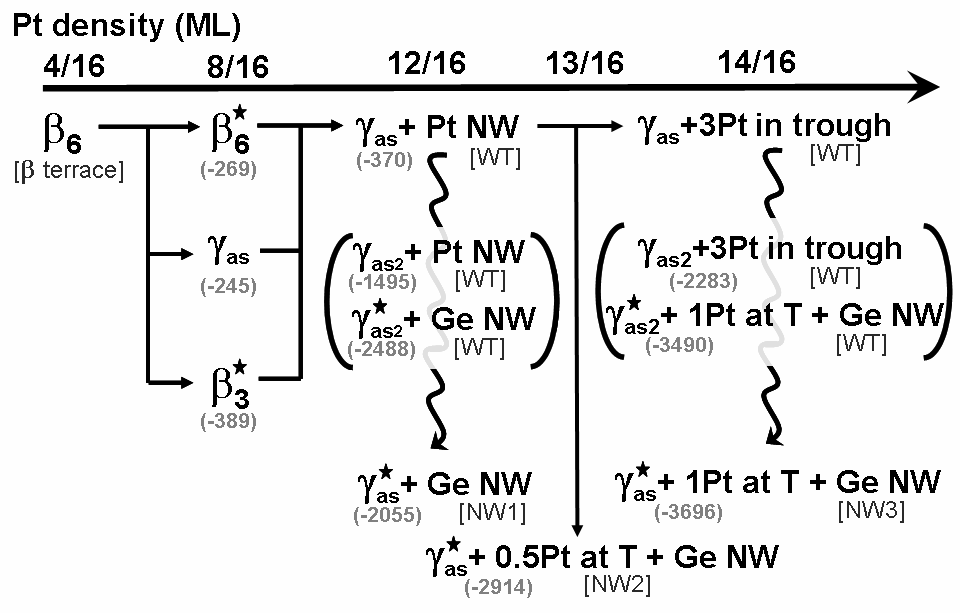}\\
\end{center}
  \caption{Schematic diagram of a possible formation path for NW arrays on
  Ge(001) with regard to increasing local Pt density, given from left to right
  in ML. Straight arrows indicate paths with increasing Pt density, curved arrows indicate possible relaxation paths during and after high temperature
  annealing. LDA formation energies are given in meV with regard to
  the Ge(001) $b(2\times1)$ surface reconstruction. WT and NW
  indicate if the pseudo-STM pictures show widened trough (WT) like
  images or nanowire (NW) like images. Formation paths are
  discussed in the text.}\label{fig:10formationdiagram}
\end{figure}
Based on a comparison of these results with experiments, a growth path as function of increasing Pt density can be suggested for the NWs. From experiments it is clear that the anneal temperature after deposition is crucial in the formation process. Firstly, the high temperature should cause the Pt atoms not to move deeper into the substrate, but rather move back to the surface. This requires the solubility of Pt in Ge to be very small, which is the case.\cite{Massalski:BAPD90} Secondly, this high temperature is needed to break the Ge surface dimers such that Pt atoms can replace Ge atoms and form surface Pt-Ge dimers. This bond breaking temperature is reported to be approximately \mbox{$1000$ K} for the Ge(001) surface.\cite{Zeng:ss02} Thirdly, in combination with the anneal time, the high annealing temperature allows for larger kinetic barriers to be crossed, allowing the WT to transform into a NW.\\
\indent Figure~\ref{fig:10formationdiagram} shows a diagram indicating possible formation paths we will discuss here. In our discussion we will assume that the high annealing temperature causes a large portion of the Pt that moved subsurface during deposition to be ejected onto the surface again. Starting with $0.25$ ML Pt atoms being ejected, and the high anneal temperature breaking the surface Ge dimers, the ejected Pt atoms can now form Pt-Ge dimers in the top layer of the surface, while the second Ge atom of the former Ge surface dimer can move away to a step edge. The formation of Pt-Ge dimers can be seen as a consequence of the fact that Pt-Ge bonds are preferred over Ge-Ge bonds, the larger bond strength allows for the Pt-Ge dimers to replace the broken Ge-Ge surface dimers. The most homogeneous way of spreading $0.25$ ML of Pt atoms in the top layer is the $\beta_{6}$-geometry, which we showed in previous work to be the geometry of the $\beta$-terrace.
\cite{Vanpoucke:prb09beta,vanpoucke:prb2008R}\\
\indent If locally $0.5$ ML of Pt is ejected onto the surface, either more Pt-Ge surface dimers can be formed, creating a surface only containing Pt-Ge dimers, or the extra Pt atoms can move to Ge positions at the bottom of the trough ($\star$-geometries). In case of a surface purely made up out of Pt-Ge dimers it was shown in Ref.~\onlinecite{Vanpoucke:prb09beta} that a $\gamma_{as}$-surface is most stable. The formation of surface Pt-Ge dimers could prohibit these Pt atoms to move subsurface. However if this barrier is not large enough or the preference of forming Pt-Ge surface dimers is too small the $\beta_{6}^{\star}$- and $\beta_{3}^{\star}$-surfaces become available which are slightly more stable. Further STM studies of the surface near the edge between the $\beta$-terrace and a NW-patch could reveal which of these three structures is present. In the filled state images the $\beta_{6}^{\star}$- and normal $\beta_{6}$-surface are nearly indistinguishable, but at positive bias, close to the Fermi level the typical triangular structures of the $\beta_{6}$-surface are absent for the $\beta_{6}^{\star}$-surface.\cite{Vanpoucke:prb09beta} This again shows that the $\beta$-terrace only contains $0.25$ ML of Pt in its surface layers.\\
\subsubsection{Solitary NWs.}
If the ejected amount of Pt increases up to $0.75$ ML, $0.5$ ML can bind to Ge atoms of the top layer transforming all surface dimers into Pt-Ge dimers forming a $\gamma_{as}$-surface. The remaining $0.25$ ML end up in the Pt lined trough where they sink in, appearing as a WT. During this process the $\gamma_{as}$-reconstruction transforms into the $\gamma_{as2}$-reconstruction, also appearing as a WT. In the next step of the NW formation, the $0.25$ ML of Pt atoms located in the Pt-lined trough exchange places with the Ge atoms at the bottom of this trough. The new structure $\gamma_{as2}^{\star}$ $+$ Ge NW, still resembles the experimentally observed WT (\textit{cf.}\ Fig.~\ref{fig:9stm_b4as_b4as2_ptL3}c and d). To elevate the Ge NW out of this trough and transform the $\gamma_{as2}^{\star}$-geometry back to the $\gamma_{as}^{\star}$-geometry a considerable amount of energy needs to be put into the system. At the end of this process a $\gamma_{as}^{\star}$-surface $+$ Ge NW is formed, we will refer to this structure as the NW$1$ structure.\\
\indent The fact that WTs are observed experimentally when only very short anneal times are used, suggests that the high anneal temperature has an important role in the formation process, however, it can not explain why no WTs, due to the $\gamma_{as2}^{\star} +$ Ge NW, are present after a longer high temperature anneal step. Because the $\gamma_{as2}$-surface is much lower in energy than the $\gamma_{as}$-surface, the high temperature anneal step will only affect their relative fraction, and the amount of $\gamma_{as}$-surface will always be smaller than the amount of $\gamma_{as2}$-surface.\cite{fn:GGAcompar} The fact that no large amounts of WT (due to an $\gamma_{as2}$-geometry) are observed in experiment must have a different reason. If we look at the $\gamma_{as2}$-geometry in Fig.~\ref{fig:4bis_gamma_as2} we note quite a strong deformation of the surface with regard to the $\gamma_{as}$-geometry (\textit{cf.}\ Fig.~\ref{fig:4geom_set2}b). The Pt-Ge dimers tilting into the substrate is the most striking feature. In the above formation path we assume the $\gamma_{as}$-surface transforms into the $\gamma_{as2}$-surface, and then needs to transform back to a $\gamma_{as}$ type surface ($\gamma_{as}^{\star}$). However, if the barrier for this $\gamma_{as}\rightarrow\gamma_{as2}$ (and corresponding $\gamma_{as}^{\star}\rightarrow\gamma_{as2}^{\star}$) transformation is too high, then the first transformation in the suggested formation path would not occur, and there would be no need for the second transformation back afterwards. The lack of WT structures in the experiments seems to support the idea that the barrier between the $\gamma_{as}$- and $\gamma_{as2}$-structures is too large to be overcome.\\
\indent As a result, the suggested formation path simplifies drastically. The $0.25$ ML of Pt atoms in the Pt lined trough of the $\gamma_{as}$ just sinks in, making the $\gamma_{as}$-surface appear in calculated STM images as a WT. Then these Pt atoms exchange positions with the Ge atoms at the bottom of the Pt lined trough, transforming the $\gamma_{as}$-surface directly into a $\gamma_{as}^{\star}$-surface, while the exchanged Ge atoms dimerise and form the NW observed in experiment. Although the exchange might be energetically favorable, kinetic barriers could limit this exchange, making the high temperature anneal step crucial.\\
\indent This NW structure shows a very good agreement with the experimentally observed NWs (\textit{cf.}\ Fig.~\ref{fig:9stm_b4as_b4as2_ptL3}a and b). The double peaked NW dimer images are present in the filled state image, while a single peak dimer image is shown for the empty state image. The wire image is located nicely centered in the trough between the QDRs and symmetric bulges are visible as in experiment. In low temperature experiments however, a $4\times1$ periodicity along the NW appears for the NWs in the patches. For the NW at the edge of a NW-patch or for a solitary NW this
$4\times1$ periodicity does not appear.\\
\indent The unit cell used in the calculations above has a $4\times2$ surface periodicity, with a single NW dimer ($2\times1$) on top, making it too small to observe this $4\times1$ periodicity. To be able to make a comparison with experiment two calculations with a large super cell ($4\times8$ surface cell with four NW dimers on top) are carried out. In the first case the NW$1$ geometry is build without modifications, while in the second case the Ge NW dimers are tilted according to the observed $4\times1$ periodicity. After relaxation of both structures, the formation energies are calculated and pseudo-STM pictures are generated. The formation energies per $4\times2$ surface cell of both systems is the same (within the margin of error), $E_{\mathrm{f}}=-2079$ meV per $4\times2$ surface cell, as for the NW$1$ structure. Furthermore, the pseudo-STM images shows the exact same results. The NW dimers, in the cell with the tilted NW dimers, have flattened out to a tilt angle of $<0.2^{\circ}$. In correspondence to this very small tilt angle an equally small difference in the height of the two NW dimer peaks in the pseudo-STM images is observed. This indicates that the observed $4\times1$ periodicity, and the related tilting of the NW dimers, probably has a structural cause rather than being the signature for a Peierls instability.\\
\indent The above results also suggest the solitary NWs and the NWs at the edge of a NW patch to be different from those inside the patch. In experiment the difference between the array NWs and the NWs at the edge or solitary NWs presents itself in the appearance of the NW dimers. For the solitary and edge NWs symmetric NW dimers with a $2\times1$ geometry are found, while the array NWs show asymmetric NW dimers with a $4\times1$ geometry. Based on the good agreement with the experimental STM images we propose the NW$1$ geometry (\textit{cf.}\ Fig.~\ref{fig:8b4as_ptL3_nwge}) as model for
the solitary NWs and the NW at the edge of a NW-patch.\\
\indent In this model the NW consists of flat stretched Ge dimers with a bond length of $2.72$ \AA. These dimers are bound to two opposing Pt atoms in the top layer, pinning the Ge NW dimers at the center of the trough. In addition, weak bonds with a length of $3.13$ \AA\ to the Pt atoms at the bottom of the trough are present as can be seen in the charge density plots (Fig.~\ref{fig:8b4as_ptL3_nwge}b and c). The Pt-Ge dimers bound to the NW are pulled slightly inward to the trough
and stretched about $3$\% compared to the other Pt-Ge dimers.\\
\begin{figure}[!t]
\begin{center}
  \includegraphics[width=8cm,keepaspectratio]{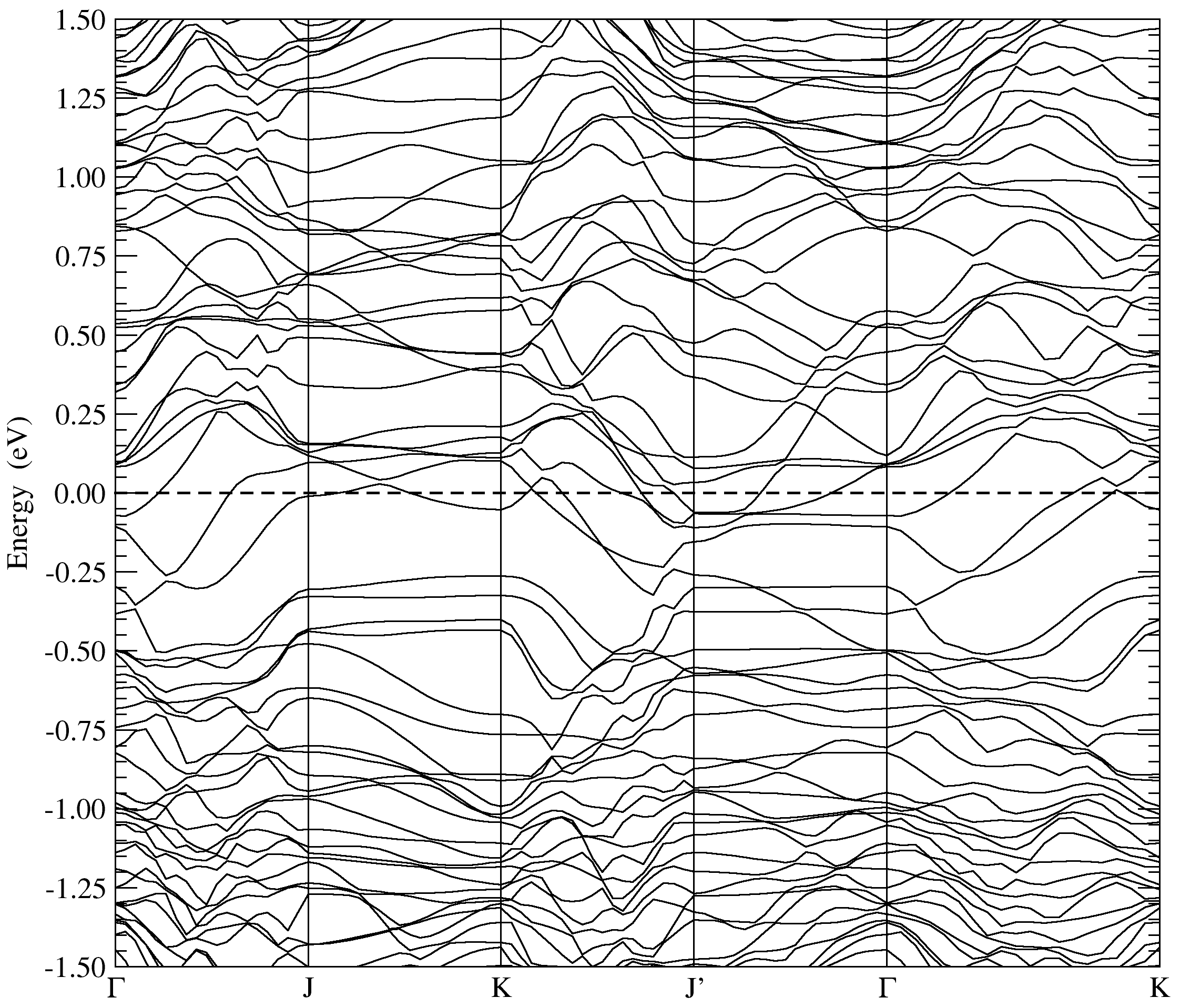}\\
\end{center}
  \caption{Band structure of the NW$1$ model shown in
  Fig.~\ref{fig:8b4as_ptL3_nwge}.
  The energy zero is given by the Fermi level.}\label{fig:16BANDS_NW1}
\end{figure}
\indent Figure~\ref{fig:16BANDS_NW1} shows the electronic band structure of the NW$1$ model along the (high) symmetry lines $\Gamma$-J-K-J$^{\prime}$-$\Gamma$-K of the surface BZ. This shows a metallic behavior, as observed experimentally.\cite{Gurlu:apl03} The bands near the Fermi level along the J-K and J$^{\prime}$-$\Gamma$ lines (perpendicular to the NW direction) show very little dispersion, while those along the $\Gamma$-J and K-J$^{\prime}$ lines (parallel to the NW direction) show a large dispersion. Close to the Fermi level the bands have a predominant Pt character. The band crossing the Fermi level near the middle of the $\Gamma$-J line (\textit{cf.}\ Fig.~\ref{fig:16BANDS_NW1}) characterizes the $\pi$-bond of the Pt-Ge dimers which are not bound to the NW, while the band below is the one related to the bonds between the NW and the surface Pt atoms. This last band comes close to the Fermi level near the J point, where the character of the band has become mainly that of the Pt trough atoms, with a largely $d_{\mathrm{xy}}$ and some $d_{\mathrm{xz}}$ or $d_{\mathrm{yz}}$ orbital character. The remaining Ge NW orbital character of this band however is still $sp_3$.\\
\indent The orbital character of the Ge NW contribution to these bands near the Fermi level is purely $sp_3$, while the Pt atoms in the trough bottom have a mainly $p_{\mathrm{z}}$ and $d_{\mathrm{z^2}}$ character and the top layer Pt atoms a $d_{\mathrm{xz}}$ and $d_{\mathrm{yz}}$ character.\\
\indent Around the middle of the J-K line there is a band crossing just above the Fermi level. The character of these crossing bands again has a large contribution from the Pt atoms at the trough bottom and the Ge NW. For both bands there is also a large contribution from two top layer Pt atoms. The interesting part here is that it are the Pt atoms of the same QDR, and each time one Pt atom (the one bound to the NW) shows a mainly $p_{\mathrm{z}}$ and $d_{\mathrm{z^2}}$ orbital character, while the second Pt atom presents a mainly $d_{\mathrm{xz}}$ and $d_{\mathrm{yz}}$ orbital character, indicating the presence of empty bonds just above the Fermi level between all top layer Pt atoms and the NW along the NW direction.\\
\begin{figure}[!t]
\begin{center}
  \includegraphics[width=8cm,keepaspectratio]{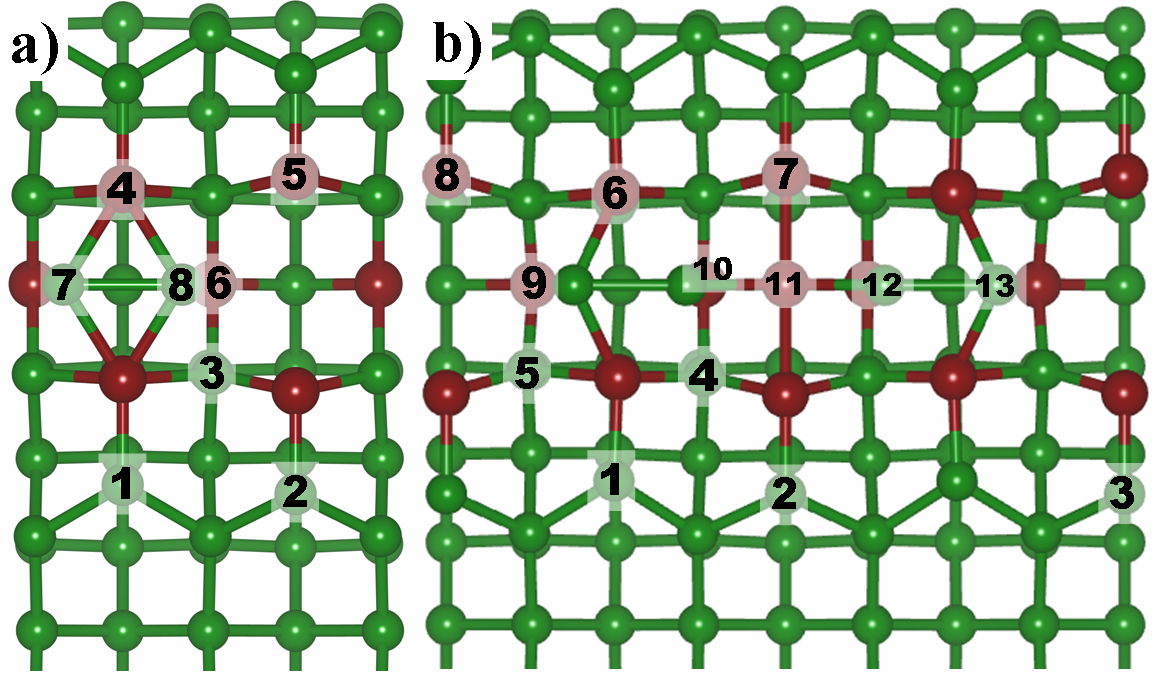}\\
\end{center}
  \caption{(color online) Ball-and-stick representations of the relaxed NW$1$ (a) and NW$2$
  (b) models indicating the (inequivalent) atoms of which the LDOS is shown in
  Fig.~\ref{fig:18DOS_NW1} and \ref{fig:19DOS_NW2}. }\label{fig:20DOS_atomslocations}
\end{figure}
\begin{figure}[!t]
\begin{center}
  \includegraphics[width=8cm,keepaspectratio]{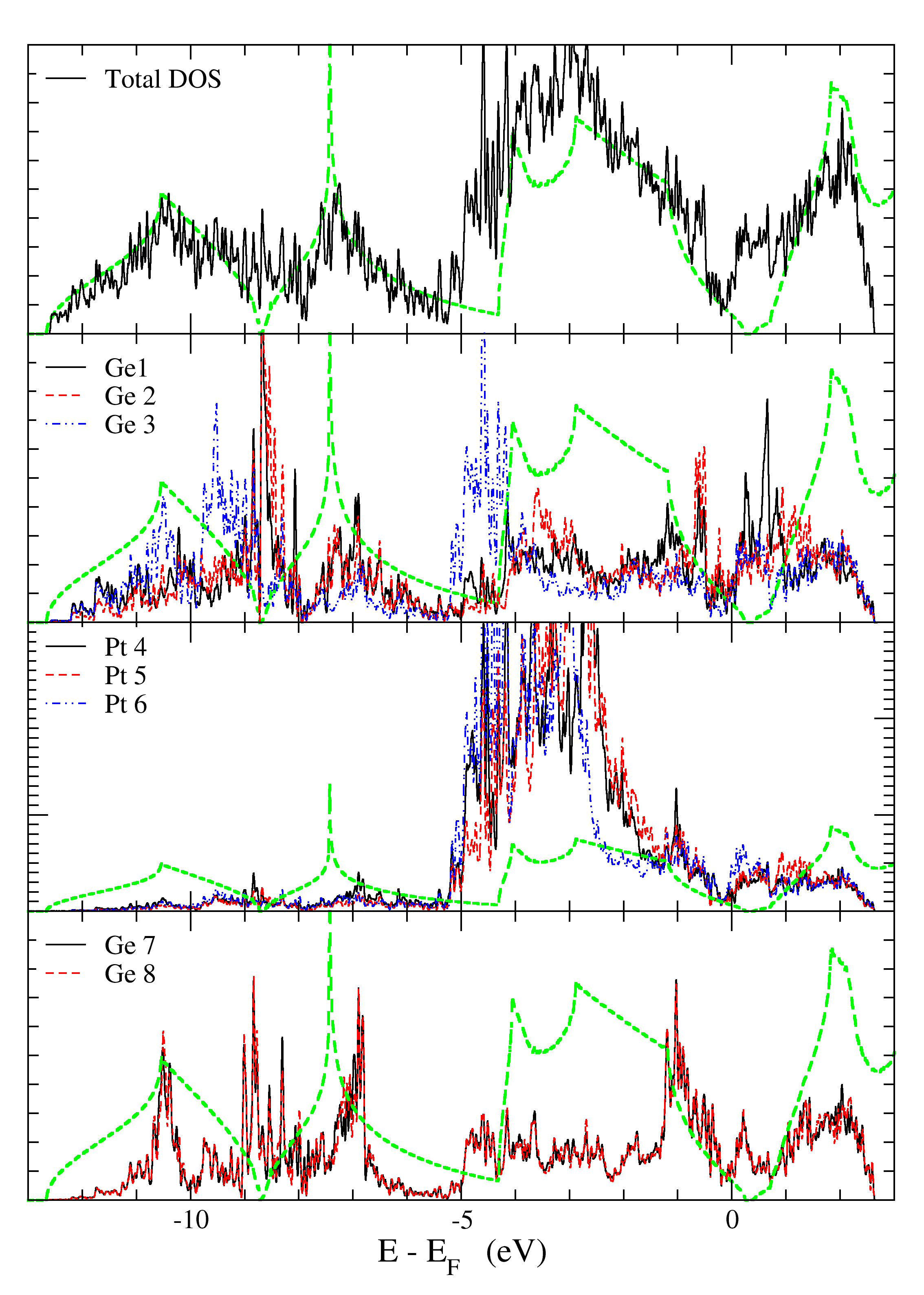}\\
\end{center}
  \caption{(color online) Total DOS of the NW$1$ structure and
  LDOS of the surface atoms as shown in Fig.~\ref{fig:20DOS_atomslocations}a. The bold dashed green line shows the DOS of Bulk Ge, it was shifted to align the BG regions.}\label{fig:18DOS_NW1}
\end{figure}
\indent Figure~\ref{fig:18DOS_NW1} shows the total density of states (DOS) of the NW$1$ model and the LDOS of the different inequivalent surface atoms, as indicated in Fig.~\ref{fig:20DOS_atomslocations}. The total DOS shows this model to be metallic, with a dip in the DOS just below the Fermi level. The general trend of the total DOS follows that of the DOS of bulk Ge quite well, with the exception of extra states in the band gap (BG) region and some states near $5$ and $9$ eV below the Fermi level. From the LDOS we learn the peaks at $-5$ eV to be Pt states, and Pt induced states in second layer Ge atoms. The extra states at $-9$ eV come from surface Ge atoms and the Ge NW atoms.\\
\indent The states in the bulk BG are most interesting, as they determine the properties of the system around the Fermi level. The LDOS of the surface atoms shows clearly that the major contribution comes from the Pt-Ge surface dimers. Only a minor contribution comes from the NW itself. Although presenting a metallic behavior, the NW (Ge atoms $7$ and $8$ in Fig.~\ref{fig:18DOS_NW1}) only shows an increased contribution just above the Fermi level, and follows the Ge bulk DOS along the edge of the valence band. The experimental observation of confined states between wires with a separation of $2.4$ nm, can thus be interpreted as the observation of the states just above the Fermi level of the Ge and the Pt atoms in the Pt-Ge surface dimers (Ge atoms $1$ and $2$, and Pt atoms $4$ and $5$).\cite{Oncel:prl05,Houselt:nanol06}
\subsubsection{Nanowire patches}
To find the geometry of the NWs inside the patches, a higher local Pt density is assumed. If $0.875$ ML of Pt is ejected, and we adopt the same argument as given in the previous paragraph, three Pt atoms will end up in the Pt lined trough.\cite{fn:PerSurfUCell} Again an exchange of two Pt atoms with the two Ge atoms in the bottom of the trough takes place, resulting in the $\gamma_{as}^{\star}$-geometry containing one Pt atom and a Ge dimer in the trough. Based on our discussion for the NW$1$ formation path, we can also assume here that the energy barrier is too high to allow for the transformation from the $\gamma_{as}$- to $\gamma_{as2}$-substrate structure. In this case, however, even if such a transformation would take place the transformation back from the $\gamma_{as2}^{\star} +$ $1$ Pt at T + Ge NW to $\gamma_{as}^{\star} +$ $1$ Pt at T + Ge NW would be energetically favorable, removing the problem that existed for the NW$1$ formation path.\\
\indent In this scenario all steps lead to energetically more favorable structures making it more likely to happen spontaneously.\cite{fn:GGAsame} The pseudo-STM images of the final structure, shown in Fig.~\ref{fig:11stm_b4as_ptL3_p2cX_GeNW}c and d, also show a NW structure, however the double peaked NW images at positive simulated bias are somewhat unsatisfying (\textit{cf.}\ Sec.~\ref{ssc:model4_npu}). Therefore we investigate a system with the average Pt density of the system just described and the NW$1$-system.\\
\indent The final structure is again a $\gamma_{as}^{\star}$-surface but now containing one additional Pt atom in the trough per two surface unit cells and one Ge dimer per surface unit cell. We will refer to this $\gamma_{as}^{\star} + 0.5$Pt $+$ Ge NW structure as NW$2$. In this NW$2$-geometry, shown in Fig.~\ref{fig:12b4as_ptL3_p2c025_nwge}a, the two Ge NW dimers are bound to the extra Pt atom in the trough. This stabilizes the NW dimers at this position in the trough and tilts the Ge NW dimers over an angle of $3.6^{\circ}$. This tilting causes a $4\times1$ periodicity along the wire. It is also clearly visible in the pseudo-STM images, where it causes a height difference of $0.36$ \AA\ between the peaks of the NW dimer in the filled state image at a simulated bias of $-0.7$ V. The empty state image shows a single peaked dimer image as is experimentally observed. Also the symmetric bulges are clearly visible, resulting in an excellent agreement between the pseudo-STM images of the NW$2$ structure and the low temperature STM images of NWs inside a NW-patch. Fig.~\ref{fig:12b4as_ptL3_p2c025_nwge}a shows the relaxed
NW$2$ structure we propose as model for the observed NWs inside a NW-patch.\\
\indent In this model a periodicity doubling occurs with regard to the NW$1$-model. The Ge dimers forming the NW are, with a bond length of $2.65$ \AA, less stretched than is the case for the NW$1$-model.
The Pt-Ge dimers of the QDR bound to the NW are pulled slightly inward to the trough and stretched by $3$\%. Just like the NW$1$-model a weak bond between the NW and the Pt atoms in the trough bottom is present, as can be seen in the charge density contours of Fig.~\ref{fig:12b4as_ptL3_p2c025_nwge}b and c. In this case the bond length with the up-Ge NW atom is $3.40$ \AA, and
$3.04$ \AA\ with the down-Ge NW atom.\\
\begin{figure}[!t]
\begin{center}
  \includegraphics[width=8cm,keepaspectratio]{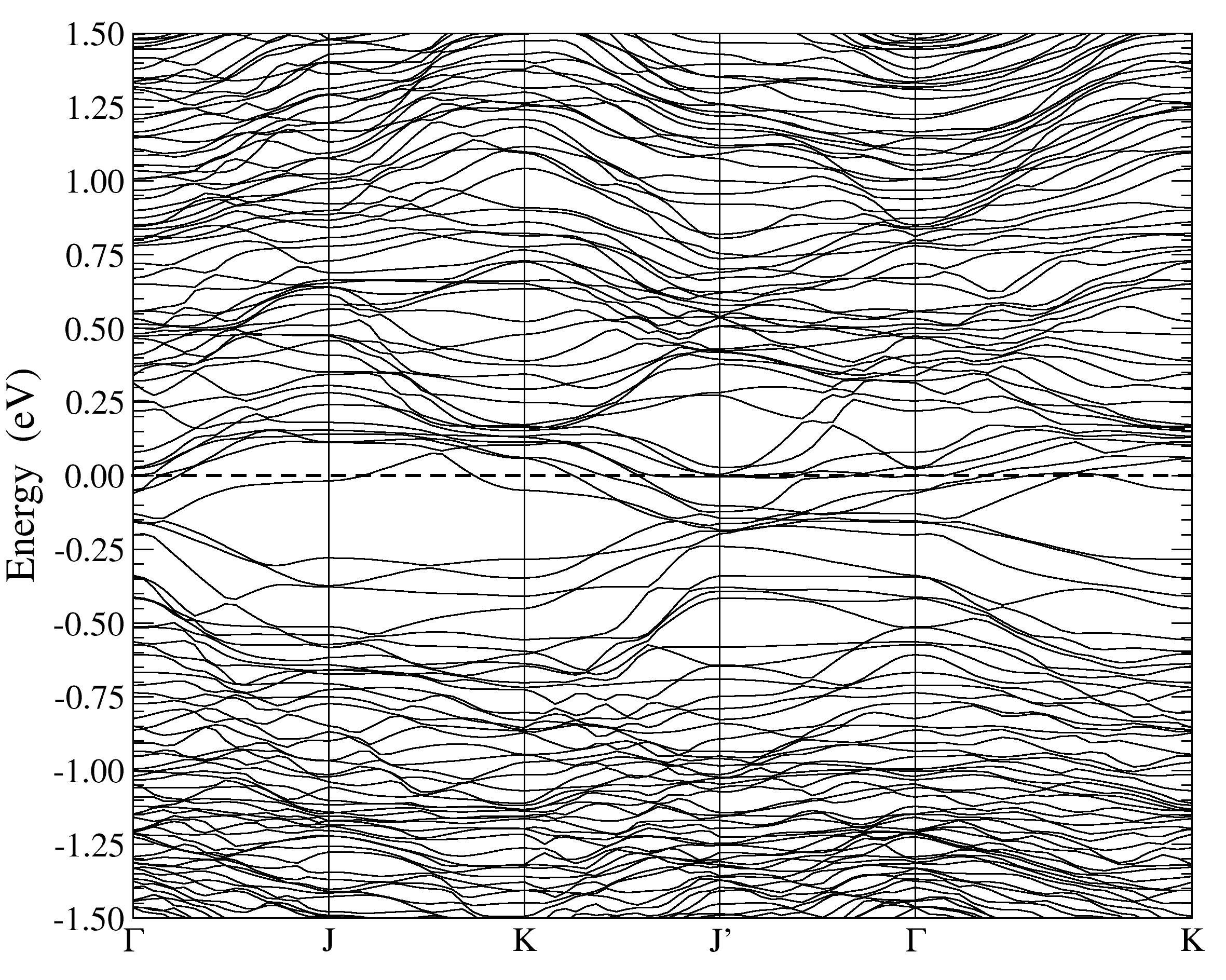}\\
\end{center}
  \caption{Band structure of the NW$2$ model shown in
  Fig.~\ref{fig:12b4as_ptL3_p2c025_nwge}.
  The energy zero is given by the Fermi level.}\label{fig:17BANDS_NW2}
\end{figure}
\indent Figure~\ref{fig:17BANDS_NW2} shows the band structure of the NW$2$-model. It shows quite similar trends as seen for the NW$1$-structure. Close to the Fermi level the bands have almost no dispersion. However, contrary to the NW$1$-model, even along the J-K and J$^{\prime}$-$\Gamma$ lines the bands close to the Fermi level show almost no dispersion. For the J$^{\prime}$-$\Gamma$ line there is even a band located roughly on top of the Fermi level along the entire line. This partially filled band has mainly a $p_{\mathrm{z}}$ orbital character and can be traced back to the Ge atoms of the surface Pt-Ge dimers bound to the NW dimers, making them responsible for the experimentally observed confined states between the
NWs in the NW patches.\cite{Oncel:prl05,Houselt:nanol06}\\
\indent The two bands crossing just above the Fermi level along the J-K line, also seen for the NW$1$-model, have a slightly different character now. For the NW$2$-model the orbital character is dominated by the contribution of the extra Pt atom in the trough (Pt $11$ in Fig.~\ref{fig:20DOS_atomslocations}b), the two Pt atoms in the top layer bound to this Pt atom (both Pt $7$ in Fig.~\ref{fig:20DOS_atomslocations}b) and the Ge atoms forming a Pt-Ge dimer with the previous two Pt atoms (both Ge $2$ in Fig.~\ref{fig:20DOS_atomslocations}b). All these atoms give a strong $z$-oriented orbital contribution to these bands, showing the presence of a conduction band just above the Fermi level directed perpendicular to the NWs, but more importantly, connecting two sides of a NW, making the electronic structure of a NW patch two dimensional ($2$D). The character of this band is even maintained along the $\Gamma$-J line, where it is located just below the Fermi level. This shows that NW patches are not purely $1$D systems, but also contain a $2$D (in plane) component, in contrast with the solitary NWs.\\
\begin{figure}[!tb]
\begin{center}
  \includegraphics[width=8cm,keepaspectratio]{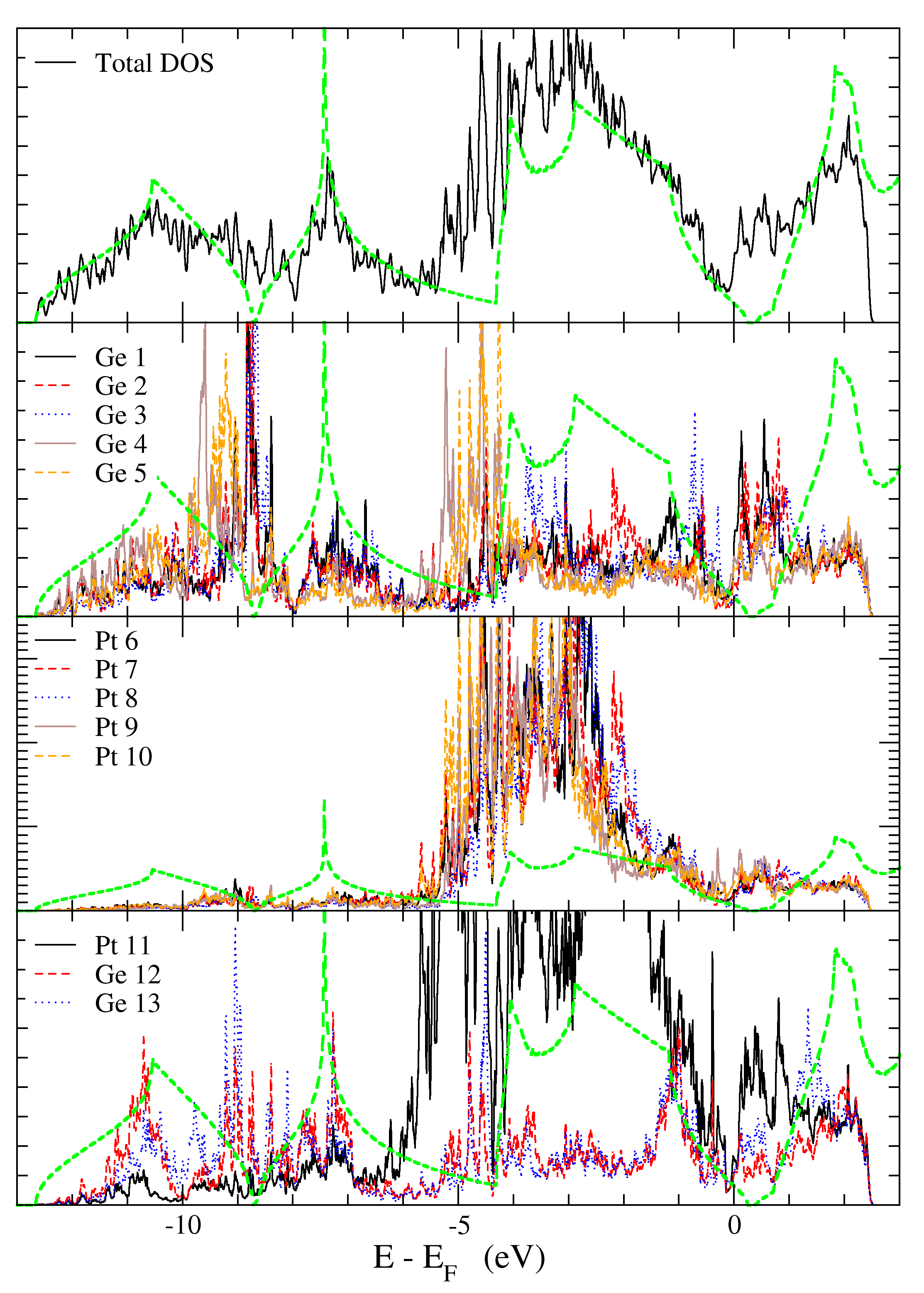}\\
\end{center}
  \caption{(color online) Total DOS of the NW$2$ structure and
    LDOS of the surface atoms as shown in
    Fig.~\ref{fig:20DOS_atomslocations}b. The bold dashed green line
    shows the DOS of Bulk Ge, it was shifted to align the BG
    regions.}\label{fig:19DOS_NW2}
\end{figure}
\indent Figure~\ref{fig:19DOS_NW2} shows the total DOS of the NW$2$-model and the LDOS of the surface atoms as indicated in Fig.~\ref{fig:20DOS_atomslocations}. The total DOS is nearly identical to the one found for the NW$1$-model, with extra states around $-5$ and $-9$ eV. These states can be attributed to the Pt and the second layer Ge atom states for the first state at $-5$ eV, and surface Ge and Ge NW atom states for the second state at $-9$ eV. The main contribution to the states located in the bulk BG comes from the Pt-Ge surface dimers, similar as was seen for the NW$1$-model. Also similar as to the NW$1$-model is the contribution of the Ge NW atoms, although in this case, the asymmetry in the geometry is also visible in the LDOS, resulting in small differences in the LDOS of the two inequivalent Ge NW atoms (\textit{cf.}\ bottom part of Fig.~\ref{fig:19DOS_NW2}). Although this LDOS does not show a BG, the LDOS of the NW atoms is significantly smaller than the LDOS of the Ge atoms in the Pt-Ge surface dimers, which explains that the experimentally observed metallicity between the wires rather than on top of the wires.\cite{Gurlu:apl03}
\begin{figure}[!tb]
\begin{center}
  \includegraphics[width=8cm,keepaspectratio]{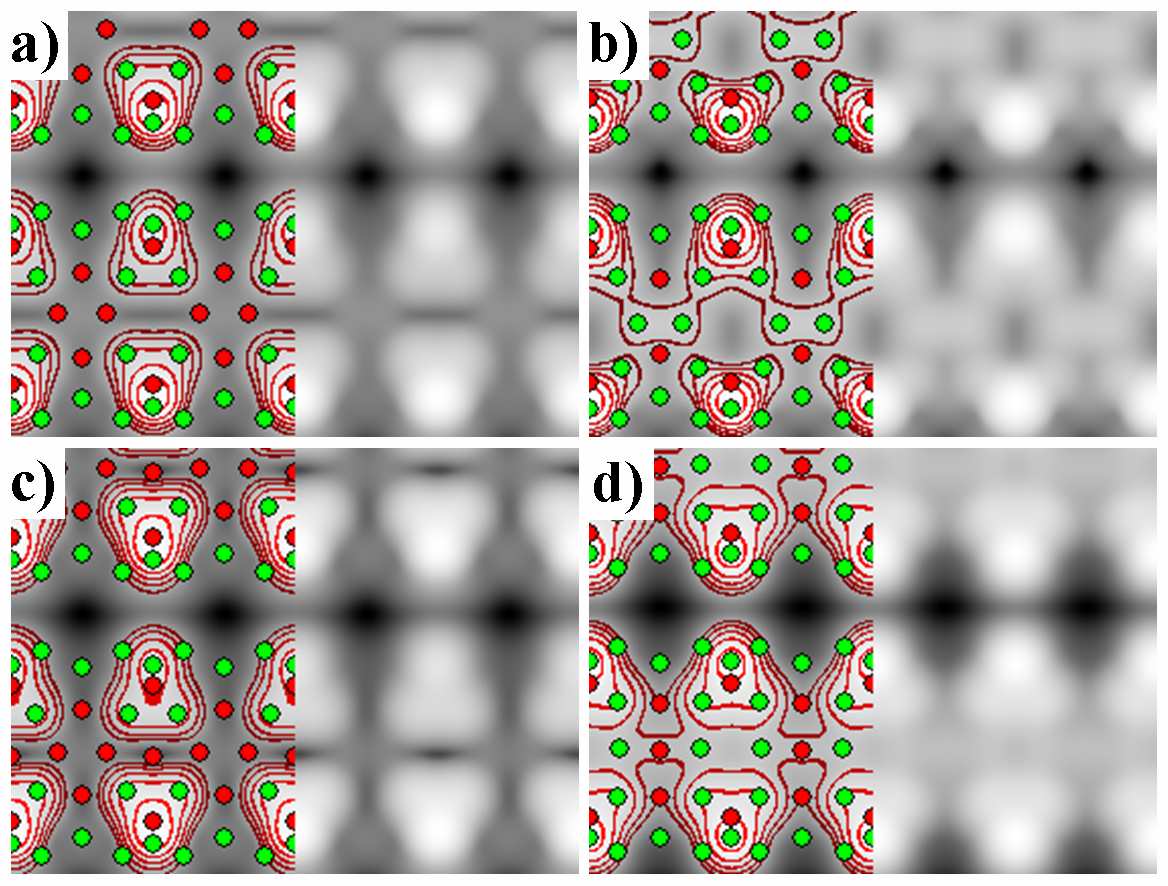}\\
\end{center}
  \caption{(color online) Calculated filled state STM images of some WT geometries indicated in Fig.~\ref{fig:10formationdiagram}, for $z=3.0$ \AA\ and a simulated bias of $-0.70$ V. The green/red (light/dark gray) discs show the positions of Ge/Pt atoms in the two top layers of the surface and the adsorbed structure. Contours are added to guide the eye and are spaced
  $0.2$ \AA\ in the z-direction.\newline
  a) $\gamma_{as2}$ + Pt NW, b) $\gamma_{as2}^{\star}$ + Ge NW,
  c) $\gamma_{as2}$ + $3$Pt atoms in the trough, d)
  $\gamma_{as2}^{\star}$ + $1$Pt at T + Ge NW.}\label{fig:13WT_STM}
\end{figure}
\subsubsection{Wide troughs as precursors to nanowires}
In Sec.~\ref{sc:results} it was indicated that there are a few geometries whose pseudo-STM images resemble the experimentally observed WTs (also indicated in Fig.~\ref{fig:10formationdiagram}) very well. Because the two proposed NW geometries require different WT structures and even multiple WT structures in a single formation path are present, we will only
look at the general properties shared by these WT geometries.\\
\indent Comparison of the structures presenting a WT pseudo-STM image shows that most have a $\gamma_{as2}$- (\textit{cf.}\ Fig.~\ref{fig:4bis_gamma_as2}) or the derived $\gamma_{as2}^{\star}$-structure. The extremely tilted surface dimers cause the Ge atoms of those dimers to stick high out of the surface. This results in the twofold increased periodicity along the QDR. Also, because of this reconstruction the apparent trough becomes much deeper, such that atoms and dimers located in this trough are not visible. They sink in too deep. Fig.~\ref{fig:13WT_STM} shows filled state images of some WT geometries indicated in Fig.~\ref{fig:10formationdiagram}. The contours show maxima directed away from the Pt lined trough giving the impression of a widening of the trough. However, small nuances are still clearly visible between the different structures. Troughs filled with Ge dimers or atoms seem to be more shallow than those filled only with Pt atoms and/or dimers. Bare $\gamma_{as2}$- and $\gamma_{as2}^{\star}$-surfaces give pseudo-STM images comparable to Fig.~\ref{fig:13WT_STM}a, but with even sharper peaks,which makes the impression of the widening of the trough even stronger. The structures shown in Fig.~\ref{fig:13WT_STM} and the bare $\gamma_{as2}^{\star}$-structure are all stable geometries with formation energies ranging from $E_{\mathrm{f}}=-1.5$ eV to $-3.5$ eV. Since all of them have a place in the proposed NW formation paths, further high precision low temperature STM experiments might be needed to identify the nuances
seen in the calculated STM images in Fig.~\ref{fig:13WT_STM}.
\begin{figure}[!tb]
\begin{center}
  \includegraphics[width=8cm,keepaspectratio]{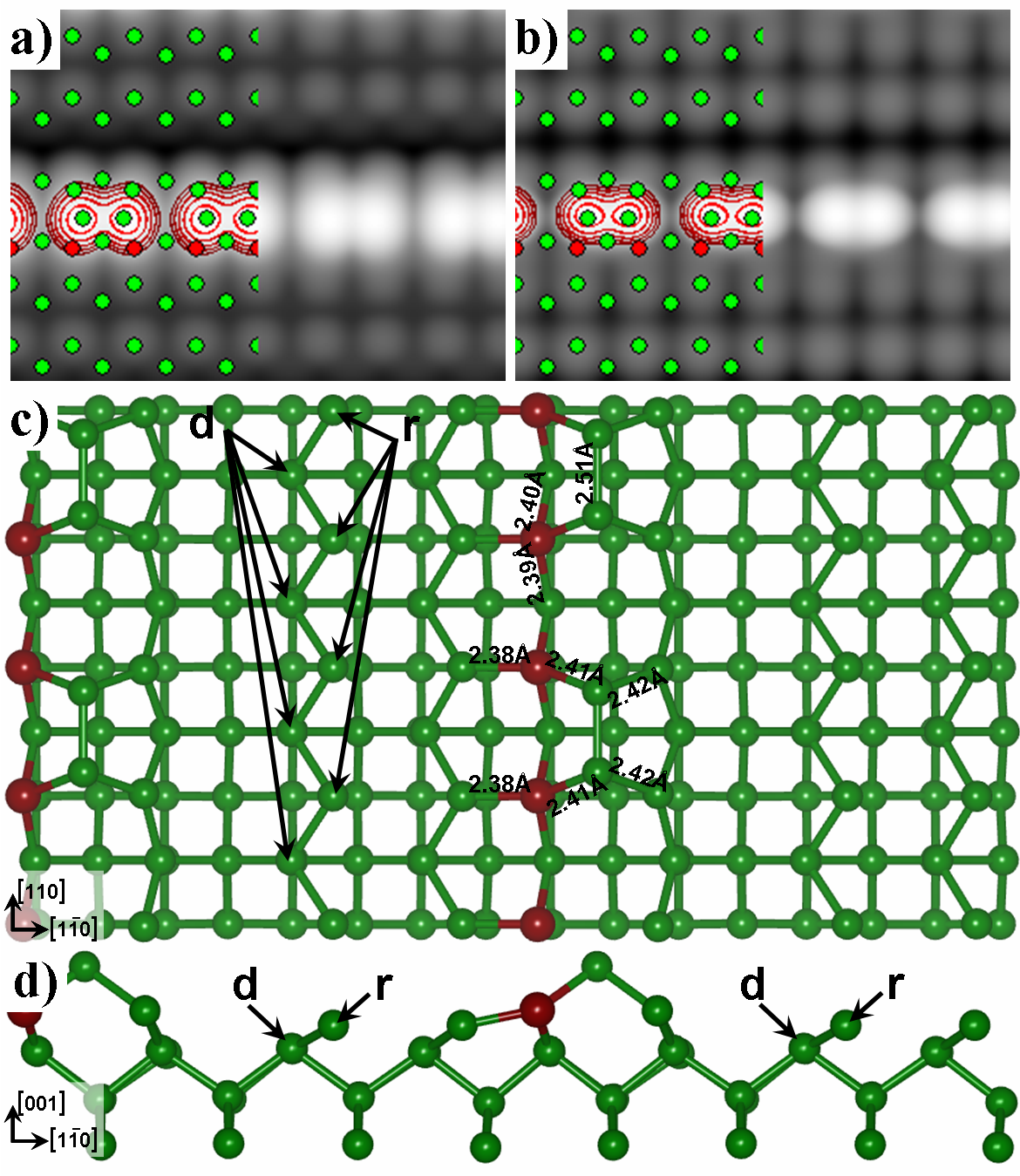}\\
\end{center}
  \caption{(color online) Calculated filled (a) and empty (b) state STM images of the Ge $b(2\times1)$ T$_{\mathrm{1d}}$ structure, for $z=3.0$ \AA\ and a
  simulated bias of $\pm$1.50 V. The green/red (light/dark gray) discs show the positions of Ge/Pt atoms in the top layers of the surface and the adsorbed
  structure. The five topmost contours are added to guide the eye. They are
  spaced $0.2$ \AA\ in the $z$-direction.\newline
  c) and d) Top and side view of the Ge $b(2\times1)$ T$_{\mathrm{1d}}$
  structure after relaxation, for comparison to figure $9$ in
  Ref.~\onlinecite{stekolnikov:prb08}. \textbf{r} and \textbf{d}
  indicate the Ge atoms that need to \textbf{d}imerize or
  be \textbf{r}emoved to have the TDC structure proposed by
  Stekolnikov \textit{et al.} Important
  bond lengths are shown for comparison.}\label{fig:14PtonGe_comp_Stekol}
\end{figure}

\subsection{Comparison to literature}\label{ssc:compar}
At the moment of writing a few models for the NWs on Ge(001) are present. The first suggested model was the one of G\"url\"u \textit{et al.}\cite{Gurlu:apl03} In their model the entire reconstruction contains $0.5$ ML of Pt atoms. $0.25$ ML of Pt are imbedded in the top layer of the Ge(001) surface creating a checkerboard pattern of Ge-Ge and Pt-Ge dimers, called the $\beta$-terrace. The remaining $0.25$ ML of Pt atoms form Pt dimers which form NWs in the troughs between the QDRs of Pt-Ge and Ge-Ge dimers. In Sec.~\ref{ssc:model1_onb6u} the first models we calculated are based on this model. Despite the simplicity of this model and the successful prediction of the $\beta$-terrace model, we clearly show in this work that this experimental model can not reproduce the experimentally observed STM images at all. This, due to the unforseen invisibility of
the Pt atoms in the trough. In addition, the energetics for NWs on a $\beta_6$ geometry are shown to be very unfavorable.\\
\indent More recently, Stekolnikov \textit{et al.} \cite{stekolnikov:prl08} published, almost simultaneously with the presenting authors,\cite{vanpoucke:prb2008R} a model where they propose a tetramer-dimer-chain (TDC) reconstruction for the studied system. Stekolnikov \textit{et al.} present a careful DFT study of a set of Ge surfaces containing $0.25$ ML of Pt. By comparing the formation energy of those structures and a comparison of calculated STM images with experimental STM images they conclude their TDC structure to represent the geometry of the observed NW arrays. Close examination of this structure, shown in Fig. $9$ of Ref.~\onlinecite{stekolnikov:prb08}, shows it to be remarkably similar to the relaxed Ge b$(2\times1)$ T$_{\mathrm{d1}}$ structure, discussed in Sec.~\ref{ssc:model1_onb6u}.\\
\indent Figures~\ref{fig:14PtonGe_comp_Stekol}a and b show the calculated filled and empty state images of our Ge b$(2\times1)$ T$_{\mathrm{d1}}$ structure. As noted by Stekolnikov \textit{et al.} a dimerised NW image is visible. However, as already noted in Sec.~\ref{ssc:model1_onb6u}, this image shows a double peak feature for both a positive and negative bias (also visible in both left columns of figure $10$ of Ref.~\onlinecite{stekolnikov:prb08}), in contradiction to the experimentally observed NWs. Furthermore, the stability of this system is unfavorable, as compared to a normal $\beta_{6}$-surface reconstruction (which also contains $0.25$ ML of Pt) by almost half an eV per surface unit cell. We find the formation energy of the Ge b$(2\times1)$ T$_{\mathrm{d1}}$ structure to be roughly $2$ eV more stable than for a structure with a Pt dimer on a Ge dimer row (\textit{cf.}\ Table~\ref{table:1formEfirstset}), in agreement with the observation of Stekolnikov \textit{et al.} Figures~\ref{fig:14PtonGe_comp_Stekol}c and d show ball and stick images of the Ge b$(2\times1)$ T$_{\mathrm{d1}}$ structure after reconstruction, indicating the necessary modifications needed to obtain the TDC structure of Stekolnikov \textit{et al.} Bond lengths are given for comparison with Fig.~$9$ in Ref.~\onlinecite{stekolnikov:prb08}. The differences between the bond lengths are of the order of $2$--$5$\%,
what could be expected when comparing LDA and GGA calculations.\\
\indent When the possible local Pt density of the NW system has to be limited to only $0.25$ ML of Pt, then this model probably gives the best agreement to the experimental STM images one can get. However quite some problems remain. The unfavorable formation energy, \textit{i.e.}\ the fact that this structure is less stable than a normal reconstructed Ge(001) surface, being the most unsettling one from a computational point of view. As can be seen in Fig.~\ref{fig:5_b_NWdomainBoundary}, the experimentally observed NW is located at the center of the trough between two QDRs, where in the TDC model proposed by Stekolnikov \textit{et al.} only at one side a QDR is present. Furthermore, the fact that only partial agreement (\textit{e.g.}\ symmetric bulges are missing and empty state pictures show double peaked NW dimer images in contrast to the experimental STM images) between experimental and calculated STM images is found also indicates this model not to be fully correct.\\
\indent Expansion to local Pt densities beyond $0.25$ ML is necessary to come to full agreement with the experiment, as we show in this paper. The experimental observation that after the deposition and annealing process a fraction of the surface is covered by $\beta$-terraces and a fraction of the $\beta$-terraces is covered with NWs combined with the model suggested for the $\beta$-terrace which already contains $0.25$ ML of Pt further suggests the evolution into larger local Pt densities.
\section{Conclusions}\label{sc:conclusion}
We have studied a large number of Pt modified Ge(001) surfaces, with Pt densities varying from $0.25$ ML up to $1.0$ ML, using \textit{ab initio} DFT calculations. Starting from simple structures with adsorbed Pt atoms and dimers on a clean Ge(001)- or $\beta_{6u}$-surface, based on the experimental observations and previous work, it was shown that these structures could not be responsible for the experimentally observed NW arrays. These calculations showed it unlikely that Pt dimers form the observed wires due to the unfavorable energetics and the large discrepancy between calculated and experimental STM images. However, the calculated STM images indicated that Ge atoms might be responsible for the observed NW images.\\
\indent In a second set of models, both Pt and Ge dimers were adsorbed on a surface reconstruction containing $0.5$ ML of Pt in the top layer. For these structures the energetics were favorable and in case of the Ge dimers also a NW image was seen in the calculated STM images. It was found that Pt dimers sink into the trough, and the possibility of Pt atoms exchanging their position with the Ge atoms located at the bottom of the trough led to a third set of models.\\
\indent The reconstructed surface for these models contained $0.75$ ML of Pt in the top layers. Again, both Pt and Ge dimers were adsorbed on these $\gamma_{as}^{\star}$- and $\gamma_{as2}^{\star}$-surfaces. The energetics of these structures are very favorable, and the calculated STM images for the Ge NW on $\gamma_{as}^{\star}$ (NW$1$) show
very good agreement with the experimentally observed NWs.\\
\indent However, one detail is still missing at this point: the experimentally observed $4\times1$ periodicity along the wires in NW-patches. To study this $4\times1$ periodicity, the obtained NW$1$ structure is modified to contain extra Pt or Ge atoms in the trough with the wire. It is found that one extra Pt atom per two surface unit cells is sufficient to introduce the observed $4\times1$ periodicity. The extra Pt atom binds to two Ge NW dimers, fixing the position of both NW dimers and tilting them to create the observed
$4\times1$ periodicity. This model for the NWs is named NW$2$.\\
\indent Reviewing the calculated structures we propose a path for the formation of NW arrays on Ge(001) as a function of the local Pt density. In this formation path, the experimentally observed WTs are linked to the $\gamma_{as2}^{(\star)}$ reconstruction. Furthermore, we conclude that the geometry of NWs inside a NW-patch is slightly different from that of the solitary NWs and the NWs at the edge of a NW-patch, as a result of a slightly different local Pt density. We propose the NW$1$-geometry as model for the solitary and patch-edge NWs, and the NW$2$-geometry as model for the NWs inside a NW patch. In all cases, the NWs adsorbed on the Surface consist of \textit{Ge atoms}.\\
\indent Finally, we make a comparison of the models present in literature and show their relation to the Pt-density dependent formation path and the NW models proposed in this paper.
\section*{Acknowledgements}
We thank Harold Zandvliet and Arie van Houselt for making their experimental STM results available and for the many stimulating discussions on this topic. This work is part of the research program of the ``Stichting voor Fundamenteel Onderzoek der Materie" (FOM) and the use of supercomputer facilities was sponsored by the "Stichting Nationale Computer Faciliteiten" (NCF), both financially supported by the ``Nederlandse Organisatie voor Wetenschappelijk Onderzoek" (NWO).



\begin{thebibliography}{48}
\expandafter\ifx\csname natexlab\endcsname\relax\def\natexlab#1{#1}\fi
\expandafter\ifx\csname bibnamefont\endcsname\relax
  \def\bibnamefont#1{#1}\fi
\expandafter\ifx\csname bibfnamefont\endcsname\relax
  \def\bibfnamefont#1{#1}\fi
\expandafter\ifx\csname citenamefont\endcsname\relax
  \def\citenamefont#1{#1}\fi
\expandafter\ifx\csname url\endcsname\relax
  \def\url#1{\texttt{#1}}\fi
\expandafter\ifx\csname urlprefix\endcsname\relax\def\urlprefix{URL }\fi
\providecommand{\bibinfo}[2]{#2}
\providecommand{\eprint}[2][]{\url{#2}}

\bibitem[{fn:({\natexlab{a}})}]{fn:Moore}
\bibinfo{note}{In the original article, a doubling every year is calculated,
  Moore later refined the period to two years, based on further development in
  the industry.}

\bibitem[{\citenamefont{Moore}(1965)}]{Moore:elec65}
\bibinfo{author}{\bibfnamefont{G.~E.} \bibnamefont{Moore}},
  \bibinfo{journal}{Electronics} \textbf{\bibinfo{volume}{38}},
  \bibinfo{pages}{8} (\bibinfo{year}{1965}).

\bibitem[{\citenamefont{Luttinger}(1963)}]{Lutt:jmp63}
\bibinfo{author}{\bibfnamefont{J.~M.} \bibnamefont{Luttinger}},
  \bibinfo{journal}{J. Math. Phys.} \textbf{\bibinfo{volume}{4}},
  \bibinfo{pages}{1154} (\bibinfo{year}{1963}).

\bibitem[{\citenamefont{Yeom et~al.}(1999)\citenamefont{Yeom, Takeda,
  Rotenberg, Matsuda, Horikoshi, Schaefer, Lee, Kevan, Ohta, Nagao
  et~al.}}]{Yeom:prl99}
\bibinfo{author}{\bibfnamefont{H.~W.} \bibnamefont{Yeom}},
  \bibinfo{author}{\bibfnamefont{S.}~\bibnamefont{Takeda}},
  \bibinfo{author}{\bibfnamefont{E.}~\bibnamefont{Rotenberg}},
  \bibinfo{author}{\bibfnamefont{I.}~\bibnamefont{Matsuda}},
  \bibinfo{author}{\bibfnamefont{K.}~\bibnamefont{Horikoshi}},
  \bibinfo{author}{\bibfnamefont{J.}~\bibnamefont{Schaefer}},
  \bibinfo{author}{\bibfnamefont{C.~M.} \bibnamefont{Lee}},
  \bibinfo{author}{\bibfnamefont{S.~D.} \bibnamefont{Kevan}},
  \bibinfo{author}{\bibfnamefont{T.}~\bibnamefont{Ohta}},
  \bibinfo{author}{\bibfnamefont{T.}~\bibnamefont{Nagao}},
  \bibnamefont{et~al.}, \bibinfo{journal}{Phys. Rev. Lett.}
  \textbf{\bibinfo{volume}{82}}, \bibinfo{pages}{4898} (\bibinfo{year}{1999}).

\bibitem[{\citenamefont{Yao et~al.}(1999)\citenamefont{Yao, Postma, Balents,
  and Dekker}}]{Yao:nat99}
\bibinfo{author}{\bibfnamefont{Z.}~\bibnamefont{Yao}},
  \bibinfo{author}{\bibfnamefont{H.~W.~C.} \bibnamefont{Postma}},
  \bibinfo{author}{\bibfnamefont{L.}~\bibnamefont{Balents}}, \bibnamefont{and}
  \bibinfo{author}{\bibfnamefont{C.}~\bibnamefont{Dekker}},
  \bibinfo{journal}{Nature} \textbf{\bibinfo{volume}{402}},
  \bibinfo{pages}{273} (\bibinfo{year}{1999}).

\bibitem[{\citenamefont{Gambardella et~al.}(2002)\citenamefont{Gambardella,
  Dallmeyer, Maiti, Malagoli, Eberhardt, Kern, and Carbone}}]{Gamb:nat02}
\bibinfo{author}{\bibfnamefont{P.}~\bibnamefont{Gambardella}},
  \bibinfo{author}{\bibfnamefont{A.}~\bibnamefont{Dallmeyer}},
  \bibinfo{author}{\bibfnamefont{K.}~\bibnamefont{Maiti}},
  \bibinfo{author}{\bibfnamefont{M.~C.} \bibnamefont{Malagoli}},
  \bibinfo{author}{\bibfnamefont{W.}~\bibnamefont{Eberhardt}},
  \bibinfo{author}{\bibfnamefont{K.}~\bibnamefont{Kern}}, \bibnamefont{and}
  \bibinfo{author}{\bibfnamefont{C.}~\bibnamefont{Carbone}},
  \bibinfo{journal}{Nature} \textbf{\bibinfo{volume}{416}},
  \bibinfo{pages}{301} (\bibinfo{year}{2002}).

\bibitem[{\citenamefont{Shen et~al.}(1997)\citenamefont{Shen, Skomski, Klaua,
  Jenniches, Manoharan, and Kirschner}}]{Shen:prb97}
\bibinfo{author}{\bibfnamefont{J.}~\bibnamefont{Shen}},
  \bibinfo{author}{\bibfnamefont{R.}~\bibnamefont{Skomski}},
  \bibinfo{author}{\bibfnamefont{M.}~\bibnamefont{Klaua}},
  \bibinfo{author}{\bibfnamefont{H.}~\bibnamefont{Jenniches}},
  \bibinfo{author}{\bibfnamefont{S.~S.} \bibnamefont{Manoharan}},
  \bibnamefont{and}
  \bibinfo{author}{\bibfnamefont{J.}~\bibnamefont{Kirschner}},
  \bibinfo{journal}{Phys. Rev. B} \textbf{\bibinfo{volume}{56}},
  \bibinfo{pages}{2340} (\bibinfo{year}{1997}).

\bibitem[{\citenamefont{Dorantes-D\'avila and Pastor}(1998)}]{DoranD:prl98}
\bibinfo{author}{\bibfnamefont{J.}~\bibnamefont{Dorantes-D\'avila}}
  \bibnamefont{and} \bibinfo{author}{\bibfnamefont{G.~M.}
  \bibnamefont{Pastor}}, \bibinfo{journal}{Phys. Rev. Lett.}
  \textbf{\bibinfo{volume}{81}}, \bibinfo{pages}{208} (\bibinfo{year}{1998}).

\bibitem[{\citenamefont{Nilius et~al.}(2002)\citenamefont{Nilius, Wallis, and
  Ho}}]{Nilius:sc02}
\bibinfo{author}{\bibfnamefont{N.}~\bibnamefont{Nilius}},
  \bibinfo{author}{\bibfnamefont{T.~M.} \bibnamefont{Wallis}},
  \bibnamefont{and} \bibinfo{author}{\bibfnamefont{W.}~\bibnamefont{Ho}},
  \bibinfo{journal}{Science} \textbf{\bibinfo{volume}{297}},
  \bibinfo{pages}{1853} (\bibinfo{year}{2002}).

\bibitem[{\citenamefont{Crain and Pierce}(2005)}]{Crain:sc05}
\bibinfo{author}{\bibfnamefont{J.~N.} \bibnamefont{Crain}} \bibnamefont{and}
  \bibinfo{author}{\bibfnamefont{D.~T.} \bibnamefont{Pierce}},
  \bibinfo{journal}{Science} \textbf{\bibinfo{volume}{307}},
  \bibinfo{pages}{703} (\bibinfo{year}{2005}).

\bibitem[{\citenamefont{Lagoute et~al.}(2006)\citenamefont{Lagoute, Liu, and
  Folsch}}]{Lagoute:prb06}
\bibinfo{author}{\bibfnamefont{J.}~\bibnamefont{Lagoute}},
  \bibinfo{author}{\bibfnamefont{X.}~\bibnamefont{Liu}}, \bibnamefont{and}
  \bibinfo{author}{\bibfnamefont{S.}~\bibnamefont{Folsch}},
  \bibinfo{journal}{Phys. Rev. B} \textbf{\bibinfo{volume}{74}},
  \bibinfo{pages}{125410} (\bibinfo{year}{2006}).

\bibitem[{\citenamefont{Lim et~al.}(2007)\citenamefont{Lim, Lee, Lee, Bae,
  Choi, Kim, Ji, Ragan, Ohlberg, Chang et~al.}}]{LimDK:nano07}
\bibinfo{author}{\bibfnamefont{D.~K.} \bibnamefont{Lim}},
  \bibinfo{author}{\bibfnamefont{D.}~\bibnamefont{Lee}},
  \bibinfo{author}{\bibfnamefont{H.}~\bibnamefont{Lee}},
  \bibinfo{author}{\bibfnamefont{S.}~\bibnamefont{Bae}},
  \bibinfo{author}{\bibfnamefont{J.}~\bibnamefont{Choi}},
  \bibinfo{author}{\bibfnamefont{S.}~\bibnamefont{Kim}},
  \bibinfo{author}{\bibfnamefont{C.}~\bibnamefont{Ji}},
  \bibinfo{author}{\bibfnamefont{R.}~\bibnamefont{Ragan}},
  \bibinfo{author}{\bibfnamefont{D.~A.~A.} \bibnamefont{Ohlberg}},
  \bibinfo{author}{\bibfnamefont{Y.~A.} \bibnamefont{Chang}},
  \bibnamefont{et~al.}, \bibinfo{journal}{Nanotechnology}
  \textbf{\bibinfo{volume}{18}}, \bibinfo{pages}{095706}
  (\bibinfo{year}{2007}).

\bibitem[{\citenamefont{Hong}(2007)}]{Hong:prb07}
\bibinfo{author}{\bibfnamefont{J.}~\bibnamefont{Hong}}, \bibinfo{journal}{Phys.
  Rev. B} \textbf{\bibinfo{volume}{76}}, \bibinfo{pages}{092403}
  (\bibinfo{year}{2007}).

\bibitem[{\citenamefont{Eigler and Schweizer}(1990)}]{Eigler:nat90}
\bibinfo{author}{\bibfnamefont{D.~M.} \bibnamefont{Eigler}} \bibnamefont{and}
  \bibinfo{author}{\bibfnamefont{E.~K.} \bibnamefont{Schweizer}},
  \bibinfo{journal}{Nature} \textbf{\bibinfo{volume}{344}},
  \bibinfo{pages}{524} (\bibinfo{year}{1990}).

\bibitem[{\citenamefont{Yanson et~al.}(1998)\citenamefont{Yanson, Bollinger,
  van~den Brom, Agrait, and van Ruitenbeek}}]{Yanson:nat98}
\bibinfo{author}{\bibfnamefont{A.~I.} \bibnamefont{Yanson}},
  \bibinfo{author}{\bibfnamefont{G.~R.} \bibnamefont{Bollinger}},
  \bibinfo{author}{\bibfnamefont{H.~E.} \bibnamefont{van~den Brom}},
  \bibinfo{author}{\bibfnamefont{N.}~\bibnamefont{Agrait}}, \bibnamefont{and}
  \bibinfo{author}{\bibfnamefont{J.~M.} \bibnamefont{van Ruitenbeek}},
  \bibinfo{journal}{Nature} \textbf{\bibinfo{volume}{395}},
  \bibinfo{pages}{783} (\bibinfo{year}{1998}).

\bibitem[{\citenamefont{G{\"u}rl{\"u} et~al.}(2003)\citenamefont{G{\"u}rl{\"u},
  Adam, Zandvliet, and Poelsema}}]{Gurlu:apl03}
\bibinfo{author}{\bibfnamefont{O.}~\bibnamefont{G{\"u}rl{\"u}}},
  \bibinfo{author}{\bibfnamefont{O.~A.~O.} \bibnamefont{Adam}},
  \bibinfo{author}{\bibfnamefont{H.~J.~W.} \bibnamefont{Zandvliet}},
  \bibnamefont{and} \bibinfo{author}{\bibfnamefont{B.}~\bibnamefont{Poelsema}},
  \bibinfo{journal}{Appl. Phys. Lett.} \textbf{\bibinfo{volume}{83}},
  \bibinfo{pages}{4610} (\bibinfo{year}{2003}).

\bibitem[{\citenamefont{Wang et~al.}(2004)\citenamefont{Wang, Li, and
  Altman}}]{Wang:prb04}
\bibinfo{author}{\bibfnamefont{J.}~\bibnamefont{Wang}},
  \bibinfo{author}{\bibfnamefont{M.}~\bibnamefont{Li}}, \bibnamefont{and}
  \bibinfo{author}{\bibfnamefont{E.~I.} \bibnamefont{Altman}},
  \bibinfo{journal}{Phys. Rev. B} \textbf{\bibinfo{volume}{70}},
  \bibinfo{pages}{233312} (\bibinfo{year}{2004}).

\bibitem[{\citenamefont{Wang et~al.}(2005)\citenamefont{Wang, Li, and
  Altman}}]{Wang:ss05}
\bibinfo{author}{\bibfnamefont{J.}~\bibnamefont{Wang}},
  \bibinfo{author}{\bibfnamefont{M.}~\bibnamefont{Li}}, \bibnamefont{and}
  \bibinfo{author}{\bibfnamefont{E.~I.} \bibnamefont{Altman}},
  \bibinfo{journal}{Surf. Sci.} \textbf{\bibinfo{volume}{596}},
  \bibinfo{pages}{126} (\bibinfo{year}{2005}).

\bibitem[{\citenamefont{Barth et~al.}(2005)\citenamefont{Barth, Constantini,
  and Kern}}]{Barth:nat05}
\bibinfo{author}{\bibfnamefont{J.~V.} \bibnamefont{Barth}},
  \bibinfo{author}{\bibfnamefont{G.}~\bibnamefont{Constantini}},
  \bibnamefont{and} \bibinfo{author}{\bibfnamefont{K.}~\bibnamefont{Kern}},
  \bibinfo{journal}{Nature} \textbf{\bibinfo{volume}{437}},
  \bibinfo{pages}{671} (\bibinfo{year}{2005}), \bibinfo{note}{and references
  therein.}

\bibitem[{\citenamefont{Eames et~al.}(2006)\citenamefont{Eames, Bonet, Probert,
  Tear, and Perkins}}]{Eames:prb06}
\bibinfo{author}{\bibfnamefont{C.}~\bibnamefont{Eames}},
  \bibinfo{author}{\bibfnamefont{C.}~\bibnamefont{Bonet}},
  \bibinfo{author}{\bibfnamefont{M.~I.~J.} \bibnamefont{Probert}},
  \bibinfo{author}{\bibfnamefont{S.~P.} \bibnamefont{Tear}}, \bibnamefont{and}
  \bibinfo{author}{\bibfnamefont{E.~W.} \bibnamefont{Perkins}},
  \bibinfo{journal}{Phys. Rev. B} \textbf{\bibinfo{volume}{74}},
  \bibinfo{eid}{193318} (\bibinfo{year}{2006}).

\bibitem[{\citenamefont{Sch{\"a}fer et~al.}(2008)\citenamefont{Sch{\"a}fer,
  Blumenstein, Meyer, Wisniewski, and Claessen}}]{Schafer:prl2008}
\bibinfo{author}{\bibfnamefont{J.}~\bibnamefont{Sch{\"a}fer}},
  \bibinfo{author}{\bibfnamefont{C.}~\bibnamefont{Blumenstein}},
  \bibinfo{author}{\bibfnamefont{S.}~\bibnamefont{Meyer}},
  \bibinfo{author}{\bibfnamefont{M.}~\bibnamefont{Wisniewski}},
  \bibnamefont{and} \bibinfo{author}{\bibfnamefont{R.}~\bibnamefont{Claessen}},
  \bibinfo{journal}{Phys. Rev. Lett.} \textbf{\bibinfo{volume}{101}},
  \bibinfo{eid}{236802} (pages~\bibinfo{numpages}{4}) (\bibinfo{year}{2008}).

\bibitem[{\citenamefont{Sch{\"a}fer et~al.}(2006)\citenamefont{Sch{\"a}fer,
  Schrupp, Preisinger, and Claessen}}]{Schafer:prb06}
\bibinfo{author}{\bibfnamefont{J.}~\bibnamefont{Sch{\"a}fer}},
  \bibinfo{author}{\bibfnamefont{D.}~\bibnamefont{Schrupp}},
  \bibinfo{author}{\bibfnamefont{M.}~\bibnamefont{Preisinger}},
  \bibnamefont{and} \bibinfo{author}{\bibfnamefont{R.}~\bibnamefont{Claessen}},
  \bibinfo{journal}{Phys. Rev. B} \textbf{\bibinfo{volume}{74}},
  \bibinfo{eid}{041404(R)} (pages~\bibinfo{numpages}{4}) (\bibinfo{year}{2006}).

\bibitem[{\citenamefont{Vanpoucke and Brocks}(2008)}]{vanpoucke:prb2008R}
\bibinfo{author}{\bibfnamefont{D.~E.~P.} \bibnamefont{Vanpoucke}}
  \bibnamefont{and} \bibinfo{author}{\bibfnamefont{G.}~\bibnamefont{Brocks}},
  \bibinfo{journal}{Phys. Rev. B} \textbf{\bibinfo{volume}{77}},
  \bibinfo{eid}{241308(R)} (pages~\bibinfo{numpages}{4}) (\bibinfo{year}{2008}).

\bibitem[{\citenamefont{Vanpoucke and Brocks}(2009)}]{Vanpoucke:prb09beta}
\bibinfo{author}{\bibfnamefont{D.~E.~P.} \bibnamefont{Vanpoucke}}
  \bibnamefont{and} \bibinfo{author}{\bibfnamefont{G.}~\bibnamefont{Brocks}},
 (\bibinfo{year}{2009}), \bibinfo{note}{submitted}.

\bibitem[{\citenamefont{Bl{\"o}chl}(1994)}]{Blochl:prb94}
\bibinfo{author}{\bibfnamefont{P.~E.} \bibnamefont{Bl{\"o}chl}},
  \bibinfo{journal}{Phys. Rev. B} \textbf{\bibinfo{volume}{50}},
  \bibinfo{pages}{17953} (\bibinfo{year}{1994}).

\bibitem[{\citenamefont{Kresse and Joubert}(1999)}]{Kresse:prb99}
\bibinfo{author}{\bibfnamefont{G.}~\bibnamefont{Kresse}} \bibnamefont{and}
  \bibinfo{author}{\bibfnamefont{D.}~\bibnamefont{Joubert}},
  \bibinfo{journal}{Phys. Rev. B} \textbf{\bibinfo{volume}{59}},
  \bibinfo{pages}{1758} (\bibinfo{year}{1999}).

\bibitem[{\citenamefont{Kresse and Hafner}(1993)}]{Kresse:prb93}
\bibinfo{author}{\bibfnamefont{G.}~\bibnamefont{Kresse}} \bibnamefont{and}
  \bibinfo{author}{\bibfnamefont{J.}~\bibnamefont{Hafner}},
  \bibinfo{journal}{Phys. Rev. B} \textbf{\bibinfo{volume}{47}},
  \bibinfo{pages}{558} (\bibinfo{year}{1993}).

\bibitem[{\citenamefont{Kresse and Furthm\"uller}(1996)}]{Kresse:prb96}
\bibinfo{author}{\bibfnamefont{G.}~\bibnamefont{Kresse}} \bibnamefont{and}
  \bibinfo{author}{\bibfnamefont{J.}~\bibnamefont{Furthm\"uller}},
  \bibinfo{journal}{Phys. Rev. B} \textbf{\bibinfo{volume}{54}},
  \bibinfo{pages}{11169} (\bibinfo{year}{1996}).

\bibitem[{\citenamefont{Monkhorst and Pack}(1976)}]{Monkhorst:prb76}
\bibinfo{author}{\bibfnamefont{H.~J.} \bibnamefont{Monkhorst}}
  \bibnamefont{and} \bibinfo{author}{\bibfnamefont{J.~D.} \bibnamefont{Pack}},
  \bibinfo{journal}{Phys. Rev. B} \textbf{\bibinfo{volume}{13}},
  \bibinfo{pages}{5188} (\bibinfo{year}{1976}).

\bibitem[{\citenamefont{Tersoff and Hamann}(1985)}]{Tersoff:prb85}
\bibinfo{author}{\bibfnamefont{J.}~\bibnamefont{Tersoff}} \bibnamefont{and}
  \bibinfo{author}{\bibfnamefont{D.~R.} \bibnamefont{Hamann}},
  \bibinfo{journal}{Phys. Rev. B} \textbf{\bibinfo{volume}{31}},
  \bibinfo{pages}{805} (\bibinfo{year}{1985}).

\bibitem[{\citenamefont{{\"O}ncel et~al.}(2005)\citenamefont{{\"O}ncel, van
  Houselt, Huijben, Hallback, G{\"u}rl{\"u}, Zandvliet, and
  Poelsema}}]{Oncel:prl05}
\bibinfo{author}{\bibfnamefont{N.}~\bibnamefont{{\"O}ncel}},
  \bibinfo{author}{\bibfnamefont{A.}~\bibnamefont{van Houselt}},
  \bibinfo{author}{\bibfnamefont{J.}~\bibnamefont{Huijben}},
  \bibinfo{author}{\bibfnamefont{A.~S.} \bibnamefont{Hallback}},
  \bibinfo{author}{\bibfnamefont{O.}~\bibnamefont{G{\"u}rl{\"u}}},
  \bibinfo{author}{\bibfnamefont{H.~J.~W.} \bibnamefont{Zandvliet}},
  \bibnamefont{and} \bibinfo{author}{\bibfnamefont{B.}~\bibnamefont{Poelsema}},
  \bibinfo{journal}{Phys. Rev. Lett.} \textbf{\bibinfo{volume}{95}},
  \bibinfo{pages}{116801} (\bibinfo{year}{2005}).

\bibitem[{\citenamefont{de~Vries et~al.}(2008)\citenamefont{de~Vries, Saedi,
  Kockmann, van Houselt, Poelsema, and Zandvliet}}]{Vriesde:apl2008}
\bibinfo{author}{\bibfnamefont{R.~J.} \bibnamefont{de~Vries}},
  \bibinfo{author}{\bibfnamefont{A.}~\bibnamefont{Saedi}},
  \bibinfo{author}{\bibfnamefont{D.}~\bibnamefont{Kockmann}},
  \bibinfo{author}{\bibfnamefont{A.}~\bibnamefont{van Houselt}},
  \bibinfo{author}{\bibfnamefont{B.}~\bibnamefont{Poelsema}}, \bibnamefont{and}
  \bibinfo{author}{\bibfnamefont{H.~J.~W.} \bibnamefont{Zandvliet}},
  \bibinfo{journal}{Appl. Phys. Lett.} \textbf{\bibinfo{volume}{92}},
  \bibinfo{pages}{174101} (\bibinfo{year}{2008}).

\bibitem[{\citenamefont{van Houselt et~al.}(2008)\citenamefont{van Houselt,
  Gnielka, Aan~de Brugh, {\"O}ncel, Kockmann, Heid, Bohnen, Poelsema, and
  Zandvliet}}]{Houselt:ss08}
\bibinfo{author}{\bibfnamefont{A.}~\bibnamefont{van Houselt}},
  \bibinfo{author}{\bibfnamefont{T.}~\bibnamefont{Gnielka}},
  \bibinfo{author}{\bibfnamefont{J.~M.~J.} \bibnamefont{Aan~de Brugh}},
  \bibinfo{author}{\bibfnamefont{N.}~\bibnamefont{{\"O}ncel}},
  \bibinfo{author}{\bibfnamefont{D.}~\bibnamefont{Kockmann}},
  \bibinfo{author}{\bibfnamefont{R.}~\bibnamefont{Heid}},
  \bibinfo{author}{\bibfnamefont{K.~P.} \bibnamefont{Bohnen}},
  \bibinfo{author}{\bibfnamefont{B.}~\bibnamefont{Poelsema}}, \bibnamefont{and}
  \bibinfo{author}{\bibfnamefont{H.}~\bibnamefont{Zandvliet}},
  \bibinfo{journal}{Surf. Sci.} \textbf{\bibinfo{volume}{602}},
  \bibinfo{pages}{1731} (\bibinfo{year}{2008}).

\bibitem[{\citenamefont{Zandvliet}()}]{Zandvliet:privComm}
\bibinfo{author}{\bibfnamefont{H.~J.~W.} \bibnamefont{Zandvliet}},
  \bibinfo{howpublished}{private communication}.

\bibitem[{\citenamefont{Fischer et~al.}(2007)\citenamefont{Fischer, van
  Houselt, Kockmann, Poelsema, and Zandvliet}}]{Fischer:prb07}
\bibinfo{author}{\bibfnamefont{M.}~\bibnamefont{Fischer}},
  \bibinfo{author}{\bibfnamefont{A.}~\bibnamefont{van Houselt}},
  \bibinfo{author}{\bibfnamefont{D.}~\bibnamefont{Kockmann}},
  \bibinfo{author}{\bibfnamefont{B.}~\bibnamefont{Poelsema}}, \bibnamefont{and}
  \bibinfo{author}{\bibfnamefont{H.~J.~W.} \bibnamefont{Zandvliet}},
  \bibinfo{journal}{Phys. Rev. B} \textbf{\bibinfo{volume}{76}},
  \bibinfo{eid}{245429} (pages~\bibinfo{numpages}{5}) (\bibinfo{year}{2007}).

\bibitem[{fn:({\natexlab{b}})}]{fn:Beta4Ge}
\bibinfo{note}{Replacing the Pt atoms by Ge atoms in the surface dimers of the
  \bsu-geometry shown in Fig.~\ref{fig:1geomonbeta6andge} gives the Ge($2\times
  1$) surface reconstruction.}

\bibitem[{\citenamefont{Niranjan et~al.}(2007)\citenamefont{Niranjan, Kleinman,
  and Demkov}}]{Niranjan:prb07}
\bibinfo{author}{\bibfnamefont{M.~K.} \bibnamefont{Niranjan}},
  \bibinfo{author}{\bibfnamefont{L.}~\bibnamefont{Kleinman}}, \bibnamefont{and}
  \bibinfo{author}{\bibfnamefont{A.~A.} \bibnamefont{Demkov}},
  \bibinfo{journal}{Phys. Rev. B} \textbf{\bibinfo{volume}{75}},
  \bibinfo{pages}{085326} (\bibinfo{year}{2007}).

\bibitem[{\citenamefont{van Houselt}(2008)}]{Houselt:thesis}
\bibinfo{author}{\bibfnamefont{A.}~\bibnamefont{van Houselt}}, Ph.D. thesis,
  \bibinfo{school}{University of Twente} (\bibinfo{year}{2008}).

\bibitem[{\citenamefont{Stekolnikov
  et~al.}(2008{\natexlab{a}})\citenamefont{Stekolnikov, Bechstedt, Wisniewski,
  Schafer, and Claessen}}]{stekolnikov:prl08}
\bibinfo{author}{\bibfnamefont{A.~A.} \bibnamefont{Stekolnikov}},
  \bibinfo{author}{\bibfnamefont{F.}~\bibnamefont{Bechstedt}},
  \bibinfo{author}{\bibfnamefont{M.}~\bibnamefont{Wisniewski}},
  \bibinfo{author}{\bibfnamefont{J.}~\bibnamefont{Schafer}}, \bibnamefont{and}
  \bibinfo{author}{\bibfnamefont{R.}~\bibnamefont{Claessen}},
  \bibinfo{journal}{Phys. Rev. Lett.} \textbf{\bibinfo{volume}{100}},
  \bibinfo{eid}{196101} (pages~\bibinfo{numpages}{4})
  (\bibinfo{year}{2008}{\natexlab{a}}).

\bibitem[{fn:({\natexlab{c}})}]{fn:throughflip}
\bibinfo{note}{It is important to note that what is described here is a static
  situation and not a dynamic process of a NW jumping from one trough to
  another and back. Since this NW flip is only a small part of the boundary
  between different NW arrays, we will not refer to it as boundary to prevent
  confusion.}

\bibitem[{\citenamefont{Massalski and Okamoto}(1990)}]{Massalski:BAPD90}
\bibinfo{author}{\bibfnamefont{T.~B.} \bibnamefont{Massalski}}
  \bibnamefont{and} \bibinfo{author}{\bibfnamefont{H.}~\bibnamefont{Okamoto}},
  \emph{\bibinfo{title}{Binary Alloy Phase Diagrams}}, vol.
  \bibinfo{volume}{1--3} (\bibinfo{publisher}{ASM},
  \bibinfo{address}{Cleveland}, \bibinfo{year}{1990}).

\bibitem[{\citenamefont{Zeng and Elsayed-Ali}(2002)}]{Zeng:ss02}
\bibinfo{author}{\bibfnamefont{X.}~\bibnamefont{Zeng}} \bibnamefont{and}
  \bibinfo{author}{\bibfnamefont{H.~E.} \bibnamefont{Elsayed-Ali}},
  \bibinfo{journal}{Surf. Sci.} \textbf{\bibinfo{volume}{497}},
  \bibinfo{pages}{373} (\bibinfo{year}{2002}).

\bibitem[{fn:({\natexlab{d}})}]{fn:GGAcompar}
\bibinfo{note}{In case a {GGA} functional is used, the $\gamma_{as}$- and
  $\gamma_{as}^{\star}$-surfaces are more stable than the $\gamma_{as2}$- and
  $\gamma_{as2}^{\star}$-surfaces As a consequence, the
  $\gamma_{as2}^{\star}\rightarrow\gamma_{as}^{\star}$ transformation in the
  formation path described is energetically favorable. Since {LDA} is known to
  overbind, and {GGA} to underbind, the real energy values will be in between
  those of the {LDA} and {GGA} calculations. For this reason we have placed the
  $\gamma_{as2}$ and $\gamma_{as2}^{\star}$ intermediate steps between brackets
  in Fig.~\ref{fig:10formationdiagram}. In the text we will follow the {LDA}
  values as a worst case scenario, and try to explain the existence of the
  experimentally observed NWs within the framework of these calculations.}

\bibitem[{\citenamefont{van Houselt et~al.}(2006)\citenamefont{van Houselt,
  {\"O}ncel, Poelsema, and Zandvliet}}]{Houselt:nanol06}
\bibinfo{author}{\bibfnamefont{A.}~\bibnamefont{van Houselt}},
  \bibinfo{author}{\bibfnamefont{N.}~\bibnamefont{{\"O}ncel}},
  \bibinfo{author}{\bibfnamefont{B.}~\bibnamefont{Poelsema}}, \bibnamefont{and}
  \bibinfo{author}{\bibfnamefont{H.}~\bibnamefont{Zandvliet}},
  \bibinfo{journal}{Nano Letters} \textbf{\bibinfo{volume}{6}},
  \bibinfo{pages}{1439} (\bibinfo{year}{2006}).

\bibitem[{fn:({\natexlab{e}})}]{fn:PerSurfUCell}
\bibinfo{note}{The number of atoms mentioned here is each time per $4\times2$
  surface unit cell.}

\bibitem[{fn:({\natexlab{f}})}]{fn:GGAsame}
\bibinfo{note}{The same is true in case GGA functionals are used.}

\bibitem[{\citenamefont{Stekolnikov
  et~al.}(2008{\natexlab{b}})\citenamefont{Stekolnikov, Furthm{\"u}ller, and
  Bechstedt}}]{stekolnikov:prb08}
\bibinfo{author}{\bibfnamefont{A.~A.} \bibnamefont{Stekolnikov}},
  \bibinfo{author}{\bibfnamefont{J.}~\bibnamefont{Furthm{\"u}ller}},
  \bibnamefont{and}
  \bibinfo{author}{\bibfnamefont{F.}~\bibnamefont{Bechstedt}},
  \bibinfo{journal}{Phys. Rev. B} \textbf{\bibinfo{volume}{78}},
  \bibinfo{pages}{155434} (\bibinfo{year}{2008}{\natexlab{b}}).

\end{thebibliography}

\end{document}